\documentclass[aps,prd,nofootinbib,superscriptaddress,onecolumn,notitlepage,preprintnumbers]{revtex4-1}

\usepackage{mathtools}
\usepackage{bbm}
\usepackage{appendix}
\usepackage{amsmath,graphicx,verbatim,epsfig}
\usepackage{amsfonts}
\usepackage{mathrsfs}  
\usepackage{epstopdf}
\usepackage[utf8]{inputenc}
\usepackage{amsmath,graphicx}
\usepackage{simplewick}
\usepackage{datetime}
\usepackage[colorlinks=true,linkcolor=blue,citecolor=blue]{hyperref}
\usepackage[usenames,dvipsnames]{color}
\usepackage{multirow}
\usepackage{subcaption}
\usepackage{tikz}
\usetikzlibrary{positioning}

\newcommand{\vect}[1]{\boldsymbol{#1}} 
\newcommand{\timeord}{{\cal T}}
\newcommand{\antitimeord}{{\cal \overline{T}}}
\newcommand{\colav}{\frac{\text{Tr}_c}{N_c}}

\newcommand{\eps}{\varepsilon}

\newcommand{\be}{\begin{equation}}
\newcommand{\ee}{\end{equation}}
\newcommand{\bea}{\begin{eqnarray}}
\newcommand{\eea}{\end{eqnarray}}
\newcommand{\balign}{\begin{align}}
\newcommand{\ealign}{\end{align}}

\newcommand\eqdef{\stackrel{\mathclap{\tiny\mbox{.}}}{=}}

\newcommand{\bg}{\begin{gather}}
\newcommand{\foma}{\end{gather}}

\newcommand{\noopsort}[1]{}

\def\slash{\rlap{/}}

\def\<{\langle}
\def\>{\rangle}

\def\g{\gamma}

\def\({\left(}
\def\[{\left[}
\def\){\right)}
\def\]{\right]}

\def\cos{\hbox{cos}}
\def\sin{\hbox{sin}}
\def\ln{\hbox{ln}}

\def\Tr{\hbox{Tr}}

\newcommand{\ben}{\begin{eqnarray}}
\newcommand{\een}{\end{eqnarray}}

\newcommand{\bef}{\begin{figure}[htb]\centering}
\newcommand{\eef}{\end{figure}}

\pdfoutput=1 

\begin{document}


\title{Introduction to transverse momentum imaging}

\author{Andrea Signori}
\thanks{Electronic address: andrea.signori@unito.it - \href{https://orcid.org/0000-0001-6640-9659}{ORCID: 0000-0001-6640-9659}} 
\affiliation{Physics Department, University of Turin, via Pietro Giuria 1, I-10133 Turin, Italy}
\affiliation{INFN, Sezione di Torino, via Pietro Giuria 1, I-10133 Turin, Italy}

\date{\today}



\begin{abstract}
This set of notes complements the lectures and recitation sessions discussed in the following graduate schools: 
HUGS at Jefferson Lab (years 2018, 2019, 2021), 
the International School and Workshop on Probing Hadron Structure at the Electron-Ion Collider at ICTS (2024), 
Frontiers in Nuclear and Hadronic Physics at GGI (2025), 
and the International Workshop and School on Hadron Structure and Strong Interactions at Nanjing University (2025). 
\end{abstract}

\maketitle


\tableofcontents

\clearpage
\part{Introduction}
\label{p:intro}

These lecture notes are essentially a written version of the lectures and recitation sessions that were delivered at the 
\href{https://www.jlab.org/stem/hugs}{HUGS} program at Jefferson Lab (years 2018, 2019, 2021), 
the \href{https://icts.res.in/program/QEICIII2024}{International School and Workshop on Probing Hadron Structure at the Electron-Ion Collider} at ICTS (2024), 
\href{https://www.ggi.infn.it/showevent.pl?id=494}{Frontiers in Nuclear and Hadronic Physics} at GGI (2025), 
and the \href{https://indico.ihep.ac.cn/event/27023/}{International Workshop and School on Hadron Structure and Strong Interactions} at Nanjing University (2025). \\

The aim of these notes is not to provide a comprehensive overview of the topic of hadron tomography in momentum space, but to provide a guide, a map, for the newcomers to find an accessible entry point and position themselves across the vast literature and research directions within this exciting research field. Training the next generation of scientists is an essential step in moving science forward. Often graduate students learn directly from research articles, which are rarely written in a pedagogical way. In this respect, providing an entry-level manual to tackle the field of hadronic imaging in momentum space is a valuable effort.  \\

The lecture notes are organized as follows. 
In Part~\ref{p:intro} we briefly introduce the distribution functions that encode information on the structure of hadrons and a selection of suggested references is given. 
In Part~\ref{p:breaking} we present the process of Deep-Inelastic Scattering (DIS) and the response functions used to parametrize the structure of the target.
In Part~\ref{p:non_coll_partons} we introduce transverse momentum distributions and discuss Semi-Inclusive DIS, as a tool to investigate the dependence on partonic transverse momentum in the description of the hadronic target and of the produced hadron.
In Part~\ref{p:symmetries} we discuss the interplay between the gauge and the discrete symmetries which give rise to spin-momentum correlations that obey generalized universality properties.
In Part~\ref{p:pheno} we briefly present some of the steps needed to establish an imaging program relying on perturbation theory, namely the concept of TMD evolution and the relevance of the emerging non-perturbative effects in TMD observables.

\section{The hadron structure landscape}
\label{s:landscape}

\begin{figure}[hb]
\begin{center}
  \includegraphics[width=0.4\textwidth]{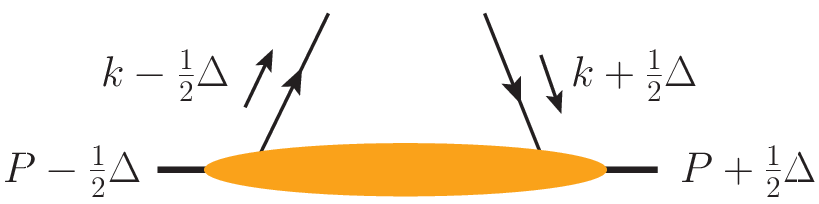}
\end{center}
\caption{Graphical representation of the parton correlation function $H(k,P,\Delta)$ defined in Eq.~\eqref{e:pcf}. The labels indicate the momentum assignments for the quarks and the hadrons.}
\label{f:corr}
\end{figure}

\begin{figure}
\centering
\includegraphics[width=0.8\textwidth]{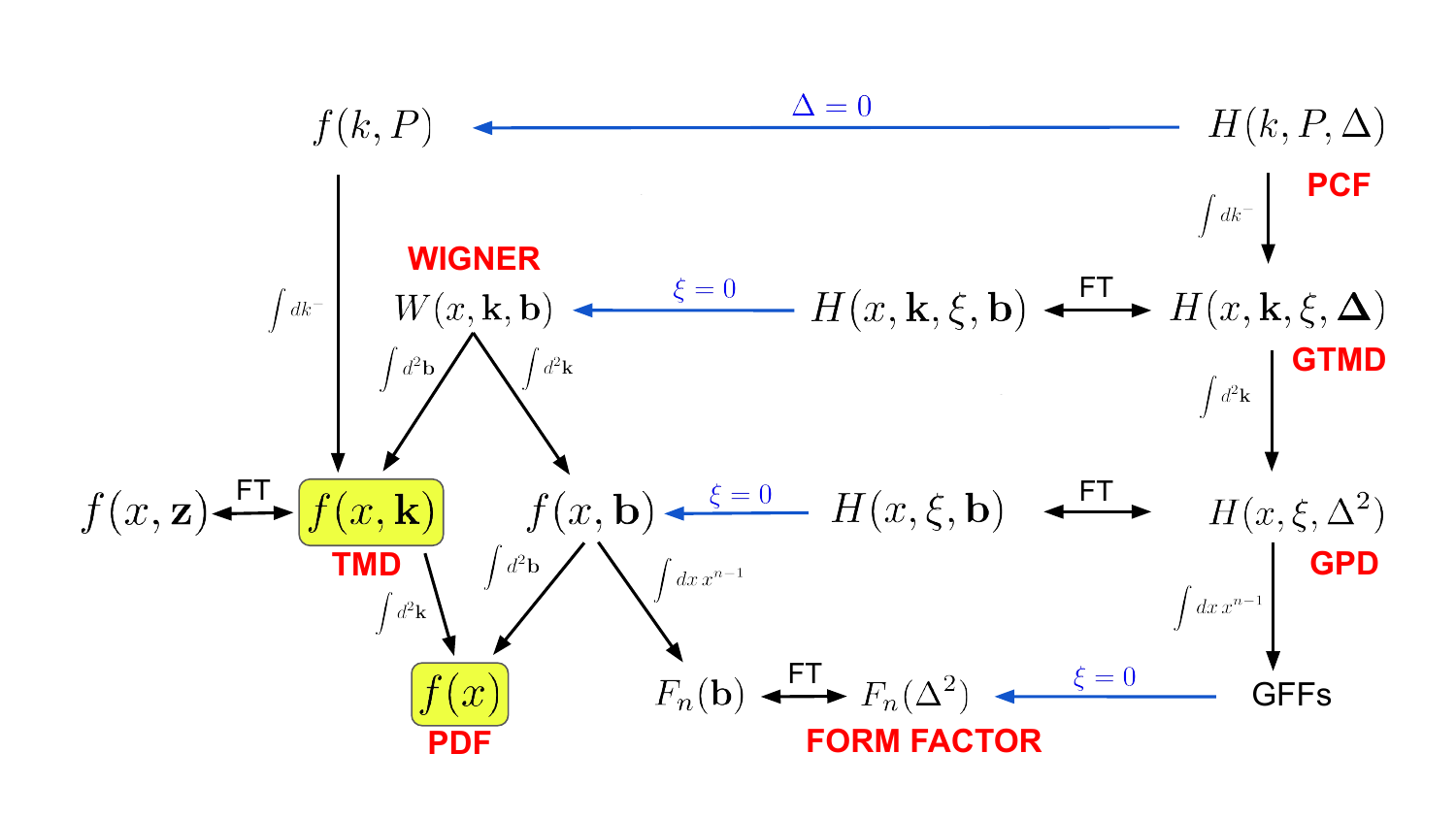}
\caption{Selected quantities that can be derived from the parton correlation function (PCF) $H(k,P,\Delta)$ defined in Eq.~\eqref{e:pcf}. Double arrows marked by ``FT'' denote a Fourier transform between $\vect{\Delta}$ and $\vect{b}$ or between $\vect{k}$ and $\vect{z}$. Fractions of plus-momentum (commonly called ``longitudinal momentum fractions'') are written as $x = k^+/P^+$ and $2\xi = -\Delta^+/P^+$. 
The scales introduced by ultraviolet renormalization or by the regulation of rapidity divergences are suppressed. As discussed in the text, the integrals cannot be taken literally but must be supplemented with a regularization procedure. Figure inspired by Ref.~\cite{Diehl:2015uka}.
Acronyms: PCF (parton correlation function), GTMD (generalized transverse momentum distribution), GPD (generalized parton distribution), GFF (generalized form factor), TMD (transverse momentum distribution), PDF (parton distribution function).}
\label{f:landscape}
\end{figure}

In these lecture notes we will focus mostly on collinear and transverse-momentum-dependent distribution functions and fragmentation functions. 
However, these quantities are part of a much broader and richer landscape of functions that characterize hadron structure and hadronization within the framework of Quantum Chromodynamics (QCD). 
At the most general level, one can introduce generalized parton correlation functions (see Eq.~\eqref{e:pcf}), which depend simultaneously on the partonic four momentum $k$, the four momentum $P$ of the hadronic target, and the four momentum $\Delta$ transferred to the final-state hadron. 
For a two-quark correlation function the definition reads~\cite{Diehl:2015uka}:
\begin{equation}
\label{e:pcf}
H(k, P, \Delta) = \int \frac{d^4z}{(2\pi)^{4}}\,
e^{i k z} \, 
\langle P + \Delta/2) |\,
\overline{\psi}(-z/2)\, \Gamma\, \psi(z/2)\, 
| P - \Delta/2 \rangle \, ,
\end{equation}
where the ket and bra are momentum eigenstates of the hadronic target, $\psi$ is the Dirac field of the struck quark, $z$ is the space-time separation conjugated to the momentum $k$, and the Dirac matrix $\Gamma$ selects a specific Dirac projection (as it will be explained in more detail in the forthcoming sections). 
For now, we do not consider the additional complications associated with the gauge invariance and the renormalization of these objects.
Fig.~\ref{f:corr} provides a graphical representation of Eq.~\eqref{e:pcf}, in which a quark with four momentum $k - \Delta/2$ is extracted from a hadron target with four momentum $P - \Delta/2$ and is then reabsorbed into the bound state transferring a four-momentum imbalance $\Delta$. 
By taking suitable limits, Fourier transforms, and projections of the parent correlation function $H(k,P,\Delta)$, one can obtain a number of distributions that appear in the context of hadron tomography. 
Fig.~\ref{f:landscape} gives a graphical representation of the connection between $H$ and these other distributions.
For example, 
setting the momentum transfer $\Delta$ to zero and then integrating over the suppressed quark momentum component $k^-$, leads to transverse-momentum-dependent distributions (TMDs), which provide a three-dimensional picture of hadrons in momentum space. 
Additionally integrating over the partonic transverse momentum $\vect{k}$ yields the simpler collinear parton distribution functions (PDFs). 
Alternatively, integrating $H$ over $k^-$ and over the transverse momentum $\vect{k}$ while retaining a finite momentum transfer leads to the generalized parton distributions (GPDs~\cite{Diehl:2003ny,Diehl:2015uka,particles9030069}). These functions describe the spatial distribution of partons inside hadrons, and are connected with impact-parameter distributions and form factors. 
Last but not least, Wigner distributions and generalized transverse-momentum-dependent distributions (GTMDs) combine information on both momentum and position~\cite{Lorce:2025aqp}. 
Fig.~\ref{f:landscape} summarizes these relations schematically, emphasizing how the different functions are interconnected through well-defined mathematical operations rather than representing independent descriptions of hadron structure.
Although the function $H(k,P,\Delta)$ in Eq.~\eqref{e:pcf} provides the most general description of the quark correlation function, no phenomenologically established process is currently known to probe it directly through a factorization theorem. Existing factorization frameworks involve only lower-dimensional projections of $H$, such as PDFs, TMDs, GTMDs, or GPDs. Thus, the partonic structure of hadrons can be investigated by combining information from different projections of the same underlying structure~\cite{Lorce:2025aqp}.

A word of caution is in order before concluding this introduction. The integrations shown schematically in Fig.~\ref{f:landscape} are formally valid for bare (unrenormalized) correlation functions. Once ultraviolet and rapidity divergences are taken into account, the definition of the physical distributions requires appropriate regularization and renormalization schemes~\cite{Collins:2011zzd}. As a consequence, some of the simple integral relations suggested by the figure receive important qualifications, and understanding how the different renormalized hadronic distributions are connected is still the subject of ongoing theoretical research~\cite{Boussarie:2023izj}.

\section{Suggested references}

As mentioned above, these notes are far from being a complete introduction to the field. Several important topics are missing or just introduced without delving into the details. 
To compensate for that, here a selection of additional resources is provided, organized by type. 

\subsection{Graduate schools lecture notes}

\begin{itemize}
    \item R. Jaffe, \href{https://arxiv.org/pdf/hep-ph/9602236.pdf}{\emph{Erice School Lecture Notes}}~\cite{Jaffe:1996zw}  
    
    \item P. Mulders, \href{http://www.nat.vu.nl/~mulders/tmdreview-vs3.pdf}{\emph{GGI Lecture Notes}}~\cite{Mulders:2015ggi} 
    
    \item C. D. Roberts, \href{https://inspirehep.net/literature/1392805}{\emph{Three Lectures on Hadron Physics}}~\cite{Roberts:2015lja} 
    
    \item V. Barone, \href{https://www.fe.infn.it/cabeo_school/2010/cabeo_school_2010.pdf}{\emph{Cabeo Lecture Notes}} 
    
    \item A. Bacchetta, \href{https://www2.pv.infn.it/~bacchett/teaching/Bacchetta_Trento2012.pdf}{\emph{Trento Lecture Notes}}~\cite{Bacchetta:2012aln} 
    
    \item \href{https://www.jlab.org/education/hugs/references}{The HUGS Pedagogical Page} 
    
\end{itemize}


\subsection{Books}

\begin{itemize}
    \item V. Barone and P. Ratcliffe, \emph{Transverse Spin Physics}
    \item J. Collins, \emph{Foundations of Perturbative QCD}~\cite{Collins:2011zzd}
    \item R. Devenish and A. Cooper-Sarkar, \emph{Deep Inelastic Scattering}
    \item T. Muta, \emph{Foundations of Quantum Chromodynamics}~\cite{Muta:2010xua}
    \item M. Peskin, D. Schroeder, \emph{An Introduction to Quantum Field Theory}~\cite{Peskin:1995ev}
\end{itemize}


\subsection{Papers and reviews}

\begin{itemize}

    \item J. C. Collins, D. E. Soper, \href{https://inspirehep.net/literature/166064}{\emph{Parton Distribution and Decay Functions}}~\cite{Collins:1981uw}

    \item \href{https://inspirehep.net/literature/2650019}{\emph{TMD Handbook}}~\cite{Boussarie:2023izj}

    \item \href{https://inspirehep.net/literature/357192}{\emph{Handbook of Perturbative QCD}}~\cite{Brock:1993sz}

    \item EPJ-A Topical Issue: \href{https://link.springer.com/journal/10050/topicalCollection/AC_628286e999d9a60c9a780398df15f93d}{\emph{The 3D Structure of the Nucleon}}

    \item M. Diehl, \href{https://inspirehep.net/literature/1408303}{\emph{Introduction to GPDs and TMDs}}~\cite{Diehl:2015uka}

    \item A. Bacchetta et al., \href{https://inspirehep.net/literature/660999}{\emph{Single Spin Asymmetries: The Trento Conventions}}~\cite{Bacchetta:2004jz}

    \item J. Collins, \href{https://inspirehep.net/literature/443368}{\emph{Light Cone Variables, Rapidity and All That}}~\cite{Collins:1997nm}

    \item A. Metz and A. Vossen, \href{https://inspirehep.net/literature/1475000}{\emph{Parton Fragmentation Functions}}~\cite{Metz:2016swz}

    \item I. Scimemi, \href{https://inspirehep.net/literature/1716549}{\emph{A Short Review on Recent Developments in TMD Factorization}}~\cite{Scimemi:2019mlf}

    \item C. D. Roberts, \href{https://inspirehep.net/literature/1094861}{\emph{Strong QCD and Dyson–Schwinger Equations}}~\cite{Roberts:2012sv}

    \item T. Horn and C. D. Roberts, \href{https://inspirehep.net/literature/1421431}{\emph{The Pion: An Enigma within the Standard Model}}~\cite{Horn:2016rip}

    
    \item X. Ji et al., \href{https://inspirehep.net/literature/1814771}{\emph{What We Know and What We Don’t Know About the Proton Spin}}~\cite{Ji:2020ena} 
    
\end{itemize}


\subsection{Experimental overviews}

\begin{itemize}
    \item J. Dudek et al., \href{https://inspirehep.net/literature/1125972}{\emph{Physics Opportunities with the 12 GeV Upgrade at Jefferson Lab}}~\cite{Dudek:2012vr}

    \item A. Accardi et al., \href{https://inspirehep.net/literature/1206324}{\emph{Electron Ion Collider: The Next QCD Frontier}}~\cite{Accardi:2012qut}

    \item R. Abdul Khalek et al., \href{https://inspirehep.net/literature/1851258}{\emph{The EIC Yellow Report}}~\cite{AbdulKhalek:2021gbh}

    \item M. Anderle et al., \href{https://inspirehep.net/literature/1847312}{\emph{Electron-Ion Collider in China}}~\cite{Anderle:2021wcy} 
    
    \item C. Aidala et al., \href{https://inspirehep.net/literature/2914859}{\emph{LHCspin: a Polarized Gas Target for LHC}}~\cite{LHCspin:2025lvj}
    
\end{itemize}

\clearpage
\part{Probing hadron structure}
\label{p:breaking}

In this part we introduce the Deep-Inelastic Scattering (DIS) process and exploit it as a tool to investigate the (un)polarized structure of the hadronic target in momentum space. 
In particular, we focus on the definition of specific structure functions, which act as response functions of the target to an external probe, and their link to parton distribution functions in the collinear factorization framework.

Between 1906 and 1913 the experiments performed by Rutherford, Geiger, and Marsden led to the discovery of the atomic nucleus. By scattering alpha particles off a gold target, it was realized that positive electric charges were condensed in a nucleus rather than being uniformly distributed.
In a similar fashion, but at much higher energies, between the 1960s and 1970s scientists at MIT and SLAC scattered leptons off hadronic targets in order to resolve the inner structure of hadronic matter. These experiments led to an evidence~\cite{Callan:1969uq} that protons were essentially made out of point-like, free, and spin 1/2 particles, named partons.

\section{Deep-Inelastic Scattering}
\label{s:DIS}

In DIS, a lepton $\ell$ with 4-momentum $l$ scatters off a hadronic target such as a nucleon $N$ with 4-momentum $P$. Then another lepton $\ell^\prime$ with 4-momentum $l^\prime$ is detected in the final state:
\begin{equation}
\label{e:DISproc}
\ell(l) + N(P) \to \ell'(l') + X \, .
\end{equation}
The 4-momentum transfer is defined as:
\begin{equation}
q = l - l' \, .
\end{equation}

\subsection{Kinematics}
\label{ss:DIS_kin}

If the target is a nucleon, there are four available independent vectors: the three 4-momenta in Eq.~\eqref{e:DISproc} and the covariant spin of the target. With the momenta, one can build the following invariants:
\begin{align}
s &= (P+\ell)^2 \, , \\
W^2 &= (P+q)^2 \, , \\
Q^2 &= -q^2 = -(\ell - \ell^\prime)^2 \, ,
\end{align}
where $s$ is the center of mass energy squared, $W$ is the invariant mass of the hadronic final states, and $Q^2>0$ quantifies how deep the hadronic target is probed ($q^2<0$ is the inviariant mass of the intermediate photon).
One can also define the following variables:
\begin{equation}
\label{e:xBj}
x_B = \frac{Q^2}{2 P \cdot q} \, ,
\end{equation}
which is the so-called Bjorken-$x$ variable, and the inelasticity of the reaction:
\begin{equation}
y = \frac{P \cdot q}{P \cdot l} \, .
\end{equation}
The deep-inelastic regime of the reaction in Eq.~\eqref{e:DISproc}, or Bjorken limit, is reached when:
\begin{equation}
\label{e:DIS_regime}
Q^2\, , P \cdot q \to \infty \ \ (\gg M^2) \, ,
\qquad
x_B \ \text{fixed} \, ,
\end{equation}
where $M$ is the mass of the hadronic target (around 1 GeV for the nucleon).

Since we are dealing with high-energy scattering events and momenta which are (almost) light-like, it is more convenient to work with light cone coordinates in place of the standard Cartesian coordinates. The relation between these two systems is outlined in App.~\ref{a:lc_coordinates} and, e.g., in Ref.~\cite{Collins:1997nm}.
For DIS, it is customary~\cite{Mulders:2016pln,Mulders:2015ggi} to identify $P$ and $q$ as the two light-like directions with which one can construct a light-cone basis.
When we refer to ``transverse'' quantities, we refer to vectors which are transverse with respect to the $P,q$ basis in the sense of Eq.~\eqref{e:gT_npnm}.

\subsection{Cross section}
\label{ss:DIS_xsec}

\begin{figure}
\centering
\includegraphics[width=5cm]{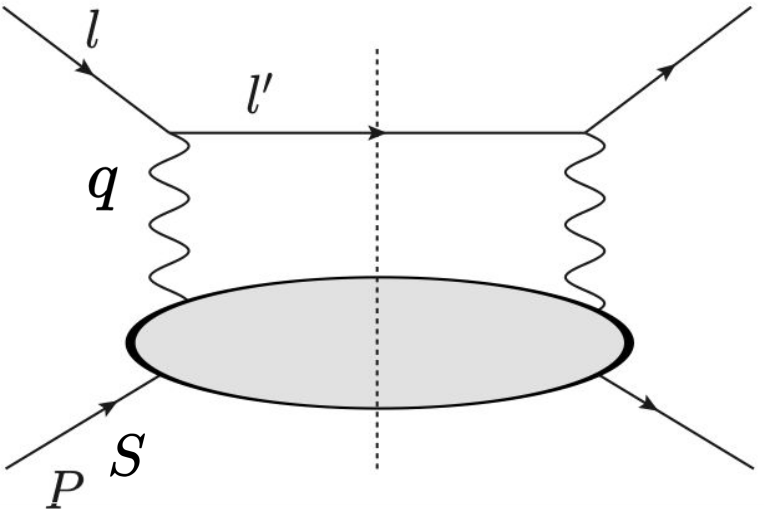}
\caption{Contraction of the leptonic and hadronic tensors in DIS. The lines crossing the cut represent on-shell states that enter the final state of the reaction.}
\label{f:LW_DIS}
\end{figure}

The DIS cross section differential with respect to the Bjorken-$x$ variable, the inelasticity $y$, and the azimuthal angle $\phi_S$ of the transverse target spin $\vect{S}_T$ (see App.~\ref{a:spin}) reads~\cite{Bacchetta:2002xd}:
\begin{equation}
\label{e:DIS_xsec}
\frac{d^3\sigma}{dx_B\, dy\, d\phi_S} =
\frac{\alpha^2\, y}{2 Q^4}\, L_{\mu\nu}(l,l',\lambda_\ell)\, 2M\, W^{\mu\nu}(P,q,S) \, .
\end{equation}
The cross section is essentially the contraction of two tensors: the leptonic tensor $L_{\mu\nu}$ and the hadronic tensor $W^{\mu\nu}$. This contraction can be pictorially represented as a cut diagram with specific Feynman rules for states that cross the cut~\cite{Cutkosky:1960sp,Zwicky:2016lka}, see Fig.~\ref{f:LW_DIS}.
The structure of Eq.~\eqref{e:DIS_xsec} remains valid for a target with different spin. In the case of a deuteron (spin 1) target, the hadronic tensor depends on an additional spin matrix $T$, as discussed in App.~\ref{a:spin} and, e.g., Refs.~\cite{Bacchetta:2000jk,Bacchetta:2002xd,Boer:2016xqr,Kumano:2024fpr,Zhao:2025vol}.

The leptonic tensor represents the leptonic part of the process. It depends on the lepton momenta and on the helicity $\lambda_\ell$ of the leptons. The term proportional to the transverse lepton polarization is of the order of the lepton mass, ${\mathcal O}(m_\ell)$, and thus it can be ignored compared to other mass scales. 
In the one-photon exchange approximation
the lepton tensor reads~\cite{Bacchetta:2006tn,Bacchetta:2002xd}: 
\begin{align}
\label{e:L}
L^{\mu\nu} (l,l^\prime,\lambda_\ell) &= 
\sum_{\lambda_\ell^\prime}\, 
\bigg( \bar{u}\big(l^\prime,\lambda_\ell^\prime\big) \gamma_\mu u\big(l,\lambda_\ell\big) \bigg)^*\, 
\bigg( \bar{u}\big(l^\prime,\lambda_\ell^\prime\big) \gamma_\nu u\big(l,\lambda_\ell\big) \bigg) = \\ 
\nonumber
& = 
2 \left( l^\mu l'^\nu + l^\nu l'^\mu - g^{\mu\nu} l \cdot l' \right) +
2 i\, \lambda_\ell\, \epsilon_{\mu\nu\rho\sigma} l^\rho l^{\prime\, \sigma}\, ,
\end{align}
where the last term refers to the contribution to the scattering process due to the longitudinal incoming lepton polarization.

\subsection{Hadronic tensor}
\label{ss:had_tensor}

The hadronic tensor, instead, encodes all the information about the hadronic part of the process. Because of the scale dependence of the QCD coupling, in a high energy scattering process involving hadrons one needs to take into account both perturbative and non-perturbative contributions to the observable. The perturbatively calculable ones are related to the hard scattering process involving partons, the non-perturbative ones are related to hadron structure and hadronization effects at scales $~{\mathcal O}(M)$.
The hadronic tensor for DIS on a spin-$1/2$ target is defined as:
\begin{equation}
\label{e:W_DIS_operator}
2 M W^{\mu\nu}(P,q,S) =
\sum_X\, \int \frac{d^3 P_X}{2E_X}\, \delta^{(4)}\big( P + q - P_X \big)\, 
\langle P| J_\mu^\dagger(0) |P_X \rangle\, 
\langle P_X| J_\nu(0) |P \rangle\, .
\end{equation}
The matrix element $\langle P_X| J_\nu(0) |P \rangle$ corresponds to the left part (with respect to the cut) of the grey blob diagram in Fig.~\ref{f:LW_DIS}. It represents the action of the current $J$ on the initial nucleon state, generating another hadronic final state with momentum $P_X$, for which an on-shell momentum integration is performed, along with a sum on all the possible final states $X$.
The current is a local operator defined as a normal-ordered product of operators, which in the electromagnetic case reads~\cite{Muta:2010xua}:
\begin{equation}
\label{e:elm_current}
J_\mu(\xi) = :\, \bar{\psi}(\xi)\, e_q\, \gamma_\mu\, \psi(\xi)\, : \, , 
\end{equation} 
where $e_q$ is the (fractional) charge of the fermion. 
In the electroweak case the current of course has a different form~\cite{Mulders:2015ggi}.

\subsection{Structure functions}
\label{ss:DIS_SFs}

At this point it is useful to introduce a specific light cone basis to describe the hadronic tensor~\cite{Mulders:1995dh,Mulders:2015ggi}. The following choice is particularly useful to decompose the hadronic tensor in structures with well defined polarization:
\begin{equation}
\label{e:tq_basis}
\hat{q}^\mu = \frac{q^\mu}{Q}\, , \ \ \ \ \ \ 
\hat{t}^\mu = \frac{2 x_B}{Q \sqrt{1+\gamma^2}}\, \bigg( P^\mu - \frac{P \cdot q}{q^2}\, q^\mu \bigg)\, ,
\end{equation}
where $\gamma = 2 M x_B/Q$ is a factor that encodes target mass corrections. 
From now on, ${\mathcal O}(\gamma^2)$ effects will be suppressed. The correct expressions including these effects can be found in e.g. Refs.~\cite{Diehl:2005pc,Bacchetta:2006tn}.
The $\hat{q}, \hat{t}$ vectors are orthogonal and normalized in the sense of a light-cone basis.
Accordingly, the symmetric and antisymmetric transverse projectors are:
\begin{equation}
g_\perp^{\mu \nu} = g_{\mu\nu} + \hat{q}^\mu \hat{q}^\nu - \hat{t}^\mu \hat{t}^\nu  \, , \ \ \ \ \ \ 
\epsilon_\perp^{\mu \nu} = \epsilon^{\mu\nu\rho\sigma}\, \hat{t}_\rho \hat{q}_\sigma \, .
\end{equation}
 
The hadronic tensor is a Lorentz matrix. Being an intrinsically non-perturbative object, it is not possible to directly compute $W^{\mu\nu}$ in perturbation theory, but the symmetries of QCD can help in shedding light on its content. In particular, considering a spin $1/2$ target, one can use the following relations~\cite{Mulders:2015ggi}:
\begin{eqnarray}
\label{e:W_herm}
\mbox{Hermiticity:} &\quad W_{\mu\nu}^*(q;P,S) = W_{\nu\mu}(q;P,S) \\
\label{e:W_parity}
\mbox{Parity:} &\quad W_{\mu\nu}(q;P,S) = \overline{W}_{\nu\mu}(\bar{q},\bar{P},-\bar{S})  \\
\label{e:W_trev}
\mbox{Time reversal:} &\quad W_{\mu\nu}^*(q;P,S) = \overline{W}_{\mu\nu}(\bar{q},\bar{P},\bar{S})  \\
\label{e:W_gauge}
\mbox{Gauge:} &\quad q_\mu\, W^{\mu\nu}(q;P,S) = q_\nu\, W^{\mu\nu}(q;P,S) = 0 \, ,
\label{e:W_transf}
\end{eqnarray}
where, in cartesian components, $\bar{p}=(p^0,-{\bf p})$. 
The previous equations encode the hermiticity property and the invariance of QCD under parity, time reversal, and gauge transformations respectively. 
Based on the previous relations, one can find the most general parametrization of $W^{\mu\nu}$ for a spin $1/2$ target based on the available four vectors. 
%
%
%
The hadronic tensor reads~\cite{Bacchetta:2006tn,Accardi:2017pmi}:
\begin{equation}
\label{e:W_DIS_spin12}
2M W^{\mu\nu} = 
\frac{1}{x_B}\, \bigg[ 
-g_\perp^{\mu\nu}\, F_{UU,T} + \hat{t}^\mu \hat{t}^\nu\, F_{UU,L} +
i S_L \epsilon_\perp^{\mu\nu}\, F_{LL} - 
i \hat{t}^{[\mu} \epsilon_\perp^{\nu ] \rho} S_\rho\, F_{LT}^{\scriptsize{\cos}\, \phi_S} 
\bigg]\, , 
\end{equation}
where the structure functions are functions of $x_B, Q^2$, and $\phi_S$ is the azimuthal angle of the transverse spin (the transverse components are calculated with respect to the light cone directions defined in Eq.~\eqref{e:tq_basis}).

The subscripts $i,j,k$ in the structure functions $F_{ij,k}$ refer to the lepton, hadron, and photon polarization respectively. Thus, $F_{UU,T}$ and $F_{UU,L}$ describe unpolarized DIS mediated by a transversely or longitudinally polarized photon, and multiply symmetric Lorentz structures. Instead, $F_{LL}$ and $F_{LT}^{\scriptsize{\cos}\, \phi_S}$ refer to the polarized case and multiply antisymmetric Lorentz structures~\cite{Jaffe:1996zw}. 
There are other widely used conventions in the literature (see e.g. Refs.~\cite{Mulders:1995dh,Jaffe:1996zw,Mulders:2015ggi}), in which the structure functions are named $F_{1,2}$, $g_{1,2}$. The relations with the structure functions used in this set of notes, up to ${\mathcal O}(\gamma^2)$ effects, are~\cite{Bacchetta:2006tn,Accardi:2017pmi}:
\begin{align}
\label{e:DIS_SFs_relations}
& F_{UU,T} = 2 x_B F_1 \, , \\
& F_{UU,L} = F_2 - 2 x_B F_1 \, , \\
& F_{LL} = 2 x_B g_1 \, , \\
& F_{LT}^{\scriptsize{\cos}\, \phi_S} = -2 x_B \gamma \big( g_1 + g_2 \big) \, ,
\end{align}

After contracting the leptonic and the hadronic tensor, the cross section in terms of the DIS structure functions reads:
\begin{align}
\label{e:DIS_xsec_SFs}
\frac{d\sigma}{dx_B\, dy\, d\phi_S} =
\frac{2\alpha^2}{x_B y Q^2}
\Bigg\{
& \left(1-y+\frac{y^2}{2}\right) F_{UU,T}(x_B,Q^2)
+ (1-y) F_{UU,L}(x_B,Q^2) \\ \nonumber
& + S_L \lambda_e\, y\left(1-\frac{y}{2}\right) F_{LL}(x_B,Q^2)
+ |\vect{S}_T| \lambda_e\, y\sqrt{1-y}\, \cos\phi_S\, F_{LT}^{\scriptsize{\cos}\, \phi_S}(x_B,Q^2)
\Bigg\}
\end{align}

In order to connect this expression of the DIS cross section with the one presented in Ref.~\cite{Bacchetta:2006tn}, one needs to properly take into accounts target mass ${\mathcal O}(\gamma^2)$ correction throughout the whole Sect.~\ref{s:DIS}. In particular, one needs to explicitly introduce these corrections in the definitions of the light-cone basis vectors $\hat{t}$ and $\hat{q}$ and propagate them to the kinematics and the calculation of the hadronic tensor in terms of structure functions.

For a spin 1 target, there are four additional structure functions emerging in the hadronic tensor and in the cross section, owing to the presence of the tensor polarization states. 
As for the spin $1/2$ case, there are different conventions to describe these additional configurations, which are extensively discussed in Refs.~\cite{Hoodbhoy:1988am,Cosyn:2024drt,Zhao:2025vol}.

\section{Operator Product Expansion}
\label{s:OPE}


Until now we just organized the cross section and the hadronic tensor in terms of structure functions, which describe the response of the target to excitations from an external electromagnetic probe. 
At this point we need to devise a connection with the partonic degrees of freedom of QCD, quarks and gluons. 
The tool that we are going to employ in the following is the Operator Product Expansion~\cite{Wilson:1969zs,Wilson:1972ee,Peskin:1995ev,Muta:2010xua,Weinberg:1996kr,Jaffe:1996zw,Mulders:2016pln} (OPE), which allows one to link structure functions to parton distribution functions (PDFs).
A conceptually similar result can be obtained with collinear factorization in perturbation theory~\cite{Brock:1993sz,Collins:2011zzd}. 
The renormalization of the involved operators is not discussed in these notes, but it is of fundamental importance in order to build a framework with predictive power. 
The PDFs introduced either via the OPE or factorization encode the intrinsic non-perturbative QCD effects for which universality properties hold between different processes~\cite{Buffing:2015tja}.

In Ref.~\cite{Wilson:1969zs} K. Wilson showed that a product of local operators $A(x)B(y)$ has an expansion, when $|x-y|$ is small, of the form:
\begin{equation}
A(x)B(y) = \sum_n C_n (x-y)\, {\mathcal O}_n(x) \ .
\label{e:wilson_ope_def}
\end{equation}
This expression is called the Operator Product Expansion (OPE) of the non-local product $A(x)B(y)$ on the basis of the local operators $O_n(x)$.
The OPE has been demonstrated by W. Zimmerman in perturbation theory~\cite{Zimmermann:1972tv} for a time-ordered product of operators (which, in turn, can be a normal-ordered products of field operators, as for a current).

\subsection{Light-cone dominance}
\label{ss:lc_dominance}

The only two processes for which an OPE on the hadronic tensor $W$ is formally proved are inclusive DIS and inclusive $e^+e^-$ annihilation~\cite{Muta:2010xua}. The reason being that in both cases the hadronic tensor contains a {\em commutator} of currents, which is related to time-ordered products of currents. 

Starting from Eq.~\eqref{e:W_DIS_operator} and by using the completeness relation on the intermediate states $|P_X \rangle$, the properties of the delta distribution, the hermiticity of the current, and translating the argument of the current, one can rewrite the DIS hadronic tensor as:
\begin{align}
\label{e:W_DIS_comm}
& 2M W^{\mu\nu}(q,P,S) = \frac{1}{2\pi}\int d^4x\ e^{i q \cdot x } \langle PS|\ [J^\mu(x),J^\nu(0)]\ |PS \rangle \ , \\ 
\label{e:comm_vs_Tord}
& \text{sign}(x^0)\ [J^\mu(x),J^\nu(0)] = T[ J^\mu(x) J^\nu(0) ] - T[ J^\mu(x) J^\nu(0) ]^\dagger \ .
\end{align}


Thanks to the presence of the commutator, it is possible to apply the OPE to the DIS hadronic tensor. In fact,
causality implies that the commutator is non-zero only in the time-like $x^2<0$ region. Moreover, the Riemann-Lebesgue lemma implies that $W$ is zero unless $x^0 \to 0$. Thus, the hadronic tensor $W$ is dominated by the $x^2 \approx 0$ region~\cite{Muta:2010xua}. This property, named {\em light-cone dominance}, allows the application of Eq.~\eqref{e:wilson_ope_def} to Eq.~\eqref{e:W_DIS_comm}.
Inclusive electron-positron annihilation is the only other case in which the OPE can be formally applied to the hadronic tensor, as this is dominated by the $x \approx 0$ space-time region~\cite{Muta:2010xua}.

\subsection{Handbag diagram}
\label{ss:handbag}

\begin{figure}
\centering
\includegraphics[width=0.95\textwidth]{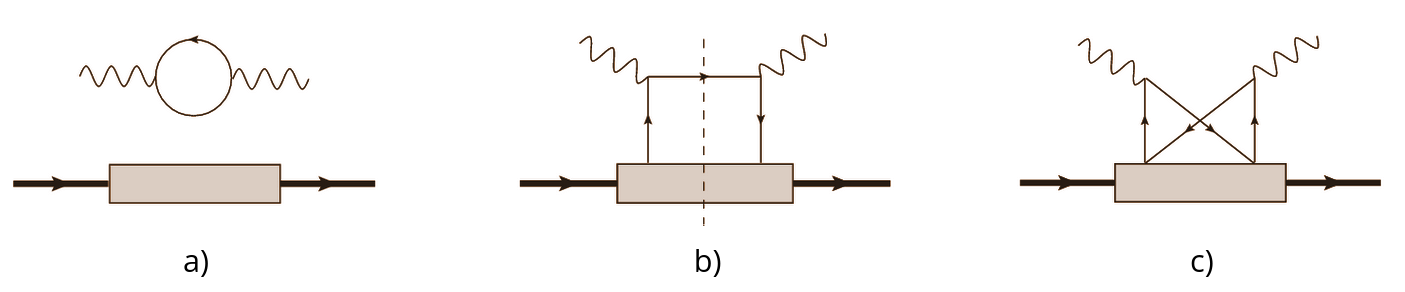}
\caption[OPE contributions in DIS]{Diagrammatic interpretations of the operators contributing to the DIS hadronic tensor. Diagram (a) represents a disconnected contribution, in which the target is not probed.    
The diagram (b) represents the so-called ``handbag diagram'', in which an (anti)quark is probed, whereas in (c), the ``cat's ear diagram'' two (anti)quarks in the target are probed in the same diagram.}
\label{f:OPE_diag}
\end{figure}

Applying the OPE to Eq.~\eqref{e:W_DIS_comm} and relying on Eq.~\eqref{e:elm_current} for the electromagnetic current, it is possible to identify different contributions~\cite{Muta:2010xua} to the DIS hadronic tensor, with a diagrammatic interpretation given in Fig.~\ref{f:OPE_diag}. 
The first one (a) is a disconnected contribution and does not represent a scattering with the target. 
The second one (b), the so-called ``handbag diagram'', coincides with probing an (anti)quark in the target and its reinsertion in the remnant at a separation $\xi$ via a vector or an axial coupling.  
The third one (c), the so-called ``cat's ear diagram'' is power suppressed and involves the interaction with two partons.

Accordingly, the DIS hadronic tensor can be approximated at leading power by the handbag diagram. Inserting in Eq.~\eqref{e:W_DIS_comm} the expression for its vector component one obtains:
\begin{align}
\label{e:W_DIS_handbag}
2M W^{\mu\nu} \approx & \sum_q e_q^2\, \frac{1}{2}\, \int d^4\xi \, e^{i k \cdot \xi}\,   
\langle PS| \bar{\psi}(0) \psi(\xi) |PS \rangle_{|_{\xi^+=\xi_T=0}}\, \gamma^\mu\, \gamma^+\, \gamma^\nu \ .
\end{align}
For the detailed calculation we refer the reader to Refs.~\cite{Jaffe:1996zw,Muta:2010xua}.
The matrix elements are evaluated at fixed light cone time $\xi^+=0$, so the time ordering on the currents eventually does not have a practical effect~\cite{Diehl:1998sm}.

\subsection{Twist}
\label{ss:twist}

The light-cone expansion for the product of two currents is of the form~\cite{Jaffe:1996zw,Muta:2010xua}:
\begin{equation}
\label{e:general_ope_curr}
j^\mu(x)j^\nu(0) = \sum_{i,n} C^{(i)}_n(x^2)\, x^{\mu_1} \cdots x^{\mu_n}\, O^{(i)}_{\mu_1 \cdots \mu_n}(0) \ ,
\end{equation}
where $O^{(i)}_{\mu_1 \cdots \mu_n}$ is an irreducible symmetric traceless tensor of rank $n$, referred to as the spin of the local operator.
The singularity structure of the coefficient functions $C^{(i)}_n$ is crucial to determine which operators contribute most, but their form depends on the structure of the currents (it is not a general property of the OPE).
In a free field theory one can perform~\cite{Muta:2010xua} a power counting of the mass dimensions in Eq.~\eqref{e:general_ope_curr} and see that the singularity structure of the coefficients is of the form:
\begin{equation}
\label{e:C_sing_struct}
C^{(i)}_n(x^2) \sim (x^2)^{-d_j -n/2 + d_O^{(i)}(n)/2} \ ,
\end{equation} 
where $d_j$ is the dimension of the current $j$ and $d_O^{(i)}(n)$ is the dimension of the local operator $O^{(i)}_{\mu_1 \cdots \mu_n}$.
The strength of the singularity for each term in the OPE is governed by $d_O^{(i)}(n) - n$, which is the {\em twist} of the local operator $O^{(i)}_{\mu_1 \cdots \mu_n}$~\cite{Gross:1971wn,Muta:2010xua}. The smaller the twist of the operator, the higher the singularity of the coefficient function and, accordingly, the more relevant the local operator will be in physical observables.

The coefficient functions that appear in the OPE of the DIS hadronic tensor are proportional to~\cite{Jaffe:1996zw}:
\begin{equation}
\label{e:tw_MQ}
\bigg( \frac{M}{Q} \bigg)^{ d_O^{(i)}(n) - n - 2} \, .
\end{equation}
This suggests that one can evaluate the relevance of local operators in the OPE looking at the power of $M/Q$ to which they contribute to the coefficient functions\footnote{This argument, based on power counting, holds true when no other mass scales appear in the theory. In an interacting theory, extra care is needed~\cite{Peskin:1995ev,Wilson:1972ee,Muta:2010xua,Weinberg:1996kr}.}. 
In this context, the cat's ear diagram can be considered a higher twist effect compared to the handbag diagram in the DIS hadronic tensor.

However, in order to correctly evaluate twist $t$ effects in a OPE, one needs to identify all contributions of twist $t$ from different local operators $i$ (with increasing spin) and resum them in a ``tower'' of fixed twist $t$. This procedure (first expanding the bilocal operators, Fourier transforming and then summing all the contributions with the same twist), though, is rather lengthy. It should be possible to introduce a concept of twist directly on the non-local operators.

Based on the simple expression in Eq.~\eqref{e:tw_MQ} for the local operators, a ``working'' definition of twist for a matrix element of a bi-local operator was proposed in Ref.~\cite{Jaffe:1996zw}, based on the order in $M/Q$ at which the element contributes to the cross section of deep inelastic processes. In practice, twist-$t$ matrix elements contributes as $(M/Q)^{t-2}$. 
From a phenomenological point of view, the working definition is more practical than looking at the OPE-based twist and allows to label immediately a specific contribution as leading or sub-leading in powers of $M/Q$. In the following we will adopt this definition.

\begin{figure}[h!]
\includegraphics[width=9cm]{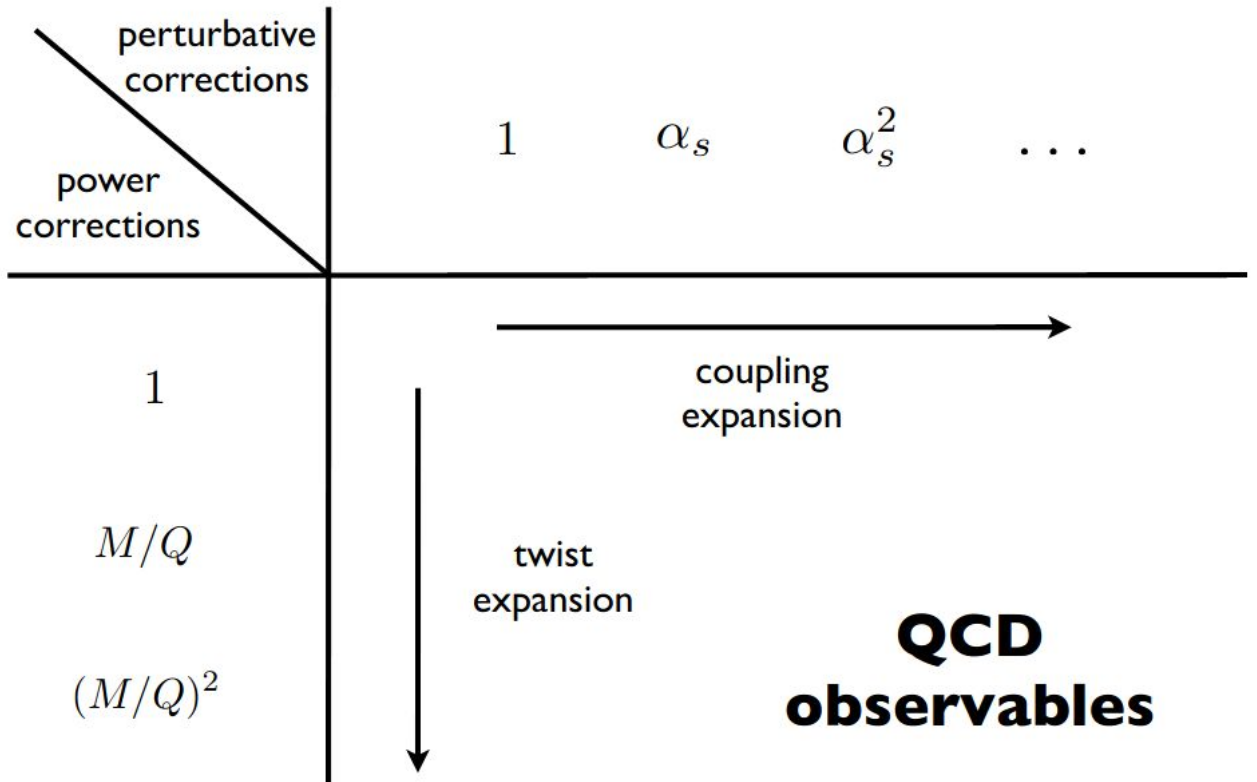}
\caption[Coupling and twist expansions in QCD]{A physical observable can be expanded in powers of the coupling constant $\alpha_s$ and in powers of $M/Q$. The twist expansion is limited by the symmetries of QCD.}
\label{f:qcd_exps}
\end{figure}

Effectively, it is possible to describe hadron structure-related observables using a double expansion, one in powers of $\alpha_s$ (the QCD coupling constant) and one in (limited) powers of $M/Q$, as illustrated in Fig.~\ref{f:qcd_exps}.
What has been presented in this Section is a characterization of the DIS hadronic tensor at leading order in $\alpha_s$ and leading twist. A discussion beyond the leading order is available, e.g., in Ref.~\cite{Brock:1993sz}.
The term ``twist'' is sometimes also used to address several other kinematic power corrections, which involve ratios of small quantities (such as masses, transverse momenta, etc.) and hard scales (typically the invariant mass of the virtual photon or the leading component of the target momentum). 

\subsection{Collinear PDFs}
\label{ss:coll_pdfs}

The right-hand side of Eq.~\eqref{e:W_DIS_handbag} can be interpred as the trace of three Dirac matrices with a collinear quark-quark {\em correlator} (or correlation function) $\Phi(x,S)$, defined as the Fourier transform of the expectation value on the target state of the fermion bilinear operator $\bar{\psi}(0) \psi(\xi)$:

\begin{align}
\label{e:W_DIS_Phi}
2M W^{\mu\nu} \approx & \sum_q e_q^2\, \frac{1}{2}\, \int d^4\xi \, e^{i k \cdot \xi}\,   
\langle PS| \overline{\psi}(0) \psi(\xi) |PS \rangle_{|_{\xi^+=\xi_T=0}}\, \gamma^\mu\, \gamma^+\, \gamma^\nu \\ 
\nonumber
= & \sum_q e_q^2\, \frac{1}{2}\, \text{Tr}\, \bigg[   
\Phi(x,S)\, \gamma^\mu\, \gamma^+\, \gamma^\nu
\bigg]\, ,
\end{align}
where
\begin{align}
\label{e:coll_phi}
\Phi_{ij}(x,S) = 
\int \frac{d\xi^-}{2\pi}\, e^{i xP^+ \xi^-}\,
\langle P,S|\, \overline{\psi}_j(0)\, \psi_i(\xi)\, | P,S \rangle 
\Big|_{\xi^+=\bm{\xi}_T=0} \, .
\end{align}
The collinear correlator $\Phi(x,S)$ is a non-perturbative Dirac matrix that encodes information about hadron structure on the light-cone ($\xi^+=\xi_T=0$). Accordingly, it can be parametrized using a basis for the linear space of Dirac matrices as~\cite{Bacchetta:2006tn}: 
\begin{align}
\label{e:Phi_coll_par_tw23}
\Phi(x,S) & =
\frac{1}{2}\, 
\bigg\{ 
f_1(x)\,\slash{n}_+ + 
g_1(x)\, S_L\, \gamma_5\, \slash{n}_+  +
h_1(x)\, \frac{[\slash{S}_T , \slash{n}_+]\, \gamma_5}{2} 
\bigg\} \\ \nonumber
& + \frac{M}{2P^+}\, 
\bigg\{
e(x) - 
e_L(x)\, i\, S_L\, \gamma_5 + 
f_T(x)\, \epsilon_T^{\rho\sigma}\, {S_T}_\rho\, \gamma_\sigma +
g_T(x)\, \gamma_5\, \slash{S}_T\, +
h_L(x)\, S_L\, \frac{[\slash{n}_+ , \slash{n}_-]}{2}\, \gamma_5 +
h(x)\, i\, \frac{[\slash{n}_+ , \slash{n}_-]}{2}\, 
\bigg\} \\ \nonumber
& + {\mathcal O}\bigg( \frac{M^2}{P^{+ 2}} \bigg) \, ,
\end{align}
where $\slash{\Gamma} \equiv \Gamma^\mu \gamma_\mu$ and thus $\slash{n}_+ = \gamma^-$.
The functions multiplying the Dirac structures in Eq.~\eqref{e:Phi_coll_par_tw23} are the collinear parton distribution functions. The first line contains the three leading twist PDFs for a spin 1/2 target (see Sec.~\ref{ss:twist}): $f_1(x)$ is the unpolarized PDF, $g_1(x)$ the helicity PDF\footnote{Not to be confused with the structure function $g_1$ mentioned in Sec.~\ref{ss:DIS_SFs}.}, and $h_1(x)$ the transversity PDF.
The second line contains the twist three PDFs, which are kinematically suppressed by a factor $M/P^+$ compared to the twist 2 ones. The twist four PDFs (third line) are simply omitted, but can be found, for example, in Ref.~\cite{Mulders:2016pln}.
Each PDF can be singled-out from the correlator $\Phi(x,S)$ by tracing it with a suitable Dirac matrix (see Sec.~\ref{ss:TMDPDFs} for more details).

For a spin-1 hadron the correlator acquires an additional dependence on the tensor spin components (see App.~\ref{a:spin}). At leading twist, there is an additional PDF $f_{1LL}(x)$. At higher twist, additional structures are available (see e.g. Refs.~\cite{Kumano:2024fpr,Zhao:2025vol}). \\

Given the fact that the hadronic tensor can be expressed both in terms of structure functions and in terms of PDFs, it is possible to write relations between these two sets of functions~\cite{Bacchetta:2006tn,Accardi:2017pmi}. Working up to twist three (namely neglecting ${\mathcal O}(\gamma^2)$ effects) and at leading order in $\alpha_s$ one has:
\begin{align}
\label{e:DIS_FT_f1}
& F_{UU,T} = x_B\, \sum_q\, e_q^2\, f_1^q(x_B)\, , \\
%
\label{e:DIS_FL_CG}
& F_{UU,L} = 0 \, , \\ 
%
\label{e:DIS_FLL_g1}
& F_{LL} = x_B\, \sum_q\, e_q^2\, g_1^q(x_B)\, , \\
%
\label{e:DIS_FLTcos_gT_h1}
& F_{LT}^{\scriptsize{\cos}\, \phi_S} = - x_B\, \sum_q e_q^2\, \gamma\, 
\bigg( g_T^q(x_B) + \frac{\tilde{m}_q}{M x_B }\, h_1^q(x_B) \bigg) \, ,
%
\end{align}
where the sums run over the partonic species $q$. 
The mass $\tilde{m}_q$ which appears in the RHS of Eq.~\eqref{e:DIS_FLTcos_gT_h1} is a contribution to the quark mass which quantifies the strength of dynamical quark-gluon-quark correlations (for more details see Refs.~\cite{Accardi:2017pmi,Accardi:2019luo,Accardi:2020iqn}).
As already mentioned, each of the previous equations receives perturbative corrections governed by powers of $\alpha_s$. 
At higher orders in $\alpha_s$ the structure functions can be written as a convolution between perturbatively calculable coefficients $C$ and the associated collinear PDFs. Schematically:
\begin{equation}
\label{e:Fi_higher_orders}
F_i(x_B,Q^2) =
\sum_q\, e_q^2\, 
\left[
C_i \otimes f^q
\right](x_B,Q^2) \, ,
\end{equation}
where both the structure functions and the PDFs have acquired a $Q^2$ scale dependence, which is a byproduct of the renormalization of ultraviolet divergencies.

Eq.~\eqref{e:DIS_FL_CG} is particularly interesting, as it coincides with the so-called Callan-Gross relation~\cite{Callan:1969uq} $2x_B F_1 = F_2$, which is a signal of the spin 1/2 nature of quarks and whose ``scaling '' properties at large $Q^2$ was interpreted as a signal of the asymptotic freedom of partons. \\

Last but not least, it is important to spend a few words about higher twist distributions. This terminology can refer to two types of higher twists: kinematical and dynamical. 
Kinematical higher twist PDFs are those which appear in Eq.~\eqref{e:Phi_coll_par_tw23} suppressed by powers of $M/Q$, for example $e(x)$. 
Dynamical higher twist PDFs are defined from three-parton matrix elements of the type $\langle P| \overline{\psi}\, A^\mu\, \psi |P\rangle$. They are usually denoted with a tilde, e.g. $\tilde{e}(x)$, and do not present any additional kinematic suppression factor~\cite{Mulders:1995dh,Bacchetta:2006tn,Mulders:2016pln}. 
The Dirac equation allows one to draw relations among twist 2 and kinematical/dynamical higher twist PDFs, which are referred to as equations of motions (EOMs)~\cite{Mulders:1995dh,Bacchetta:2006tn,Mulders:2016pln}. As an example we consider the EOM for the unpolarized sector:
\begin{equation}
\label{e:EOMs_unpolarized_PDFs}
x\, e(x) = x\, \tilde{e}(x) + \frac{m_q}{M}\, f_1(x) \, ,
\end{equation}
where $M$ is the hadron mass and $m_q$ is the current quark mass which enters the Dirac equation.
In the so-called Wandzura Wilzcek approximation~\cite{Accardi:2009au} one usually neglects the dynamical higher twists, in which case the kinematic twist 3 PDFs become proportional to the twist 2 distributions, $x\, e(x) = m_q\, f_1(x) / M$.

\section{Exercises}
\label{s:ex_probing}

\begin{itemize}

\item[1)] Derive the hermiticity condition for the hadronic tensor and its transformation under parity and time reversal invariance (see Eqs.~\eqref{e:W_herm},~\eqref{e:W_parity},~\eqref{e:W_trev})

\item[2)] Derive the parametrization of the DIS hadronic tensor given in Eq.~\eqref{e:W_DIS_spin12}

\item[3)] Relate the different parametrizations of the hadronic tensor available in Refs.~\cite{Jaffe:1996zw,Bacchetta:2006tn,Accardi:2017pmi} and derive the relations among the structure functions (see Eqs.~\eqref{e:DIS_SFs_relations}) including target mass corrections (e.g. do not drop terms proportional to $\gamma$)

\end{itemize}

\clearpage
\part{Non-collinear partons}
\label{p:non_coll_partons}

In Part~\ref{p:breaking} of these notes we introduced (at least partially) the tools to probe hadron structure in 1D momentum space, via deep-inelastic scattering and collinear PDFs.
In this Part~\ref{p:non_coll_partons} we are going to generalize the previous results, introducing transverse-momentum-dependent (TMD) parton distribution and fragmentation functions, which are the maps that can be used to image the structure of hadrons and of the hadronization mechanism in a 3D momentum space. 
One of the scattering process that we can use to access TMD distributions is Semi-Inclusive DIS (SIDIS), a generalization of DIS in which one hadron is detected in the final state. 
Measuring one hadron in the final state is necessary in order to access (collinear or TMD) fragmentation functions (FFs). Additionally measuring the transverse momentum of the detected hadron is essential in order to study the transverse momentum dependence of the partons active in the process via TMD distributions (TMDs).

\section{Transverse Momentum Distributions}
\label{s:TMDs}

\subsection{TMD PDFs}
\label{ss:TMDPDFs}

We start by introducing a more general version of the correlation function defined in Eq.~\eqref{e:coll_phi}, named unintegrated correlator. This is a function of partonic/hadronic four momenta and of the target spin.
The unintegrated $\Phi$ quark distribution correlator is defined as~\cite{Bacchetta:2006tn,Mulders:1995dh,Echevarria:2016scs}:
\begin{equation}
\label{e:unint_phi}
\Phi_{ij}(k,P,S) = \int \frac{d^4 \xi}{(2\pi)^4} e^{i k \cdot \xi}\, \text{Tr}_c 
\langle PS|  
\timeord \big[ \overline{\psi}_j(0) W_2(0,\infty) \big] \, 
\antitimeord \big[ W_1(\infty,\xi) \psi_i(\xi) \big] \, 
|PS \rangle  \, .
\end{equation}
In Eq.~\eqref{e:unint_phi} 
$k$ is the quark four momentum, $P$ is the hadronic four momentum, 
$S$ is the covariant spin vector. 
The operators $\timeord$, $\antitimeord$ represent time ordering\footnote{See Ref.~\cite{Collins:2011zzd} Sec. 6.9.4 for the role of time ordering in uncut amplitudes vs the insertion of a sum on intermediate on-shell states $\sum_X |X\rangle \langle X|$.} and anti-time ordering (respectively)~\cite{Diehl:1998sm} and $W_{1,2}$ are the gauge link operators (Wilson lines) that render the $\Phi$ operator gauge invariant (see also Part~\ref{p:symmetries}).
One should also keep in mind that only the connected contributions to $\Phi$ need to be considered~\cite{Jaffe:1983hp,Collins:2011zzd}.

The TMD parton distribution correlator is defined as:
\begin{align}
\label{e:TMD_Phi}
\Phi_{ij}(x,k_T;S) & \equiv \int_{-\infty}^{+\infty} dk^-\, dk^+\, \delta(k^+ - xP^+)\, \Phi(k, P) = 
\int dk^-\, \text{Disc} [ \Phi(k, P) ]_{\begin{subarray}{l} k^+=xP^+ \\  \end{subarray}} \\
\nonumber 
& = \int \frac{d\xi^- d^2\xi_T}{(2\pi)^3}\, e^{i k \cdot \xi}\,
\langle P,S|\, \overline{\psi}_j(0)\, \psi_i(\xi)\, | P,S \rangle 
\Big|_{\begin{subarray}{l}
\xi^+=0 \\
k^+=xP^+
\end{subarray}} \, ,
\end{align}
where $x=p^+/P^+$ is the parton light cone momentum fraction. 
A further integration over $k_T$ yields the collinear distribution correlator introduced in Eq.~\eqref{e:coll_phi}:
\begin{align}
\label{e:coll_Phi_2}
\Phi_{ij}(x;S) = \int d^2k_T\, \Phi(x,k_T) 
= \int \frac{d\xi^-}{2\pi}\, e^{i k \cdot \xi}\,
\langle P,S|\, \overline{\psi}_j(0)\, \psi_i(\xi)\, | P,S \rangle 
\Big|_{\begin{subarray}{l}
\xi^+=0 \\
\xi_T=0 \\
k^+=xP^+
\end{subarray}} \, .
\end{align}
The Dirac projections of collinear or TMD correlators are defined as:
\begin{equation}
\label{e:proj_Phi}
\Phi^{[\Gamma]}(x,\cdots) \equiv \text{Tr} \bigg[ \Phi(x,\cdots) \frac{\Gamma}{2} \bigg] \, ,
\end{equation}
where $\Gamma$ is a generic Dirac matrix. For example, $\Gamma=\gamma^+$ is associated with the unpolarized distributions, whereas the matrices associated with the other TMDs can be found in e.g. Ref.~\cite{Bacchetta:2006tn}.
In Eq.~\eqref{e:proj_Phi} the operator ``Tr'' without subscripts indicates a Dirac trace, whereas the $1/2$ factor is justified in Section 6.7 and 6.8 of Ref.~\cite{Collins:2011zzd}. 
In the spirit of Feynman rules, the Dirac trace operator ``$\text{Tr}$'' corresponds to the {\it sum} over the polarization states of the quark in the final state of the diagram that represents $\Phi(k,P,S)$ (see Fig.~\ref{f:OPE_diag}(b)). 
The analogous color trace $\text{Tr}_c$, corresponding to a sum over the color configurations, is included in Eq.~\eqref{e:unint_phi}. 
Assuming that the correlators have the standard analiticity properties of the scattering amplitudes, the integration over the suppressed momentum component $k^-$ in the definition of the TMD correlator, Eq.~\eqref{e:TMD_Phi}, can be performed by complex contour deformation. Depending on the value of $x$, one can then replace the integral of the unintegrated correlator by the integral of its $s$-channel or $u$-channel discontinuity, denoted by ``Disc''. These correspond, respectively to a quark distribution ($0 \leq x \leq 1$), and to an antiquark distribution ($-1 \leq x \leq 0$)~\cite{Jaffe:1983hp,Boer:1998im,Diehl:2003ny,Gamberg:2010uw,Diehl:1998sm}. 

As for the collinear case, also the TMD correlator $\Phi(x,k_T;S)$ is a non-perturbative Dirac matrix that encodes information about hadron structure at $\xi^+=0$. Accordingly, it can be parametrized using a basis for the linear space of Dirac matrices relying on the hermiticity property of $\Phi$ and on the invariance of QCD under parity and time reversal transformations (see also Part~\ref{p:symmetries}). 
Here we present the parametrization of the TMD correlator up to twist 3 (considering the operational definition of twist~\cite{Jaffe:1996zw})~\cite{Bacchetta:2006tn}: 
\begin{align}
\label{e:TMD_Phi_tw23_param}
\Phi(x,k_T) & = \frac{1}{2}\, \biggl\{ 
f_1 \slash{n}_+ 
- {f_{1T}^\perp}\, \frac{\eps_T^{\rho \sigma} k_{T\rho}^{} S_{T\sigma}^{}}{M} \, \slash{n}_+ 
+ g_{1s} \gamma_5\slash{n}_+
\nonumber \\[0.2em] & \quad \qquad
+h_{1T}\,\frac{\bigl[\slash{S}_T, \slash{n}_+ \bigr]\gamma_5}{2}
+ h_{1s}^\perp \,\frac{\bigl[\slash{k}_T, \slash{n}_+ \bigr]\gamma_5}{2 M}
+i \, {h_1^\perp} \frac{ \bigl[\slash{k}_T, \slash{n}_+ \bigr]}{2M}
\biggr\} 
\nonumber \\[0.2em] & \quad
+ \frac{M}{2 P^+}\,\biggl\{ 
e 
- i \,{e_s} \,\gamma_5 
- {e_{T}^\perp}\, \frac{\eps_T^{\rho \sigma} k_{T\rho}^{}
  S_{T\sigma}^{}}{M} 
\nonumber \\ & \quad \qquad \qquad
+ f^\perp\, \frac{\slash{k}_T}{M}
- {f_T'}\,\epsilon_T^{\rho\sigma} \gamma_\rho^{} S_{T \sigma}^{}
- {f_s^{\perp}}\,\frac{\eps_T^{\rho \sigma}
    \g_{\rho}^{} k_{T \sigma}^{}}{M}
\nonumber \\ & \quad \qquad \qquad
+ g_T'\, \gamma_5 \slash{S}_T
+ g_s^{\perp} \gamma_5 \frac{\slash{k}_T}{M}
- {g^\perp} \g_5\,\frac{\eps_T^{\rho \sigma} 
    \g_{\rho}^{} k_{T \sigma}^{}}{M}
\nonumber \\[0.2em] & \quad \qquad  \qquad
+ h_s\,\frac{[\slash{n}_+, \slash{n}_-]\gamma_5}{2}
+ h_T^{\perp}\,\frac{\bigl[\slash{S}_T, \slash{k}_T \bigr]\gamma_5}{2 M}
+ i \, {h} \frac{ \bigl[\slash{n}_+, \slash{n}_- \bigr]}{2} 
\biggr\} .
\end{align}
The distribution functions on the RHS depend on $x$ and $k_T^2$,
except for the functions with subscript $s$, where a shorthand notation is used~\cite{Mulders:1995dh}:
\begin{equation} 
\label{e:TMDs_sub_s}
g_{1s}(x, k_T) 
= S_L\, g_{1L}(x, k_T^2) - \frac{k_T \cdot S_T}{M}\, g_{1T}(x, k_T^2)\, ,
\end{equation} 
and so forth for the other functions.
These functions can be organized according to the polarization state of the quark and of the nucleon target they refer to. The classification is summarized in Tab.~\ref{t:TMD_PDFs}, where Tab.~\ref{t:tw2_TMD_PDFs}, in particular, refers to the twist 2 case, whereas Tab.~\ref{t:tw3_TMD_PDFs} deals with the twist three case. The color code that is usually included in this kind of tables is omitted at this stage, and it will be introduced in Part~\ref{p:symmetries}, after discussing the time reversal transformations. 
The first eight distributions of Eq.~\eqref{e:TMD_Phi_tw23_param} are the twist two TMD PDFs, and the next 
16 distributions are referred to as twist three TMD PDFs. In principle there are other eight functions of twist
four, which have been omitted here and can be found in Ref.~\cite{Goeke:2005hb}.
The 10 functions $f_{1T}^{\perp}$, $h_1^{\perp}$, $e_L^{}$, $e_T^{}$,
$e_T^{\perp}$, $f_T'$, $f_L^{\perp}$, $f_T^{\perp}$, $g^{\perp}$, $h$
are odd under time reversal transformations (T-odd)~\cite{Boer:1997nt,Goeke:2005hb} (see Part~\ref{p:symmetries}).
This nomenclature for the distribution functions follows closely that of
Ref.~\cite{Mulders:1995dh}, sometimes referred to as ``Amsterdam notation.''
A number of other notations exist in the literature for some of the
distribution functions, see e.g. Refs.~\cite{Ralston:1979ys,Barone:2001sp,Idilbi:2004vb,Ebert:2021jhy,Rodini:2023plb}.  
Leading twist TMDs have been widely discussed also by Anselmino et al in Refs.~\cite{Anselmino:1995vq,
Anselmino:1996qx,Anselmino:2005sh}. 
A dictionary to relate some of the notations is provided in Ref.~\cite{Anselmino:2005sh}.

A similar classification is available for gluon TMD PDFs, with the difference that the gluon correlation function is a Lorentz matrix and not a Dirac one. A detailed treatment of gluon TMDs can be found in e.g. Ref.~\cite{Boer:2016xqr} and references therein.

For a spin one hadron, given the additional dependence of the correlator on the tensor spin components, there are in total 18 TMD PDFs at leading twist. Their description can be found in e.g. Refs.~\cite{Bacchetta:2000jk,vanDaal:2016glj,Zhao:2025vol}.

\renewcommand{\arraystretch}{1.5}
\begin{table}[t]
\centering

\begin{subtable}{0.48\textwidth}
\centering
\begin{tabular}{c|c|c|c|c|}
\multicolumn{2}{c}{}&\multicolumn{3}{c}{quark pol.}\\
\cline{2-5}
& & U & L & T \\ \cline{2-5}
\multirow{3}{*}{\rotatebox{90}{nucleon pol.}} 
&{U} & $f_1$   & & $h_{1}^\perp$ \\
\cline{2-5}
& {L} & & $g_{1L}$ & $h_{1L}^\perp$ \\
\cline{2-5}
& {T} & $f_{1T}^\perp$ & $g_{1T}$ & $h_{1T}$, $h_{1T}^{\perp}$ \\
\cline{2-5} 
\end{tabular}
\caption{Twist-2 TMD PDFs}
\label{t:tw2_TMD_PDFs}
\end{subtable}
\hfill
\begin{subtable}{0.48\textwidth}
\centering
\begin{tabular}{c|c|c|c|c|}
\multicolumn{2}{c}{}&\multicolumn{3}{c}{quark pol.}\\
\cline{2-5}
& & U & L & T \\ \cline{2-5}
\multirow{3}{*}{\rotatebox{90}{nucleon pol.}} 
&{U} & $f^{\perp}$ & $g^{\perp}$ & $h$, $e$ \\
\cline{2-5}
& {L} & $f_{L}^\perp$ & $g_{L}^{\perp}$ & $h_L$, $e_L$ \\
\cline{2-5}
& {T} & $f_T'$, $f_{T}^\perp$ & $g_T'$, $g_{T}^{\perp}$ & $h_T$, $h_{T}^{\perp}$, $e_T$, $e_{T}^\perp$ \\
\cline{2-5}
\end{tabular}
\caption{Twist-3 TMD PDFs}
\label{t:tw3_TMD_PDFs}
\end{subtable}

\caption{Transverse-momentum-dependent quark distribution functions at twist-2 and twist-3, organized by polarization state. Columns refer to the quark polarization, rows to the nucleon polarization.}
\label{t:TMD_PDFs}

\end{table}
\renewcommand{\arraystretch}{1.5}
\begin{table}[t]
\centering

\begin{subtable}{0.48\textwidth}
\centering
\begin{tabular}{c|c|c|c|c|}
\multicolumn{2}{c}{}&\multicolumn{3}{c}{Dirac matrix}\\
\cline{2-5}
& & $\gamma^+$ & $\gamma^+\, \gamma_5$ & $i\, \sigma^{\alpha +}\, \gamma_5$ \\ \cline{2-5}
\multirow{3}{*}{\rotatebox{90}{nucleon pol.}} 
&{U} & $f_1$   & & $h_{1}^\perp$ \\
\cline{2-5}
& {L} & & $g_{1L}$ & $h_{1L}^\perp$ \\
\cline{2-5}
& {T} & $f_{1T}^\perp$ & $g_{1T}$ & $h_{1T}$, $h_{1T}^{\perp}$ \\
\cline{2-5} 
\end{tabular}
\caption{Twist-2 TMD PDFs}
\label{t:tw2_TMD_PDFs_Dirac}
\end{subtable}
\hfill
\begin{subtable}{0.48\textwidth}
\centering
\begin{tabular}{c|c|c|c|c|c|c|c|}
\multicolumn{2}{c}{} & \multicolumn{6}{c}{Dirac matrix} \\
\cline{2-8}
& & $\gamma^\alpha$ & $\gamma^\alpha\, \gamma^5$ & $i\, \sigma^{+-}\, \gamma_5$ & $i\, \sigma^{\alpha\beta}\, \gamma_5$ & $i\, \gamma_5$ & $\mathbf{1}$ \\
\cline{2-8}
\multirow{3}{*}{\rotatebox{90}{nucleon pol.}} 
& {U} & $f^{\perp}$ & $g^{\perp}$ & & $h$ & & $e$ \\
\cline{2-8}
& {L} & $f_{L}^\perp$ & $g_{L}^{\perp}$ & $h_L$ & & $e_L$ & \\
\cline{2-8}
& {T} & $f_T'$, $f_{T}^\perp$ & $g_T'$, $g_{T}^{\perp}$ & $h_T$ & $h_{T}^{\perp}$ & $e_T$ & $e_{T}^\perp$ \\
\cline{2-8}
\end{tabular}
\caption{Twist-3 TMD PDFs}
\label{t:tw3_TMD_PDFs_Dirac}
\end{subtable}
\caption{TMD PDFs organized by polarization of the parent hadron (rows) and Dirac matrices (columns). At twist 2 there is a one-to-one correspondence between the quark polarization state and a corresponding Dirac structure. At twist 3, in particular for transversely polarized quarks, the situation is more involved. The Dirac matrices mentioned in these tables correspond to the $\Gamma$ structures entering in Eq.~\eqref{e:proj_Phi} to calculate a specific distribution function. The corresponding table for TMD FFs can be obtained interchanging plus and minus directions in the light-cone basis ($n_+ \leftrightarrow n_-$).}
\label{t:TMD_PDFs_Dirac}

\end{table}

\subsection{TMD FFs}
\label{ss:TMDFFs}

As a consequence of the confinement of the strong force, any individual parton struck in a high-energy scattering process and extracted from its parent hadron must transform into at least one hadron -- in technical language, it must ``hadronize''.
During this process, a struck light quark, such as an up, down or strange quark, initially propagates as a high-energy but nearly massless colored particle radiating gluons. Before being observed, however, this system of colored, nearly massles particles will turn into a number of massive, color-neutral hadrons such as pions, kaons and protons. Hadronization is thus quite clearly and tightly connected to parton propagation, color charge neutralization, and dynamical generation of the mass, spin, and size of hadrons. However, the exact details of this parton-to-hadrons transition are poorly known. 
For this reason, it can be useful to introduce a mathematical way to describe the hadronization mechanism in momentum space, via the so-called parton fragmentation functions (FFs).
Refs.~\cite{Metz:2016swz,Boer:2008fr} provide an overview of the FFs and of electron-positron annihilation into hadrons, one of the standard processes used to study hadronization and FFs. \\

As for parton distribution functions, we introduce an unintegrated correlator, whose projections and Dirac traces are used to introduce collinear and TMD fragmentation functions.
The unintegrated correlator describing the fragmentation of a quark into a single hadron (or ``single-inclusive" correlator) is defined as~\cite{Mulders:1995dh,Goeke:2003az,Bacchetta:2006tn,Meissner:2007rx,Mulders:2016pln,Metz:2016swz,Echevarria:2016scs}:
\begin{align}
  \label{e:1hDelta_corr}
  & \Delta^h_{ij}(k,P,S) = \sum_X \int \frac{d^4 \xi}{(2\pi)^4} e^{i k \cdot \xi} \colav \langle 0| 
  \timeord \big[ W_1(\infty,\xi) \psi_i(\xi) \big] \, 
  |P S X \rangle 
  \langle P S X| \, 
  \antitimeord \big[ \overline{\psi}_j(0) W_2(0,\infty) \big] |0\rangle\, ,
\end{align}
where $k$ is the quark's four-momentum, $h$ is an identified hadron with four-momentum $P$ and spin $S$, and $X$ represents the quantum numbers of all unobserved hadrons in the final state. 

The fragmentation process can be studied either in the parton or in the hadron frame~\cite{Collins:2011zzd,Levelt:1993ac}. The Lorentz transformation between these frames and its consequences are discussed in detail in Ref.~\cite{Accardi:2020iqn}\footnote{See also Ref.~\cite{Mulders:2016pln} and Section~12.4.1 in Ref.~\cite{Collins:2011zzd}.}. 
In the parton frame, defined such that the parton's transverse momentum $\vect{k}_T=0$, one can interpret the fragmentation correlator as the probability density (in a loose sense) for the quark to fragment into a hadron of a given flavor $h$ and momentum $P$, with $\vect{P}_T$ generically non zero \cite{Collins:2011zzd,Metz:2016swz}. 
The parton frame, however, turns out not to be convenient in derivations of factorization theorems and calculations of semi-inclusive cross sections: the partonic momenta are integrated over, and the parton frame axes are not fixed (see Chap. 12 in Ref.~\cite{Collins:2011zzd}). In this case, it is preferable to utilize the hadron frame, where the experimentally observable hadron's 3-momentum determines the $z$ direction, so that $\vect{P}_T=0$. In this frame, it is the quark's transverse momentum that, in general, has a non-zero value.

The TMD parton fragmentation correlator in the parton frame is defined as:
\begin{equation}
\label{e:TMD_Delta}
\Delta(z,P_T;S) \equiv \int_{-\infty}^{+\infty} \frac{dk^+}{2z}\, \Delta(k, P)_{\begin{subarray}{l} P^-= zk^- \\  \end{subarray}}
= \int \frac{dk^+}{2z}\, \text{Disc}\, [ \Delta(k, P) ]_{\begin{subarray}{l} P^-= zk^- \\  \end{subarray}} \, ,
\end{equation}
with $z=P^-/k^-$ is the hadron's fractional momentum in the dominant momentum direction. The explicit expression reads:
\begin{equation}
\label{e:1hDelta_TMDcorr}
\Delta^h_{ij}(z,P_T) = \frac{1}{2z} \int \frac{d \xi^+ d^2 \vect{\xi}_T}{(2\pi)^3} e^{\i k^- \xi^+} 
\colav\, \text{Disc}\, \langle 0| 
W_1(\infty,\xi) \psi_i(\xi) (a_h^\dagger a_h) \overline{\psi}_j(0) W_2(0,\infty)  
|0 \rangle_{\begin{subarray}{l} \xi^-=0 \\ k^- = P^-/z \end{subarray}}  \ . 
\end{equation}

Moreover, by integrating over the hadronic transverse momentum one obtains a collinear matrix element~\cite{Bacchetta:2006tn}:
\begin{equation}
\label{e:1hDelta_integrated}
\Delta^h(z) = \int d^2 \vect{P}_T\, \Delta^h(z,P_T) \, ,
\end{equation}
where the collinear fragmentation correlator $\Delta^h(z)$ is defined as
\begin{equation}
\label{e:1hDelta_coll_op}
\Delta^h_{ij}(z) = \frac{z}{2} \int \frac{d \xi^+ }{2\pi} e^{\i \xi^+ P^-/z} 
\colav\, \text{Disc}\, \langle 0| 
W_1(\infty,\xi) \psi_i(\xi) (a_h^\dagger a_h) \overline{\psi}_j(0) W_2(0,\infty)  
|0 \rangle_{\begin{subarray}{l} \xi^-=\vect{\xi}_T=0 \\ P_T=0 
\end{subarray}}  \ . 
\end{equation}

The Dirac projections are defined as:
\begin{equation}
\label{e:proj_int_Delta}
\Delta^{[\Gamma]}(z,P_T)\ \equiv\ \text{Tr} \bigg[ \Delta(z,P_T) \frac{\Gamma}{2} \bigg] \, ,
\end{equation}
%
where the $1/2$ factor in Eq.~\eqref{e:proj_int_Delta} arises in the same way as in Eq.~\eqref{e:proj_Phi}. 
In Eq.~\eqref{e:1hDelta_TMDcorr}, the $1/z$ factor comes from the normalization of the hadronic states (see Ref.~\cite{Collins:1981uw} and Section 12.4 in Ref.~\cite{Collins:2011zzd}). 
The Dirac trace operator in eq.~\eqref{e:proj_int_Delta} receives an additional $1/2$ factor which is explicitly included in Eq.~\eqref{e:1hDelta_TMDcorr}, since in this case one needs to calculate an {\it average} over the quark polarizations in the initial state. 
On the same footing, a color trace Tr$_c/N_c$, that in the same way corresponds to an average over the parton's color configurations, is already included in the definition of the unintegrated correlator $\Delta(k,P)$. 

As for the TMD PDFs, the TMD fragmentation correlator can be parametrized using a basis for the linear space of Dirac matrices relying on the hermiticity property and on the invariance of QCD under parity and time reversal transformations. As mentioned in Part~\ref{p:symmetries}, time reversal is more involved in the hadronization case and needs special care~\cite{Boer:2003cm}. 
The parametrization of the TMD hadronization correlator up to twist 3 can be obtained from Eq.~\eqref{e:TMD_Phi_tw23_param} with the following replacements\footnote{The change of sign of the tensor $\eps_T$ is due to the exchange $n_+ \leftrightarrow n_-$ in its definition.}:
\begin{align}
\label{e:Phi_Delta_replacements}
n_+ &\leftrightarrow n_-,
&
\eps_T &\to -\eps_T,
&
P^+ &\to P_h^-, 
&
M & \to M_h,
&
x & \to 1/z \, .
\end{align} 
Moreover, the distribution functions are replaced with the corresponding
fragmentation functions, namely $f$ is replaced with $D$ and all other
letters are capitalized.
The result is:
\begin{align}
\label{e:TMD_Delta_tw23_param}
\Delta(z,P_T) & = \frac{1}{2}\, \biggl\{ 
D_1 \slash{n}_- 
- {D_{1T}^\perp}\, \frac{\eps_T^{\rho \sigma} P_{T\rho}^{} S_{T\sigma}^{}}{z\,M_h} \, \slash{n}_- 
+ G_{1s} \gamma_5\slash{n}_-
\nonumber \\[0.2em] & \quad \qquad
+H_{1T}\,\frac{\bigl[\slash{S}_T, \slash{n}_- \bigr]\gamma_5}{2}
- H_{1s}^\perp \,\frac{\bigl[\slash{P}_T, \slash{n}_- \bigr]\gamma_5}{2 z\, M_h}
- i \, {H_1^\perp} \frac{ \bigl[\slash{P}_T, \slash{n}_- \bigr]}{2 z\, M_h}
\biggr\} 
\nonumber \\[0.2em] & \quad
+ \frac{M_h}{2 P_h^-}\,\biggl\{ 
E 
- i \,{E_s} \,\gamma_5 
- {E_{T}^\perp}\, \frac{\eps_T^{\rho \sigma} P_{T\rho}^{}
  S_{T\sigma}^{}}{z\,M_h} 
\nonumber \\ & \quad \qquad \qquad
- D^\perp\, \frac{\slash{P}_T}{z\,M_h}
+ {D_T'}\,\epsilon_T^{\rho\sigma} \gamma_\rho^{} S_{T \sigma}^{}
- {D_s^{\perp}}\,\frac{\eps_T^{\rho \sigma}
    \gamma_{\rho}^{} P_{T \sigma}^{}}{z\,M_h}
\nonumber \\ & \quad \qquad \qquad
+ G_T'\, \gamma_5 \slash{S}_T
- G_s^{\perp} \gamma_5 \frac{\slash{P}_T}{z\,M_h}
- {G^\perp} \gamma_5\,\frac{\eps_T^{\rho \sigma} 
    \gamma_{\rho}^{} P_{T \sigma}^{}}{z\,M_h}
\nonumber \\[0.2em] & \quad \qquad  \qquad
+ H_s\,\frac{[\slash{n}_-, \slash{n}_+]\gamma_5}{2}
- H_T^{\perp}\,\frac{\bigl[\slash{S}_T, \slash{P}_T \bigr]\gamma_5}{2 z\, M_h}
+ i \, {H} \frac{ \bigl[\slash{n}_-, \slash{n}_+ \bigr]}{2} 
\biggr\} \, ,
\end{align}
where the TMD FFs are functions of $z$ and $P_T^2$ in the parton frame. 
The subscript $s$ notation follows again from Eq.~\eqref{e:TMDs_sub_s}.

Also the TMD FFs can be organized according to the polarization state of the quark and of the produced hadron. The classification is summarized in Tab.~\ref{t:TMD_FFs}, where Tab.~\ref{t:tw2_TMD_FFs}, in particular, refers to the twist 2 case, whereas Tab.~\ref{t:tw3_TMD_FFs} deals with the twist three case. The color code that is usually included in this kind of tables is omitted at this stage, and it will be introduced in Part~\ref{p:symmetries}, after discussing the time reversal transformations. \\

Last but not least, it is important to remark that equations of motion relations (see Sec.~\ref{ss:twist}) can be derived also for fragmentation functions. 
Considering the relations between twist-2 and twist-3 fragmentation functions in the parton frame for unpolarized hadrons one has:
\begin{align}
  \label{e:eom_E}
  & E^h = \widetilde{E}^h + z \frac{m}{M_h} D^h_1 \\
  \label{e:eom_H}
  & H^h = \widetilde{H}^h - \frac{\vect{P}_T^2}{z M_h^2} H_1^{\perp\, h} \\
  \label{e:eom_Dp}
  & D^{\perp\, h} = \widetilde{D}^{\perp\, h} + z D^h_1 \\
  \label{e:eom_Gp}
  & G^{\perp\, h} = \widetilde{G}^{\perp\, h} + z \frac{m}{M_h} H_1^{\perp\, h} \ ,
\end{align}
where the functions with a tilde parametrize the twist-3 $\widetilde{\Delta}_A^\alpha$ quark-gluon-quark correlator~\cite{Bacchetta:2006tn}, and $m$ is the current mass of the specific quark considered. These relations, that are a consequence of the Dirac equation for the quark field, have been originally presented in the hadron frame~\cite{Tangerman:1994bb,Mulders:1995dh} (see also Ref.~\cite{Bacchetta:2006tn}). In Ref.~\cite{Accardi:2020iqn} their transformation to the parton frame is discussed.

\renewcommand{\arraystretch}{1.5}
\begin{table}
\centering

\begin{subtable}{0.48\textwidth}
\centering
\begin{tabular}{c|c|c|c|c|}
\multicolumn{2}{c}{}&\multicolumn{3}{c}{quark pol.}\\
\cline{2-5}
& & U & L & T \\ \cline{2-5}
\multirow{3}{*}{\rotatebox{90}{hadron pol.}} 
&{U} & $D_1$ & & $H_{1}^\perp$ \\
\cline{2-5}
& {L} & & $G_{1L}$ & $H_{1L}^\perp$ \\
\cline{2-5}
& {T} & $D_{1T}^\perp$ & $G_{1T}$ & $H_{1}$, $H_{1T}^{\perp}$ \\
\cline{2-5} 
\end{tabular}
\caption{Twist-2 TMD fragmentation functions}
\label{t:tw2_TMD_FFs}
\end{subtable}
\hfill
\begin{subtable}{0.48\textwidth}
\centering
\begin{tabular}{c|c|c|c|c|}
\multicolumn{2}{c}{}&\multicolumn{3}{c}{quark pol.}\\
\cline{2-5}
& & U & L & T \\ \cline{2-5}
\multirow{3}{*}{\rotatebox{90}{hadron pol.}} 
&{U} & $D^{\perp}$ & $G^{\perp}$ & $H$, $E$ \\
\cline{2-5}
& {L} & $D_{L}^\perp$ & $G_{L}^{\perp}$ & $H_L$, $E_L$ \\
\cline{2-5}
& {T} & $D_{T}$, $D_{T}^\perp$ & $G_{T}$, $G_{T}^{\perp}$ & $H_T$, $H_{T}^{\perp}$, $E_T$, $E_{T}^\perp$ \\
\cline{2-5}
\end{tabular}
\caption{Twist-3 TMD fragmentation functions}
\label{t:tw3_TMD_FFs}
\end{subtable}

\caption{Transverse-momentum-dependent quark fragmentation functions at twist-2 and twist-3, organized by polarization state. Columns refer to the quark polarization, rows to the nucleon polarization.}
\label{t:TMD_FFs}

\end{table}

\section{The Drell-Levy-Yan relation}
\label{s:DLY}

We now present a relation based on the crossing symmetry for Feynman amplitudes which allows one to connect the operators that define the parton distribution and fragmentation functions. 
This relation suggests that PDFs and FFs are essentially one and the same function, evaluated on different supports in the light cone momentum fraction~\cite{Gamberg:2010uw}. 
A similar and related relation exists between the structure functions in lepton-hadron scattering and lepton pair annihilation, the so-called Drell-Levy-Ian (DLY) relation~\cite{Drell:1969jm,Drell:1969wd,Drell:1969wb}, which predates QCD. 
In Ref.~\cite{Ito:2009zc} it was shown that the relation between collinear parton distribution and fragmentation functions in the unpolarized case can be derived from the original DLY relation. Accordingly, we use the term ``DLY relation'' to indicate both the original relation between the structure functions and its more recent form between distribution and fragmentation functions. 
Here we generalize the analysis of Ref.~\cite{Ito:2009zc} to different Dirac structures working at the level of the unintegrated quark distribution and fragmentation correlators. We will also comment on some of the limitations of this approach.

\subsection{Unpolarized hadrons}
\label{ss:DLY_unpol_h}

By applying the crossing symmetry of Feynman amplitudes~\cite{Peskin:1995ev} on the unintegrated fragmentation correlator in Eq.~\eqref{e:1hDelta_corr}, one can exchange $h(P)$ in the initial (final) state with $\overline{h}(-P)$ in the final (initial) state. 
In order to neglect the complications associated with the spin~\cite{Peskin:1995ev}, we consider a spin 0 hadron. The argument can be however generalized to higher spin states. 
In this case one obtains: 
\begin{equation}
\label{e:DLY_unintegrated}
N_c\, \Delta_{ij}^{a \to h}(k,P) = \overline{\Phi}_{ij}^{\, a/\overline{h}}(-k,-P)\, ,
\end{equation}
where the $\overline{\Phi}$ correlator is defined in Eq.~\eqref{e:Phibar} (including the Wilson lines), the superscripts specify the quark/hadron combination, $k$ is the parton momentum, and $P$ is the hadron momentum.

By integrating both sides of Eq.~\eqref{e:DLY_unintegrated} over the partonic four momentum as in Eqs.~\eqref{e:coll_Phi_2},~\eqref{e:1hDelta_integrated} and by tracing with a specific Dirac matrix $\Gamma$ we can obtain a relation between the collinear fragmentation and distribution correlators: 
\begin{equation}
\label{e:DLY_coll}
\frac{2 N_c}{z}\, \Delta^{a \to h\, [\Gamma]}(z \equiv 1/x) = 
\overline{\Phi}^{\, a/\overline{h}\, [\Gamma]}(x) \, ,
\end{equation}
where at the LHS one can identify $x$ with $1/z$. 

At this point, relying on the charge conjugation relations described in Sec.~\ref{ss:charge}, one can show that:
\begin{equation}
\label{e:DLY_coll_antiq}
\overline{\Phi}^{\, a/\overline{h}\, [\Gamma]}(x) = 
\pm \Phi^{c, \, \overline{a}/\overline{h}\, [\Gamma]}(x) = \pm \Phi^{a/h\, [\Gamma]}(x) \, ,
\end{equation}
where in the first equality the $\pm$ sign depends on the specific Dirac structure $\Gamma$ considered (see Eqs.~\eqref{e:sign_plus_PhicGamma},\eqref{e:sign_minus_PhicGamma}) whereas the second equality means that a distribution for a specific quark/antiquark $a$ in a hadron $h$ is equal to the distribution for the associated anti-particle $\bar{a}$ in an antihadron $\bar{h}$ (see Eq.~\eqref{e:Phi_Phic_corr} and Ref.~\cite{Collins:2011zzd}). 
Combining Eqs.~\eqref{e:DLY_coll} and~\eqref{e:DLY_coll_antiq} one obtains a relation between the projections of the quark distribution and fragmentation collinear correlators:
\begin{equation}
\label{e:DLY_coll_final}
\frac{2 N_c}{z}\, \Delta^{a \to h\, [\Gamma]}(z \equiv 1/x) = \pm \Phi^{a/h\, [\Gamma]}(x) \, .
\end{equation}
As we mentioned earlier, this relation suggests that PDFs and FFs are one and the same function evaluated in different kinematic regions. 
In particular, the distribution functions obtained for $|x|<1$ are related to the fragmentation functions in the unphysical region $|z|>1$, and, vice-versa, the fragmentation functions obtained for $|z|<1$ are related to the distribution functions in the unphysical region $|x|>1$. 
A similar relation can be obtained for the transverse-momentum-dependent distribution and fragmentation correlators, by avoiding the integration on the partonic transverse momentum~\cite{Jakob:1997wg}. 

By calculating the relevant Dirac traces, one obtains relations that allow one to calculate the fragmentation functions for $|z|<1$ from the corresponding distribution functions in the unphysical region $|x|=|1/z|>1$. Considering only the unpolarized and the scalar PDFs:
\begin{align}
\label{e:DLY_coll_gp}
[\, \Gamma = \gamma^+\, ]\ \ \ \ \ & D_1(z) = + \frac{z}{2 N_c}\, f_1(1/z) \, , \\
\label{e:DLY_coll_id}
[\, \Gamma = \mathbb{I}\, ]\ \ \ \ \ & E(z) = - \frac{z}{2 N_c}\, e(1/z) \, . 
\end{align}

\subsection{Limitations}
\label{ss:DLY_limitations}

At this point some remarks are in order. 
First of all, these relations are presented at the bare level in the context of QCD and are, in general, broken by renormalization (evolution) effects in perturbative QCD~\cite{Ito:2009zc}. Thus, Eq.~\eqref{e:DLY_coll_gp} is of little use in the context of perturbative QCD-based phenomenological extractions of PDFs and FFs from experimental data (also in light of the fact that phenomenological extractions probe only the $0 < x,z < 1$ regions).  
Moreover, in the context of perturbative QCD, the evolution of a PDF and FF associated to the same Dirac structure cannot in general be related by analytic continuation, apart from some specific cases for which the so-called Gribov-Lipatov reciprocity relation holds true~\cite{Belitsky:1997ay,Henneman:2001ev,Ito:2009zc,Kiselev:2025lpb,Kartvelishvili:1977pi}. 
Nonetheless, in a non-renormalizable effective theory such as the Nambu--Jona-Lasinio (NJL) model these breaking effects are absent, and the validity of the DLY relations can be assumed at least at the scale at which the model calculation is performed. 

One should also note that Eq.~\eqref{e:DLY_coll} is obtained by applying the crossing symmetry on the identified hadron states in $\Delta(k,P)$ and by using the completeness relation for the unobserved states. These two steps combined effectively prevent the exchange of ``in'' and ``out'' states that generates the time reversal-odd (T-odd) components in $\Delta(z)$~\cite{Boer:2003cm}. Thus, obtaining the collinear FFs from the collinear PDFs via the DLY relations, forces one to consider only the T-even structures, at least at the collinear level.

Additionally, the DLY does not establish any relation between the moments of the PDFs and the FFs, since these are calculated in different kinematic regions for the light-cone fractions $x$ and $z$. 

Finally, the FFs obtained via the DLY relations do not satisfy the momentum and isospin sum rules~\cite{Ito:2009zc}. This can be explicitly checked, for example, in the NJL model of QCD. The reason is that hadronization is a dynamical multi-step process, that cannot be approximated by a single quark-antiquark-hadron vertex. Accordingly, the ``elementary'' FFs obtained by using the DLY relations on PDFs need to be embedded in a multi-step stastical description of hadronization (e.g. the ``NJL-jet'' model described in Ref.~\cite{Ito:2009zc,Bentz:2016rav}). By doing that, one obtains expressions that comply with the momentum and isospin sum rules.    

\section{Semi-Inclusive DIS}
\label{s:SIDIS}


Now that we have introduced the TMD correlation functions, we need to introduce one of the scattering processes whose cross section can be written as a function of these correlators. 
Closely following the treatment given in Ref.~\cite{Bacchetta:2006tn}, we consider one-particle inclusive, or semi-inclusive, deep inelastic scattering (SIDIS), where a lepton with momentum $l$ scatters off a hadron target (a nucleon $N$ in this case) with momentum $P$ and a lepton and one hadron are detected in the final state, with momenta $l^\prime$ and $P_h$ respectively:
\begin{equation}
\label{e:SIDIS_reaction}
\ell(l) + N(P) \to \ell'(l') + h(P_h) + X \, .
\end{equation}

\subsection{Kinematics}
\label{ss:SIDIS_kinematics}

We work in the one-photon exchange approximation
and neglect the lepton mass. We denote by $M$ and $M_h$ the respective
masses of the nucleon and of the hadron $h$.  
As usual we define $q =l - l'$ and $Q^2 = - q^2$ and introduce the variables
\begin{align}
\label{xyz}
x_B &= \frac{Q^2}{2\,P\cdot q},
&
y &= \frac{P \cdot q}{P \cdot l},
&
z &= \frac{P \cdot P_h}{P\cdot q},
&
\gamma &= \frac{2 M x}{Q} .
\end{align}
%
We use $P$ and $q$ to build a light cone basis, and neglect ${\mathcal O}(\gamma^2)$ effects. The expressions with the full kinematics including the ${\mathcal O}(\gamma^2)$ corrections are given in Ref.~\cite{Bacchetta:2006tn}.
Following the Trento conventions~\cite{Bacchetta:2004jz} the azimuthal
angle $\phi_h$ of the outgoing hadron is defined as:
\begin{align}
  \label{phi-h-def}
\cos\, \phi_h &= - \frac{l_\mu P_{h \nu}\, g_\perp^{\mu\nu}}{%
  \sqrt{l_\perp^2\, P_{h \perp}^2}} \, ,
&
\sin\, \phi_h &= - \frac{l_\mu P_{h \nu}\, \epsilon_\perp^{\mu\nu}}{%
  \sqrt{l_\perp^2\, P_{h \perp}^2}} \, ,
\end{align}
where 
$l_\perp^\mu = g_\perp^{\mu\nu} l^{}_\nu$ and 
$P_{h \perp}^\mu = g_\perp^{\mu\nu} P^{}_{h \nu}$ 
are the transverse components of $l$ and $P_h$ with respect to the photon momentum. 
The tensors
\begin{align}
\label{e:SIDIS_Pq_basis}
g_\perp^{\mu\nu} &= g_{}^{\mu\nu} 
  - \frac{q^\mu P^\nu + P^\mu q^\nu}{P\cdot q} \, ,
\\
\epsilon_\perp^{\mu\nu} &= \epsilon_{}^{\mu\nu\rho\sigma}
   \frac{P_\rho\, q_\sigma}{P\cdot q} \, 
\end{align}
are the symmetric and antisymmetric projectors in the perpendicular plane.
The covariant spin vector $S$ of the target is decomposed as
\begin{align}
  \label{Sperp-def}
S^\mu &= S_\parallel\, 
   \frac{P^\mu - q^\mu M^2/(P\cdot q)}{M}
   + S^\mu_\perp \,,
&
S_\parallel
  &= M\, \frac{S\cdot q}{P\cdot q} \,,
&
S^\mu_\perp &= g_\perp^{\mu\nu} S^{}_{\nu} \, .
\end{align}
Its azimuthal angle $\phi_S$ is defined in analogy to $\phi_h$ in
Eq.~\eqref{phi-h-def}, with $P_h$ replaced by $S$. 
The helicity of the lepton beam is denoted by $\lambda_e$. 
For sake of simplicity we consider the case where the detected hadron $h$ has spin zero or where its polarization is not measured.

\subsection{Cross section}
\label{ss:SIDIS_cross_section}

Assuming single photon exchange, the lepton-hadron cross section can
be expressed in terms of structure functions~\cite{Bacchetta:2006tn}, similarly to what was done in Eq.~\eqref{e:DIS_xsec_SFs}. The result including the ${\mathcal O}(\gamma^2)$ corrections is:
\begin{align}
\label{e:SIDIS_xsec_SFs}
\lefteqn{\frac{d\sigma}{dx_B \, dy\, d\psi \,dz\, d\phi_h\, d P_{h\perp}^2}
=}
\nonumber \\ & \quad 
\frac{\alpha^2}{x_B y Q^2}\,
\frac{y^2}{2\,(1-\varepsilon)}\,  \biggl( 1+\frac{\gamma^2}{2x_B} \biggr)\,
\Biggl\{
F_{UU ,T}
+ 
\varepsilon\, F_{UU ,L}
+ 
\sqrt{2\,\varepsilon (1+\varepsilon)}\,\cos\, \phi_h\, F_{UU}^{\scriptsize{\cos}\, \phi_h}
\nonumber \\  & \quad \qquad
+ \varepsilon\, \cos(2\phi_h)\, F_{UU}^{\scriptsize{\cos}\, 2\phi_h}
+ \lambda_e\, \sqrt{2\,\varepsilon (1-\varepsilon)}\, \sin\phi_h\, F_{LU}^{\scriptsize{\sin}\, \phi_h}
\phantom{\Bigg[ \Bigg] }
\nonumber \\  & \quad \qquad
+ S_\parallel\, \Bigg[ 
\sqrt{2\, \varepsilon\, (1+\varepsilon)}\, \sin\phi_h\, F_{UL}^{\scriptsize{\sin}\, \phi_h}
+  
\varepsilon\, \sin(2\phi_h)\, F_{UL}^{\scriptsize{\sin}\, 2\phi_h}
\Bigg]
\nonumber \\  & \quad \qquad
+ 
S_\parallel\, \lambda_e\, \Bigg[
\sqrt{1-\varepsilon^2}\, F_{LL}
+
\sqrt{2\,\varepsilon\, (1-\varepsilon)}\, \cos\phi_h\, F_{LL}^{\scriptsize{\cos}\, \phi_h}
\Bigg]
\nonumber \\  & \quad \qquad
+ |\bm{S}_\perp|\, \Bigg[
\sin(\phi_h-\phi_S)\, \Bigl(
F_{UT ,T}^{\scriptsize{\sin}\, (\phi_h -\phi_S)}
+ \varepsilon\, F_{UT ,L}^{\scriptsize{\sin}\, (\phi_h -\phi_S)}
\Bigr)
\nonumber \\  & \quad  \qquad \qquad
+ \varepsilon\, \sin(\phi_h+\phi_S)\, F_{UT}^{\scriptsize{\sin}\, (\phi_h +\phi_S)}
+ \varepsilon\, \sin(3\phi_h-\phi_S)\,F_{UT}^{\scriptsize{\sin}\, (3\phi_h -\phi_S)}
\phantom{\Bigg[ \Bigg] }
\nonumber \\  & \quad \qquad \qquad
+ \sqrt{2\,\varepsilon\, (1+\varepsilon)}\, \sin\phi_S\, F_{UT}^{\scriptsize{\sin}\,  \phi_S }
+ \sqrt{2\,\varepsilon\, (1+\varepsilon)}\, \sin(2\phi_h-\phi_S)\, F_{UT}^{\scriptsize{\sin}\, (2\phi_h -\phi_S)}
\Bigg]
\nonumber \\  & \quad \qquad 
+ |\bm{S}_\perp|\, \lambda_e\, \Bigg[ 
\sqrt{1-\varepsilon^2}\, \cos(\phi_h-\phi_S)\, F_{LT}^{\scriptsize{\cos}\, (\phi_h -\phi_S)}
+\sqrt{2\,\varepsilon\, (1-\varepsilon)}\, \cos\phi_S\, F_{LT}^{\scriptsize{\cos}\, \phi_S}
\nonumber \\  & \quad \qquad \qquad
+ \sqrt{2\,\varepsilon\, (1-\varepsilon)}\, \cos(2\phi_h-\phi_S)\, F_{LT}^{\scriptsize{\cos}\, (2\phi_h - \phi_S)}
\Bigg] \Biggr\},
\end{align}
where $\alpha$ is the fine structure constant and the structure
functions on the RHS depend on $x_B$, $Q^2$, $z$ and
$P_{h\perp}^2$.  
The angle $\psi$ is the azimuthal angle of $\ell'$
around the lepton beam axis with respect to an arbitrary fixed
direction, which in case of a transversely polarized target we choose
to be the direction of $S$. The corresponding relation between $\psi$
and $\phi_S$ is given in Ref.~\cite{Diehl:2005pc}. In deep inelastic
kinematics one has $d\psi \approx d\phi_S$. 
The first and second subscript of the above structure functions indicate the
respective polarization of beam and target, whereas the third
subscript in $F_{UU,T}$, $F_{UU,L}$ and
$F_{UT ,T}^{\scriptsize{\sin}\, (\phi_h -\phi_S)}$,
$F_{UT ,L}^{\scriptsize{\sin}\, (\phi_h-\phi_S)}$ 
specifies the polarization of
the virtual photon.  
Note that longitudinal or transverse target
polarization refer to the photon direction here. The conversion to
the experimentally relevant longitudinal or transverse polarization
with respect to the lepton beam direction is given in
Ref.~\cite{Diehl:2005pc}. 
The ratio $\varepsilon$ of longitudinal and transverse photon flux in 
Eq.~\eqref{e:SIDIS_xsec_SFs} is given by
\begin{align}
\label{e:eps_def}
\varepsilon &= \frac{1-y -\frac{1}{4} \gamma^2 y^2}{1-y
  +\frac{1}{2} y^2 +\frac{1}{4} \gamma^2 y^2} ,
\end{align}  

\subsection{Hadronic tensor}
\label{ss:W_SIDIS}

The cross section in Eq.~\eqref{e:SIDIS_xsec_SFs} can be also expressed as the contraction
of a hadronic and a leptonic tensor,
\begin{equation}
\label{e:SIDIS_xsec_L_W}
\frac{d\sigma}{dx_B\,dy\, d\psi\, dz\, d\phi_h\, dP_{h \perp}^2} 
= \frac{\alpha^2 y}{8 z Q^4}\, 2 M W^{\mu \nu}\,
  L_{\mu \nu} \, , 
\end{equation}
where the leptonic tensor is given in Eq.~\eqref{e:L} and the hadronic tensor for SIDIS is defined as:
\begin{equation}
\label{e:W_SIDIS}
2M W^{\mu\nu} = \frac{1}{(2 \pi)^3}\, \sum_{X}\,  
\int \frac{d^3 \vect{P}_{X}} {2 P_{X}^0}\, \delta^{(4)} \bigl( q+P-P_{X}-P_h \bigr)\,
\langle P| J^{\mu}(0) | h, {X}\rangle \langle h, {X}| 
   J^{\nu}(0)|P\rangle ,
\end{equation}
where $J^\mu(\xi)$ is the electromagnetic current divided by the
elementary charge (see also Eq.~\eqref{e:elm_current}). The final state is summed over the hadrons' polarization. Fig.~\ref{f:W_SIDIS} gives a pictorial representation of Eq.~\eqref{e:W_SIDIS}.
\begin{figure}
\centering
\includegraphics[width=5cm]{figures/W_SIDIS}
\caption{This figure gives a diagrammatic interpretation of the hadronic tensor $W^{\mu\nu}$ for SIDIS, where an electromagnetic current $J$ acts on a hadron with momentum $P$ and produces another hadron with momentum $P_h$ and a remnant. 
The left-hand part of the diagram is multiplied by the corresponding conjugated Feynman amplitude. Together, they produce a cut diagram. 
The lines crossing the cut represent on-shell states that enter the final state of the reaction.}
\label{f:W_SIDIS}
\end{figure}
\begin{figure}
\centering
\includegraphics[width=5cm]{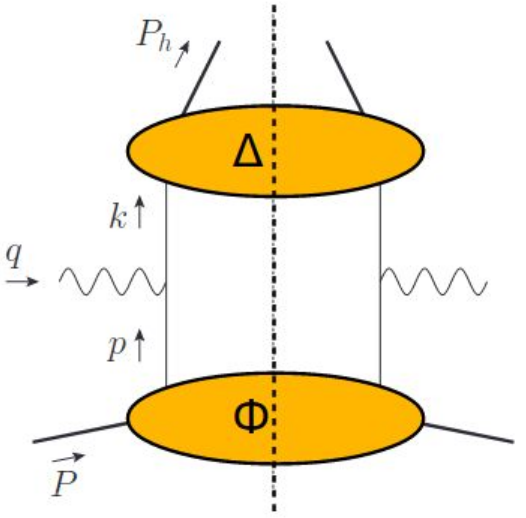}
\caption{Leading twist and tree level approximation for the SIDIS hadronic tensor as a trace of the TMD parton correlation functions $\Phi(x,p_T)$ and a $\Delta(z,k_T)$.
A virtual photon probes a quark in the hadronic target (via a parton distribution function), which then fragments into an observed hadron $h$ (via a parton fragmentation function).}
\label{f:W_SIDIS_tw2}
\end{figure}

In Sec.~\ref{s:OPE} we learned how to factorize the hadronic tensor for inclusive DIS in terms of collinear parton distribution functions and hard scattering coefficients, thanks to the possibility of applying the operator product expansion.
In this case the OPE cannot be formally applied, because the presence of the detected hadron $h$ in the final state prevents one from writing the operator structure in Eq.~\eqref{e:W_SIDIS} as a commutator of currents.
However, the diagrammatic approach~\cite{Politzer:1980me} allows one to systematically study the cross section of the process at tree level, introducing a twist expansion in powers of $M/Q$ as discussed in Sec.~\ref{ss:twist}. This approach essentially combines the two-parton and multi-parton hadronic matrix elements(such as $\Phi$ and $\Delta$) formally introduced via the OPE, even when the operator product expansion itself cannot be applied (e.g. in SIDIS, non-inclusive $e^+e^-$ annihilation, and Drell-Yan). For a review of the diagrammatic approach see e.g. Ref.~\cite{Pijlman:2006vm} and references therein.

Nowadays the factorization of the SIDIS cross section in terms of TMD parton distribution and fragmentation functions is known both at twist 2 (e.g. Ref.~\cite{Collins:2011zzd}) and at twist 3 (e.g. Refs.~\cite{Bacchetta:2019qkv,Vladimirov:2021hdn,Rodini:2023plb,Ebert:2021jhy}). In general, when a factorization theorem exists in perturbative QCD, its tree level version corresponds to the results obtained with the diagrammatic approach. The results in perturbation theory, however, automatically have a much stronger predictive power, which is rooted in the renormalization of the operators involved.

With the diagrammatic approach one can express the hadronic tensor as a trace of a TMD parton distribution correlator with a TMD fragmentation correlator, together with the Dirac matrices associated to the quark-photon hard vertices. 
For this purpose (and similarly to what is done in TMD factorization), we work in a frame where the incoming and the outgoing hadrons are collinear (in which the transverse components are denoted with a $T$ subscript), which is different from the $P$, $q$ frame used in Sec.~\ref{ss:SIDIS_cross_section} and Eqs.~\eqref{e:SIDIS_Pq_basis} (where the transverse components are denoted with a $\perp$ subscript).
Details on the relation between the two choices can be found in~\cite{Mulders:1995dh,Boer:2003cm}. 
Transverse and longitudinal components differ by sub-leading twist terms, up to ${\mathcal O}(Q^{-1})$~\cite{Bacchetta:2006tn}.

In this T frame, the corresponding expression of the hadronic tensor is~\cite{Bacchetta:2006tn,Mulders:1995dh,Boer:2003cm}
\begin{align}
\label{e:W_SIDIS_twist2}
2M W^{\mu\nu} 
= 2 z\, \sum_a\, e_a^2 \, 
\int d^2 \vect{p}_T\, d^2 \vect{k}_T\, \delta^{(2)}(\vect{p}_T + \vect{q}_T - \vect{k}_T) \,
\Tr\, \biggl\{ \Phi^a (x,p_T)\, \gamma^\mu\, \Delta^a(z,k_T)\, \gamma^\nu\, \biggr\} + 
{\mathcal O}(Q^{-1}) \, ,
\end{align}
where the sum runs over the quark and antiquark flavors $a$, and $e_a$ denotes the fractional charge of the struck quark or antiquark. Fig.~\ref{f:W_SIDIS_tw2} is a pictorial representation of the leading twist part in Eq.~\eqref{e:W_SIDIS_twist2}.
The higher twist corrections are related to diagrams that involve multi-parton correlators (dynamical higher twists) and power suppressed kinematic corrections.
For example, in Eq.~\eqref{e:W_SIDIS_twist2} the two-dimensional delta distribution allows one to express the transverse momentum of the photon $\vect{q}_T$ as the sum of the partonic transverse momentum of the struck quark $\vect{p}_T$ with the transverse momentum of the hadronizing quark $\vect{k}_T$:
\begin{equation}
\label{e:kT_pT_qT_Phperp}
\vect{p}_T + \vect{k}_T = \vect{q}_T = 
-\frac{\vect{P}_{h\perp}}{z} + {\mathcal O}(\vect{q}_T^2/Q^2) \, .
\end{equation}
The Lorentz transformation between the ``$T$'' and ``$\perp$'' frames~\cite{Collins:2011zzd}, generates the second part of Eq.~\eqref{e:kT_pT_qT_Phperp}, 
which explicitly relates the photon transverse momentum $\vect{q}_T$ and the measured transverse momentum $\vect{P}_{h\perp}$~\cite{Bacchetta:2022awv}. 
The kinematic higher twist corrections ${\mathcal O}(\vect{q}_T^2/Q^2)$ can be neglected in the limit $Q^2 \gg \vect{q}_T^2$, but they could be in principle sizeable when $Q$ is not very large (i.e. a few GeV, which is the case of the current fixed-target experiments performed by the HERMES, COMPASS, and CLAS collaborations) or in the region where the transverse momentum is comparable with $Q$ (which is again relevant at the current fixed-target experiments).

\subsection{Structure functions}
\label{ss:SIDIS_SFs}

The possibility to compute the hadronic tensor as a trace involving the TMD parton distribution and fragmentation correlators allows one to express the structure functions which appear in the fully differential cross section Eq.~\eqref{e:SIDIS_xsec_SFs} as convolutions of TMD PDFs and TMD FFs.
The complete result for the 18 SIDIS structure functions is available in Ref.~\cite{Bacchetta:2006tn}. Here we limit ourselves to some specific cases.

To present the results in a compact form, 
we introduce the unit vector $\hat{\vect{h}}=\vect{P}_{h \perp}/|\vect{P}_{h\perp}|$ and
the notation
\begin{equation}
\label{e:convolution_f_D}
{\cal C}\bigl[ w\, f\, D \bigr]
= x_B\,
\sum_a\, e_a^2\, \int d^2 \vect{p}_T\,  d^2 \vect{k}_T\,
\delta^{(2)}\bigl(\vect{p}_T - \vect{k}_T - \vect{P}_{h \perp}/z \bigr)
\, w(\vect{p}_T,\vect{k}_T)\, 
f^a(x_B,p_T^2)\, D^a(z,k_T^2) \, ,
\end{equation}
where $w(\vect{p}_T,\vect{k}_T)$ is an arbitrary function and the sum runs over quarks and antiquarks.

The leading twist structure functions are  essentially a convolution of one twist 2 TMD PDF with one twist 2 TMD FF. Their expressions (at leading order) read:
\begin{align} 
\label{e:F_UUT}
F_{UU ,T} & = {\cal C}\bigl[ f_1\, D_1 \bigr] \, ,
\\
\label{e:F_UUcos2phi}
F_{UU}^{\scriptsize{\cos}\,  2\phi_h}
& 
= {\cal C}\biggl[
   - \frac{2\, \bigl( \hat{\vect{h}} \cdot \vect{k}_T \bigr)
   \,\bigl( \hat{\vect{h}} \cdot \vect{p}_T \bigr)
    -\vect{k}_T \cdot \vect{p}_T}{M M_h}\,
    h_{1}^{\perp}\, H_{1}^{\perp} \biggr] \, ,
\\
\label{e:F_ULsinphi} 
F_{UL}^{\scriptsize{\sin}\,  2\phi_h}
&
 = {\cal C}\biggl[
   -\frac{2\,\bigl( \hat{\vect{h}} \cdot \vect{k}_T \bigr)
   \,\bigl( \hat{\vect{h}} \cdot \vect{p}_T \bigr)
   -\vect{k}_T \cdot \vect{p}_T}{M M_h}\,
    h_{1L}^{\perp}\, H_{1}^{\perp}\biggr] \, ,
\\
F_{LL} & = {\cal C}\bigl[ g_{1L}\,  D_1\bigr] \, , 
\\
\label{e:FUT_Sivers}
F_{UT ,T}^{\scriptsize{\sin}\, (\phi_h -\phi_S)} 
& = 
{\cal C}\biggl[-\frac{\hat{\vect{h}} \cdot \vect{p}_T}{M} f_{1T}^{\perp}  D_1 \biggr],
\\
\label{e:FUT_Collins}
F_{UT}^{\scriptsize{\sin}\, (\phi_h +\phi_S)} 
& = 
{\cal C}\biggl[-\frac{\hat{\vect{h}} \cdot \vect{k}_T}{M_h} h_{1} H_1^{\perp}\biggr],
\\
\label{e:FUT_sin_3}
F_{UT}^{\scriptsize{\sin}\, (3\phi_h -\phi_S)}
& = 
   {\cal C}\biggl[
   \frac{2\, \bigl(\hat{\vect{h}} \cdot \vect{p}_T \bigr)\, 
        \bigl( \vect{p}_T \cdot \vect{k}_T \bigr)
   +\vect{p}_T^2\, \bigl(\hat{\vect{h}} \cdot \vect{k}_T \bigr)
   -4\, (\hat{\vect{h}} \cdot \vect{p}_T)^2\, (\hat{\vect{h}} \cdot \vect{k}_T)}{2 M^2 M_h}\, 
    h_{1T}^{\perp}\, H_1^{\perp} \biggr] \, ,
\\
\label{e:FLT}
F_{LT}^{\scriptsize{\cos}\, (\phi_h -\phi_S)} 
& = 
{\cal C}\biggl[ \frac{\hat{\vect{h}} \cdot \vect{p}_T}{M}\, g_{1T}\, D_1 \biggr] \, . 
\end{align} 
The previous equations create a direct connection between specific modulations of the SIDIS cross section with the TMDs, and effectively set the basis for a transverse momentum imaging program based on SIDIS measurements. 

The higher twist structure functions, instead, feature convolutions which include one leading twist distribution and one higher twist distribution: 
\begin{align}
\label{e:FUU_cosphi}
F_{UU}^{\scriptsize{\cos}\, \phi_h}
& = 
\frac{2M}{Q}\, {\cal C}\biggl[
   - \frac{\hat{\vect{h}} \cdot \vect{k}_T}{M_h}
   \biggl(x_B\,  h\, H_{1}^{\perp} 
   + \frac{M_h}{M}\, f_1\, \frac{\tilde{D}^{\perp}}{z} \biggr)
   - \frac{\hat{\vect{h}} \cdot \vect{p}_T}{M}\,
     \biggl(x_B\, f^{\perp}\, D_1
   + \frac{M_h}{M}\, h_{1}^{\perp}\, \frac{\tilde{H}}{z} \biggr) \biggr] \, ,
\\
\label{e:FLU_sin_phih}
F_{LU}^{\scriptsize{\sin}\, \phi_h} 
& = 
\frac{2M}{Q}\, {\cal C}\biggl[ 
- \frac{\hat{\vect{h}}\cdot \vect{k}_T}{M_h}
   \biggl(x_B\,  e\, H_1^{\perp} 
   +\frac{M_h}{M}\, f_1\frac{\tilde{G}^{\perp}}{z}\biggr)
   + 
   \frac{\hat{\vect{h}}\cdot \vect{p}_T}{M}
   \biggl(x_B\, g^{\perp }\,  D_1\, 
   + \frac{M_h}{M}\, h_1^{\perp}\, \frac{\tilde{E}}{z} \biggr)\biggr] \, .
\end{align}
Eq.~\eqref{e:FUU_cosphi} refers to the so-called Cahn effect~\cite{Cahn:1978se}. Eq.~\eqref{e:FLU_sin_phih}, instead, can be investigated via a beam spin asymmetry~\cite{CLAS:2025asy}.
Their expressions reveal how difficult it is to single out specific higher twist distributions, since these typically enter physical observables mixed with other effects.

\section{Exercises}
\label{s:ex_SFs}

The following exercises can be performed by hand or by using tools such as Mathematica (with FeynCalc\footnote{\href{https://feyncalc.github.io/}{https://feyncalc.github.io/}}) and Python (with Sympy\footnote{\href{https://www.sympy.org/}{https://www.sympy.org/}}).

Consider Ref.~\cite{Bacchetta:2012aln}:
\begin{itemize}

\item[1)] starting from eq. (3.85) in those notes, derive the Dirac projections of the TMD correlator outlined in eqs. (3.88) - (3.90)

\item[2)] compute the Sivers structure function (in particular its angular dependence) in SIDIS. 
notes. 
[Suggested steps: define the hadronic tensor in terms of TMD correlators, define the
leptonic tensor, contract them, etc.]

\item[3)] do the same for other structure functions in SIDIS.

\end{itemize}

\clearpage
\part{The role of symmetries}
\label{p:symmetries}

The symmetries of QCD play an essential role in the definition of the functions that describe hadron structure and hadronization in momentum space.
These symmetries constrain the number of structures that can be used to parametrize a certain correlation function.
In particular, the interplay between the time reversal symmetry and the $SU(3)_c$ gauge symmetry generates the so-called generalized universality of a specific class of TMDs~\cite{Buffing:2015tja}.
The time-reversal odd (T-odd) functions, in fact, acquire a calculable process dependence because of the gauge links that enter the associated operators. The following material is partly based on Ref.~\cite{Mulders:2016pln}.

\section{Discrete symmetries}
\label{s:discrete}

To get started, let's first consider the role of discrete symmetries, using correlation functions that do not contain any gauge link. 
The role of the gauge symmetry will be considered in Sec.~\ref{s:gauge} and Sec.~\ref{s:discrete_gauge}. 
Thus, the quark-quark unintegrated correlator for a spin 1/2 hadron is defined as:
\begin{equation}
\Phi_{ij}(k,P,S) = \frac{1}{(2\pi)^4}\int d^4\xi\ e^{i\,k\cdot \xi}  
\langle P,S \vert \overline \psi_j(-\xi/2) \psi_i(\xi/2) \vert P,S \rangle \ ,
\label{e:qq_unint_corr}
\end{equation}
where a summation over color indices is implicit. 

Let us first introduce the following notation for a generic four vector:
\begin{equation}
\label{e:tilde_def}
a^\mu = (a^0, \vect{a}) \, , \ \ \ \ \ \ \ \ \ \ \tilde{a}^\mu = (a^0, - \vect{a}) \, ,
\end{equation}
where the bold face represents the three spacial components. 
Accordingly, positions ($z$), momenta ($P$), spins ($S$), light-cone vectors ($n_{\pm}$, see App.~\ref{a:lc_coordinates}), and Dirac structures ($\psi$, $\gamma$) change in the following way under parity and time reveresal transformations:
\begin{itemize}

\item parity:
\begin{align}
\label{e:P_rules}
& z^\mu \longrightarrow \tilde{z}^\mu \\ \nonumber
& P^\mu \longrightarrow \tilde{P}^\mu \\ \nonumber
& S^\mu \longrightarrow S^\mu \equiv -\tilde{S}^\mu \ \ \ (\text{since}\ S^\mu = (0,\vec{S})\ \ \text{by definition}) \\ \nonumber
& n_{\pm} \longrightarrow n_{\mp} \\ \nonumber
& \psi(\xi) \longrightarrow {\mathscr P}\,\psi(\xi) {\mathscr P}^\dagger = 
  \Lambda_{\mathscr P}\, \psi(\tilde \xi)\, , \ \ \ \Lambda_{\mathscr P} = \gamma^0 \\ \nonumber
& \gamma^\mu \longrightarrow {\mathscr P}\, \gamma^\mu {\mathscr P}^\dagger = \Lambda_{\mathscr P}\, \gamma^\mu\, \Lambda_{\mathscr P}^\dagger
\end{align}

\item time-reversal:
\begin{align}
\label{e:T_rules}
& z^\mu \longrightarrow -\tilde{z}^\mu \\ \nonumber
& P^\mu \longrightarrow \tilde{P}^\mu \\ \nonumber
& S^\mu \longrightarrow \tilde{S}^\mu \\ \nonumber
& n_{\pm} \longrightarrow -n_{\mp} \\ \nonumber
& \psi(\xi) \longrightarrow {\mathscr T}\,\psi(\xi) {\mathscr T}^\dagger = 
  \Lambda_{\mathscr T}\, \psi(-\tilde \xi)\, , \ \ \ \Lambda_{\mathscr T} = -i\gamma_5 C = i\gamma^1 \gamma^3 \\ \nonumber
& \gamma^\mu \longrightarrow {\mathscr T}\, \gamma^\mu {\mathscr T}^\dagger = \Lambda_{\mathscr T}\, \gamma^\mu\, \Lambda_{\mathscr T}^\dagger 
  = \gamma_\mu^*
\end{align}

\end{itemize}

\subsection{Hermiticity}
\label{ss:dagger}
Calculating the hermitian conjugate of the quark-quark correlator in Eq.~\eqref{e:qq_unint_corr} is a useful step to approach the transformation under parity and time reversal: 
\begin{align}
(\Phi^\dagger)_{ij} = \Phi_{ji}^\ast & = 
\frac{1}{(2\pi)^4}\int d^4\xi\ e^{-i\,k\cdot \xi} 
\langle P,S\vert\psi^\dagger_k(-\xi/2)(\gamma_0)_{ki}\psi_j(\xi/2)\vert P,S\rangle^\ast
\nonumber \\ & = 
\frac{1}{(2\pi)^4}\int d^4\xi\ e^{-i\,k\cdot \xi} 
\langle P,S\vert\psi^\dagger_j(\xi/2)(\gamma_0)_{ik}\psi_k(-\xi/2)\vert P,S\rangle
\nonumber \\ & = 
\frac{1}{(2\pi)^4}\int d^4\xi\ e^{-i\,k\cdot \xi} 
\langle P,S\vert\overline\psi_l(\xi/2)(\gamma_0)_{lj} (\gamma_0)_{ik}\psi_k(-\xi/2)\vert P,S\rangle
\nonumber \\ & = 
\frac{1}{(2\pi)^4}\int d^4\xi\ e^{i\,k\cdot \xi} 
\langle P,S\vert\overline\psi_l(-\xi/2)(\gamma_0)_{lj} (\gamma_0)_{ik}\psi_k(\xi/2)\vert P,S\rangle
\nonumber \\ & = 
(\gamma_0)_{ik}\,\Phi_{kl}\,(\gamma_0)_{lj} \ ,
\label{e:hermiticity}
\end{align}
where from the third to the fourth line the change of variable $\xi \to -\xi$ has been used. 

\subsection{Parity}
\label{ss:parity}
The invariance of the QCD lagrangian under parity transformation implies for the Dirac fermions that ${\mathscr P}\,\psi(\xi){\mathscr P}^\dagger = \gamma_0\psi(\tilde \xi)$, see Eq.~\eqref{e:P_rules}. Accordingly, for the correlator we have:
\begin{align}
\Phi_{ij}(k,P,S) & = 
\frac{1}{(2\pi)^4}\int d^4\xi\ e^{i\,k\cdot \xi} 
\langle P,S\vert{\mathscr P}^\dagger{\mathscr P}\overline\psi_j(-\xi/2) {\mathscr P}^\dagger{\mathscr P}\psi_i(\xi/2){\mathscr P}^\dagger{\mathscr P}\vert P,S\rangle
\nonumber \\ & = 
\frac{1}{(2\pi)^4}\int d^4\xi\ e^{i\,k\cdot \xi} 
\langle \tilde P,-\tilde S\vert\overline\psi_l(-\tilde{\xi}/2)(\gamma_0)_{lj} (\gamma_0)_{ik}\psi_k(\tilde{\xi}/2)\vert \tilde P,-\tilde S\rangle
\nonumber \\ & = 
\frac{1}{(2\pi)^4}\int d^4\xi\ e^{i\,\tilde k\cdot \xi} 
\langle \tilde P,-\tilde S\vert\overline\psi_l(-\tilde{\xi}/2)(\gamma_0)_{lj} (\gamma_0)_{ik}\psi_k(\tilde{\xi}/2)\vert \tilde P,-\tilde S\rangle
\nonumber \\ & = 
(\gamma_0)_{ik}\,\Phi_{kl}(\tilde k,\tilde P,-\tilde S)\,(\gamma_0)_{lj} \ ,
\label{e:parity}
\end{align}
where we have used $k\cdot \xi = \tilde k\cdot \tilde \xi$ and we have changed the integration variable from $\xi$ to $\tilde{\xi}$.

\subsection{Time reversal}
\label{ss:trev}
The invariance of the QCD lagrangian under time-reversal transformation implies for the Dirac fermions that ${\mathscr T}\,\psi(\xi){\mathscr T}^\dagger = \Lambda_{\mathscr T}\, \psi(-\tilde \xi)$, with $\Lambda_{\mathscr T} = -i\gamma_5 C = i\gamma^1\gamma^3$, see Eq.~\eqref{e:T_rules}. 
The operator ${\mathscr T}$ is anti-unitary, namely:
\begin{equation}
\langle x| {\mathscr T}^\dagger {\mathscr T} |y \rangle = \langle x|y \rangle^\ast \ .
\label{e:def_antiun}
\end{equation}
Accordingly, for the correlator we have:
\begin{align}
\Phi^\ast_{ij}(k,P,S) & = 
\frac{1}{(2\pi)^4}\int d^4\xi\ e^{-i\,k\cdot \xi} 
\langle P,S\vert \overline\psi_j(-\xi/2) \psi_i(\xi/2) \vert P,S\rangle^\ast
\nonumber \\ & = 
\frac{1}{(2\pi)^4}\int d^4\xi\ e^{-i\,k\cdot \xi} 
\langle P,S\vert \overline\psi_j(-\xi/2) {\mathscr T}^\dagger{\mathscr T} \psi_i(\xi/2) \vert P,S\rangle
\nonumber \\ & = 
\frac{1}{(2\pi)^4}\int d^4\xi\ e^{-i\,k\cdot \xi} 
\langle P,S\vert{\mathscr T}^\dagger{\mathscr T}\overline\psi_j(-\xi/2) {\mathscr T}^\dagger{\mathscr T}\psi_i(\xi/2){\mathscr T}^\dagger{\mathscr T}\vert P,S\rangle
\nonumber \\ & = 
\frac{1}{(2\pi)^4}\int d^4\xi\ e^{-i\,k\cdot \xi} 
\langle \tilde P,\tilde S\vert \overline{\psi}_l(\tilde{\xi}/2)(-i\gamma_5 C)_{lj} (-i\gamma_5 C)_{ik} \psi_k(-\tilde{\xi}/2) \vert \tilde P,\tilde S\rangle
\nonumber \\ & = 
\frac{1}{(2\pi)^4}\int d^4\xi\ e^{i\, \tilde{k}\cdot \tilde{\xi} } 
\langle \tilde P,\tilde S\vert \overline{\psi}_l(-\tilde{\xi}/2)(-i\gamma_5 C)_{lj} (-i\gamma_5 C)_{ik} \psi_k(\tilde{\xi}/2) \vert \tilde P,\tilde S\rangle
\nonumber \\ & = 
(-i\gamma_5 C)_{ik}\,\Phi_{kl}(\tilde k,\tilde P,\tilde S)\,(-i\gamma_5 C)_{lj} \ ,
\label{e:trev}
\end{align}
where from the first to the second line we used the antiunitary nature of the time-reversal operator (Eq.~\eqref{e:def_antiun}) and we applied the same property twice from the second to the third line. 
Also, we have used $ -i k \cdot \xi = i \tilde{k} \cdot (-\tilde{\xi}) $ and we have changed the integration measure from $\xi$ to $\tilde{\xi}$.

\subsection{Charge conjugation}
\label{ss:charge}

Charge-conjugation properties are discussed in, e.g., Ref.~\cite{Boer:1998im,Mulders:2016pln} for a quark-quark correlator and in Ref.~\cite{Boer:2016xqr} for a gluon-gluon correlator. \\

The charge conjugation symmetry in QCD can be implemented via the unitary 
operator ${\cal C}$ such that:
\begin{align}
\label{e:C_unitary}
& \langle h_1| {\cal C}^\dagger\, {\cal C} |h_2 \rangle = \langle h_1|h_2 \rangle\, , \\
\label{e:C_state}
& {\cal C} |h(P) \rangle = | \overline{h}(P) \rangle\, , \\
\label{e:psic_def}
& \psi^c(x) = {\cal C}\, \psi(x)\, {\cal C}^\dagger = C\, \overline{\psi}^T(x) \, ,
\end{align}
where the ${\cal C}$ operator transforms the state of a hadron $h$ with momentum $P$ into the state of an antihadron $\overline{h}$ with the same momentum, and transforms the $\psi$ operator representing a quark with flavor $a$ into the $\psi^c$ operator describeing an antiquark $\bar{a}$. 
The matrix $C$ acting in the Dirac space satisfies 
\begin{equation}
\label{e:C_Dirac_prop}
C\, \gamma_\mu^T\, C^\dagger = -\gamma_\mu \, ,    
\end{equation}
and can be chosen as 
\begin{equation}
\label{e:C_Dirac_def}
C = i\, \gamma^2\, \gamma^0 = -C^\dagger = -C^{-1} \, . 
\end{equation}

The unintegrated correlator for an antiquark $\bar{a}$ in a hadron $h$ is defined as Eq.~\eqref{e:qq_unint_corr} by replacing $\psi$ with its charge-conjugated version $\psi^c$ defined in Eq.~\eqref{e:psic_def}. In the following we report, for convenience, the definition of $\Phi(k,P)$ together with the antiquark correlator $\Phi^c(k,P)$ and the auxiliary correlator $\overline{\Phi}(k,P)$~\cite{Tangerman:1994eh,Mulders:2016pln,Ralston:1979ys}: 
\begin{align}
\label{e:Phi} 
\Phi_{ij}(k,P) & = 
\int \frac{d^4 \xi}{(2\pi)^4}\, e^{i k \cdot \xi}\, 
\mathrm{Tr}_{c} \langle h(P)| 
\overline{\psi}_j(-\xi/2)\, \psi_i(\xi/2)\, 
| h(P) \rangle\, , \\
\label{e:Phic}
\Phi^c_{ij}(k,P) & = 
\int \frac{d^4 \xi}{(2\pi)^4}\, e^{i\, k \cdot \xi}\, 
\mathrm{Tr}_{c} \langle h(P)| 
\overline{\psi_j^c}(-\xi/2) \psi^c_i(\xi/2)\,  
| h(P) \rangle\, , \\
\label{e:Phibar}
\overline{\Phi}_{ij}(k,P) & = 
\int \frac{d^4 \xi}{(2\pi)^4}\, e^{-i\, k \cdot \xi}\,  
\mathrm{Tr}_{c} \langle h(P)| 
\psi_i(\xi/2)\, \overline{\psi}_j(-\xi/2)\, 
| h(P) \rangle\, .
\end{align}

By using the unitarity of ${\cal C}$ and its action on the states and the spinors (Eqs.~\eqref{e:C_unitary},~\eqref{e:C_state},~\eqref{e:psic_def}), one can show that the correlator for a quark $a$ in a hadron $h$ is equal to the correlator for an antiquark $\bar{a}$ in an antihadron $\bar{h}$:
\begin{equation}
\label{e:Phi_Phic_corr}
\Phi^{a/h}(k,P) = \Phi^{c\, ,\, \bar{a}/\bar{h}}(k,P) \, ,
\end{equation}
where in this case the superscripts represent the quark/hadron flavor. 
As a consequence, the same holds true for TMD and collinear parton distribution functions~\cite{Tangerman:1994eh,Collins:2011zzd}. 
By using Eq.~\eqref{e:psic_def}, instead, it is possible to show that the relation between $\Phi^c(k,P)$ and $\overline{\Phi}(k,P)$ is:
\begin{equation}
\label{e:Phic_Phibar}
\Phi^c(k,P) = -C\, \overline{\Phi}^T(k,P)\, C^\dagger\, .
\end{equation}

As for $\Phi(k,P)$, it is possible to define the collinear (and TMD) version of the unintegrated correlators $\Phi^c(k,P)$ and $\overline{\Phi}(k,P)$ by integrating over the partonic momentum components and fixing $k^+ = xP^+$:
\begin{align}
\label{e:Phic_x}
& \Phi^c(x) = \int d^4 k\, \delta\big(k^+ - x P^+ \big)\, \Phi^c(k,P)\, , \\ 
\label{e:Phibar_x} 
& \overline{\Phi}(x) = \int d^4 k\, \delta\big(k^+ - x P^+ \big)\, \overline{\Phi}(k,P)\, .
\end{align}
The collinear PDFs $\overline{f}_{1,3}$, $\overline{e}$, etc. are defined as Dirac traces (see Eq.~\eqref{e:proj_Phi}) of the charge-conjugated correlators, $\Phi^{c\, [\Gamma]}(x)$. 
To this extent, the relative sign ($\pm$) between $\overline \Phi^{[\Gamma]}$ and $\Phi^{c[\Gamma]}$ needs to be taken into account~\cite{Tangerman:1994eh,Mulders:2016pln}. This sign depends on the transformation properties of the considered Dirac structure, $\Gamma = \mp C\, \Gamma^T\, C^\dagger$.
Explicitly one has: 
%
\begin{align}
\label{e:sign_plus_PhicGamma}
\Phi^{c[\Gamma]}(x) & = + \overline{\Phi}^{[\Gamma]}(x)  \qquad \mbox{for} \ \Gamma = \gamma^\pm \, , \\
\label{e:sign_minus_PhicGamma}
\Phi^{c[\Gamma]}(x) & = - \overline{\Phi}^{[\Gamma]}(x) \qquad \mbox{for} \ \Gamma = \mathbb{I} \, . 
\end{align}
Moreover, by using the anticommutation relations for the fermion fields at spacelike and timelike separations (see App.~\ref{ss:anticomm} and Ref.~\cite{Mannheim:2020rod}) one obtains\footnote{This argument generalizes the ones given in Ref.~\cite{Jaffe:1983hp,Collins:2011zzd} for $f_1(x)$ based on the anticommutation relations for the good components of the quark fields and on the fact that the PDFs are defined from connected matrix elements.}
\begin{equation}
\label{e:Phibar_x_Phi_x}
\overline{\Phi}(x) = - \Phi(-x) \, .
\end{equation}
Eqs.~\eqref{e:sign_plus_PhicGamma},~\eqref{e:sign_minus_PhicGamma},~\eqref{e:Phibar_x_Phi_x} allow one to obtain relations between a specific projection of $\Phi(x)$ and $\Phi^c(x)$, namely between quark and antiquark PDFs.  
 The $\gamma^\pm$ projections result in: 
\begin{equation}
\label{e:C_even}
\overline f_{1,3}(x) = - f_{1,3}(-x) \, ,
\end{equation}
while the $\mathbb{I}$ projection results in: 
\begin{equation}
\label{e:C_odd}
\overline e(x) = e(-x) \, .
\end{equation}
In particular, one can show how the $x<0$ domain for the quark PDFs is related to physical ($x>0$) antiquark distributions. 
Summarizing: the traces of $\Phi(x)$ define the standard $f_{1,3}(x)$, $e(x)$ with $x \in [-1,1]$, whereas the traces of $\Phi^c(x)$ define the distributions $\overline{f}_{1,3}$, $\overline{e}(x)$ with $x \in [-1,1]$.   
Thanks to Eq.~\eqref{e:C_even} and~\eqref{e:C_odd} it is possible to convert an integral over $x \in [-1,1]$ for a specific PDF $f$ to an integral over $x \in [0,1]$ (the physical domain for the quark and antiquark distributions) of the combination of $f$ and $\overline{f}$.

Let us now make a digression on the notation used in this Section versus a more common one in the literature. 
The fermion field $\psi$ already contains information on {\em both} quark $q$ and anti-quark $\overline{q}$ for a specific flavor $f$. 
For this reason, a flavor subscript (rather than a quark $q$ or anti-quark $\overline{q}$ one) should be added to the correlation functions like $\Phi$, $\Phi^c$ and their Dirac projections (e.g. $f_1$ and $\overline{f}_1$), because they all contain contributions from quark and anti-quarks. 
At this point some confusion might arise, due to the fact the $\overline{f}_1$ notation is rarely used in the literature and to the fact that quark and anti-quark contributions for a specific flavor $f$ are mixed in these objects.

The physical ({\em i.e.} with $x>0$) quark distribution for a flavor $f$ ($f_1^{q_f}$) is defined as:
\begin{equation}
\label{e:def_f1_quark_physical}
\Phi_f^{[\gamma^+]}(x)_{\vert x>0} \equiv {f_1}_f(x)_{\vert x>0}\ \dot{=}\ f_1^{q_f}(x) \, , 
\end{equation}
whereas the physical ({\em i.e.} with $x>0$) anti-quark distribution for a flavor $f$ ($\overline{f}_1^{\overline{q}_f}$) is defined as:
\begin{equation}
\label{e:def_f1_antiquark_physical}
\Phi_f^{c\, [\gamma^+]}(x)_{\vert x>0} \equiv {\overline{f}_1}_f(x)_{\vert x>0}\ \dot{=}\ \overline{f}_1^{\overline{q}_f}(x) \, \longrightarrow\, f_1^{\overline{q}_f}(x) \, . 
\end{equation}
The arrow is to indicate that usually in literature the physical antiquark PDF is indicated with the shorthand notation $f_1^{\overline{q}_f}$, rather than we the more lengthy $\overline{f}_1^{\overline{q}_f}(x)$. In these notes we will also use the more compact notation, but one must keep in mind that the physical anti-quark PDF is associated to the correlation function $\Phi^c$ rather than $\Phi$. 
The same holds true for the scalar projection $\Gamma = \mathbb{I}$.



The relations presented in this Section for collinear PDFs can be generalized to the TMD case and to the fragmentation correlator in a straightforward manner following the same strategy.

\section{Gauge symmetry}
\label{s:gauge}

Part of the following Section is taken from Refs.~\cite{Signori:2016lvd,Mulders:2016pln}. 
In order to deal with gauge invariant correlators, we need to introduce the concept of {\em gauge link}. It arises from the geometrical notion of parallel transport, which defines a way of comparing elements from different linear spaces by a connection.

The transport of a vector is defined ``parallel'' if the scalar products are conserved while the vector is transported along a path in spacetime. 
This provides a specific way of moving fields along a path, defining an isomorphism between the spaces where the fields live.
The parallel transport equation for a fermion field $\psi$ along the curve $c$ in spacetime parametrized by $t \in \mathbb{R}$ is:
\begin{equation}
D_{\dot{c}}\ \psi (x(t)) = 0 \ ,\ \ t \in I \subset \mathbb{R} \ .
\label{e:par_transp}
\end{equation}
The solution of Eq.~\eqref{e:par_transp} is given in terms of a path-ordered exponential factor involving the gauge connection:
\begin{align}
\nonumber \psi_{\beta}^\prime(x(t)) &= \mathbb{P} \exp \bigg\{ -i g \int_0^t ds \frac{dx^\mu}{ds} A_\mu^a(x(s)) T_{\beta \alpha}^a \bigg\} \ \psi_\alpha(x(0)) \\ 
&\equiv U_{\beta \alpha}(x(t),x(0))\ \psi_\alpha(x(0)) \ .
\label{e:par_transp_psi}
\end{align}
$U_{\beta \alpha}(x(t),x(0))$ is a {\em gauge link}, or {\em Wilson line}. In the case of QCD, it is a matrix in color space and $T^a$ are the Gell-Mann matrices.
A gauge link $U(x_2,x_1)$ connecting $x_1$ to $x_2$, under local gauge transformations ${\cal U}(x)$ behaves as~\cite{Peskin:1995ev,Muta:2010xua}: 
\begin{equation}
U^\prime(x_2,x_1;A^\prime) = {\cal U}(x_2)\ U(x_2,x_1;A)\ {\cal U}^\dagger(x_1) \ .
\label{e:transf_links}
\end{equation}
As a result, the operator
\begin{equation}
\overline{\psi}_j(0)\ U(0,\xi)\ \psi_i(\xi)
\label{e:local_opearator}
\end{equation}  
is invariant under $SU(3)$ gauge transformations.
This nicely shows the geometrical interpretation of the concepts underlying particle physics, in particular gauge invariance.
A similar argument holds for the matrix elements involved in gluon-gluon correlators, with the difference that we need two links to ensure gauge invariance:
\begin{equation}
F^{\mu \nu}(0)\ U(0,\xi)\ F^{\rho \sigma}(\xi)\ U^\prime(\xi,0) \ .
\label{e:link_structure_gluons}
\end{equation}
This is because gluons live in the octet representation of $3 \otimes 3^\ast$ and, accordingly, we need two color matrices to connect each of the factor representation $3$ and $3^\ast$~\cite{Buffing:2015tja}.

\subsection{Eikonal quarks}
\label{ss:eikonal_q}
Let's consider field combinations denoted as
\begin{equation}
%
U_{\pm}^{[n_-]}\psi(p^+) 
\equiv \int_{-\infty}^{+\infty} d\xi^-\ e^{i\,p^+\cdot\xi^-}
\,U_{[\pm\infty,\xi^-]}^{[n_-]}\psi(\xi^-),
\end{equation}
where $\xi^- = \xi\cdot P$ and $p^+ = p\cdot n$, which are the only relevant components that we want to consider here. 
The superscript $[n_-]$ indicates that we are considering only the piece of the gauge link along the $n_-$ direction. The subscripts $\pm$ refer to the future or past pointing links. 
For the link $U_-$ explicitly one has (with $A^+ = A\cdot n$):
\begin{equation}
%
U_-^{[n_-]}\psi(p^+) =
\sum_{N=0}^\infty (-i)^N\int_{-\infty}^\infty d\xi^- 
\int_{-\infty}^\xi d\eta^-_N \int_{\eta^-_N}^\xi d\eta^-_{N-1}
\ldots\int_{\eta^-_2}^\xi d\eta^-_1
\ A^+(\eta^-_N)\ldots A^+(\eta^-_1)\psi(\xi)\,e^{i\,p^+\xi^-},
\end{equation}
in which the arguments run between $-\infty<\eta^-_N < \eta^-_{N-1}< \ldots < \eta^-_1 < \xi^-$, implemented through $\theta(\eta^-_{N-1}-\eta^-_N)\ldots\theta(\eta^-_1-\eta^-_2) \theta(\xi^--\eta^-_1)$.
One can show that~\cite{Mulders:2016pln}:
\begin{align}
\label{e:eikonal_quarks}
& U_-^{[n_-]} \psi(p^+) \sim \psi(p^+) + g_s \int_{-\infty}^{+\infty} \frac{dp_1^+}{2\pi} \frac{A^+(p_1^+)}{p_1^+ + i\epsilon}\ \psi(p^+ - p_1^+) \ , \\ 
\nonumber 
& U_+^{[n_-]} \psi(p^+) \sim \psi(p^+) + g_s \int_{-\infty}^{+\infty} \frac{dp_1^+}{2\pi} \frac{A^+(p_1^+)}{p_1^+ - i\epsilon}\ \psi(p^+ - p_1^+) \ . 
\end{align}

\subsection{Link structure in DIS-like and Drell-Yan-like processes}
\label{ss:link_dep}

\begin{figure}
\centering
\includegraphics[width=0.90\textwidth]{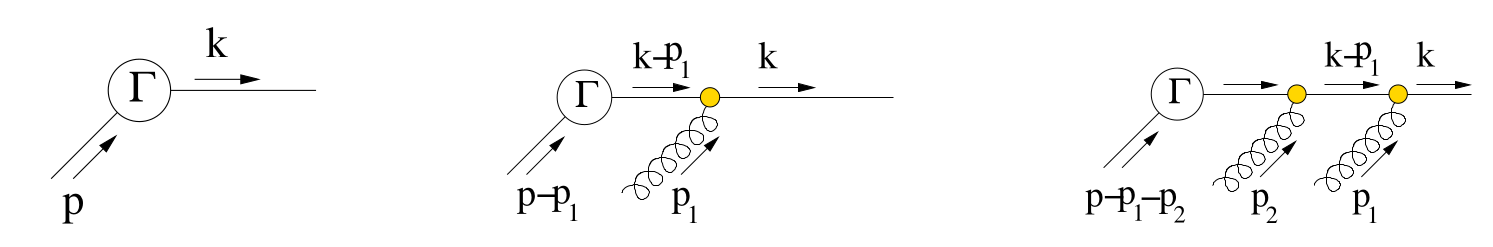}
\caption[Diagrams considered in the calculation of gauge links]{Diagrams contributing to the gauge link for DIS-like processes up to order $g_s^2$. Figure from Ref.~\cite{Mulders:2016pln}}
\label{f:link_diagrams}
\end{figure}

We want to consider a hard process and see which gluon fields need to be resummed to get a gauge-invariant correlator including a gauge link.
At this stage we neglect non-abelian effects.
Working with momentum representation of the fields~\cite{Mulders:2016pln}, we consider only the collinear term in the $A^\mu(p)$ insertion, namely $A^{n_-}(p)p^\mu/(p \cdot n_-)$. 
Furthermore, let's first consider the situation in which the hard process is a simple (constant) vertex (like a $\gamma^\alpha$ in deep inelastic scattering). 
The diagrams contributing to a DIS-like configuration up to ${\mathcal O}(g_s^2)$ are displayed in Fig.~\ref{f:link_diagrams}.
%
Here we consider the first two diagrams only. For including the third diagram and the transverse degrees of freedom too, see Ref.~\cite{Mulders:2016pln}. 
The first term is trivial:
\begin{equation}
A_0^{\text{DIS}} = \overline\psi(k)\Gamma\psi(p) \ ,
\label{e:A0_DIS}
\end{equation}
where the $\psi(p)$ and $\overline \psi(k)$ are fields belonging to the correlators of initial (momentum $p$) and final state quark (momentum $k$) respectively. 

Let us now consider $\{p,k\}$ as the light-cone basis (see App.~\ref{a:lc_coordinates}). In this way we have $p_1 \cdot k = p_1^+$. 
The first contribution to the gauge link is:
\begin{equation}
A_1^{\text{DIS}} = \int \frac{d^4p_1}{(2\pi)^4}\ \overline\psi(k) (ig_s \gamma_\mu)\ \frac{A^k(p_1)\ p_1^\mu}{p_1 \cdot k}\ \frac{i (\slash k - \slash p_1)}{(k-p_1)^2 +i\epsilon}\ \Gamma\ \psi(p - p_1) \ .
\label{e:A1_DIS}
\end{equation}
The numerator becomes $\slash p_1(\slash k -\slash p_1)$ =
$\slash p_1\slash k$ = $\{\slash k,\slash p_1\}$ = $2\,k\cdot p_1$. The added
term is zero since $\overline \psi(k)\slash k \approx 0$. Thus one has:
\begin{align}
A_1^{\text{DIS}} &= \overline\psi(k) \int \frac{d^4p_1}{(2\pi)^4} 
\ \frac{A^k(p_1)}{(-k\cdot p_1+i\epsilon)}
\,\Gamma\,\psi(p-p_1) \ .
\end{align}
Comparing to Eq.~\eqref{e:eikonal_quarks} and considering only the collinear momentum components (for the treatment including all components see Ref.~\cite{Mulders:2016pln}), we can see that:
\begin{equation}
A_0^{\text{DIS}} + A_1^{\text{DIS}} \sim \overline{\psi}(k^+)\ U_+^{[n_-]}\ \Gamma\ \psi(p^+) \ .
\label{e:DIS_link}
\end{equation}
Eq.~\eqref{e:DIS_link} means that the gauge-link structure entering DIS-like processes is the future pointing link (hadron remnant feeling the color force of a parton in the final state). 
Following the same procedure, it is possible to show that the gauge-link structure entering Drell-Yan-like processes is the past pointing link (hadron remnant feeling the color force of a parton in the initial state):
\begin{equation}
A_0^{\text{DY}} + A_1^{\text{DY}} \sim \overline{\psi}(k^+)\ U_-^{[n_-]}\ \Gamma\ \psi(p^+) \ .
\label{e:DY_link}
\end{equation}
This implies that the parton distribution functions are process dependent, but the interplay between the time-reversal symmetry and the gauge symmetry makes the process dependence calculable (a simple sign change for the quark Sivers and the quark Boer-Mulders TMD PDFs).


\subsection{Discrete transformations of the gauge link}
\label{ss:transf_link}

\begin{figure}
\centering
\begin{tabular}{ccc}
\includegraphics[width=0.3\textwidth]{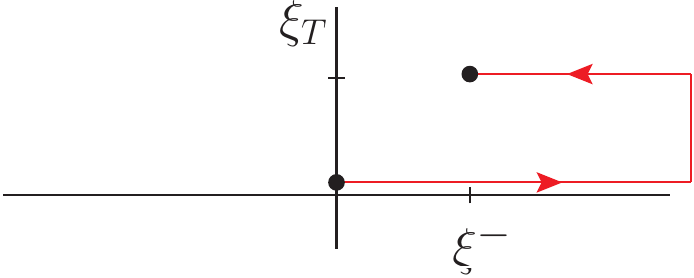}
& \hspace{2cm} &
\includegraphics[width=0.3\textwidth]{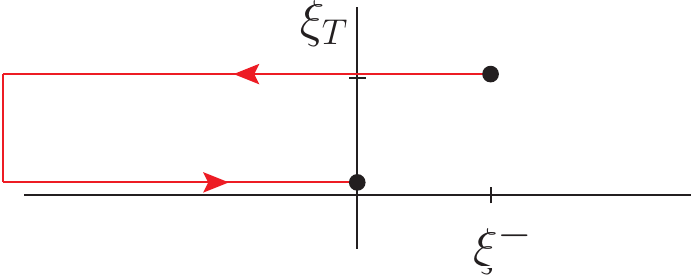}
\\
(a) & & (b)
\\ 
\end{tabular}
\caption{Representations of staple-like Wilson lines running from $0$ to $\xi$ in the hyperplane defined by the minus component and the transverse components of $\xi$. The future-pointing Wilson line $U_+(\xi,0)$ (a) and the past-pointing Wilson line $U_-(\xi,0)$ (b). Figures from Ref.~\cite{Buffing:2011mj}.}
\label{f:staples}
\end{figure}

We have defined $U^{[n_-]}$ to be a link running along the light-cone minus component in spacetime and for which the $A$ field is projected along the $n_-$ direction. A notation with similar meaning is introduced for $U^{[T]}$. 
The integral over the minus component in spacetime can be further decomposed in two consecutive integrations, one from $a^-$ to $\pm \infty$ and another from $\pm \infty$ to $b^-$. In this way, it is possible to introduce the so-called {\em staple-like} links (see Fig.~\ref{f:staples}):
\begin{equation}
U_{\pm}(b,a) \eqdef U^{[n_-]}(\pm \infty,a^-)\ U^{[T]}(b_T,a_T)\ U^{[n_-]}(b^-,\pm \infty) \, ,
\label{e:staple_links}
\end{equation} 
where the adopted notation means that the link runs from $a$ to $b$, or, in general, from the second to the first point in brackets. 

There are two crucial points here: the first one is that the path followed in spacetime to bridge the non-locality is determined by the hard scattering process (in particular, $U_-$ contributes to DY, whereas $U_+$ enters the SIDIS cross section, as briefly outlined in Sec.~\ref{ss:link_dep}). The second one is that the path dependence mixes with the discrete symmetries. 
%
%
A time reversal transformation interchanges the staple links $U_+$ and $U_-$, whereas parity and hermiticity do not. In practice:
\begin{align}
\label{e:U_herm}
& U_{\pm}(a,b)^\dagger = U_{\pm}(b,a) \\ 
\label{e:U_parity}
& {\mathscr P} U_{\pm}(a,b) {\mathscr P}^\dagger = U_{\pm}(\tilde a,\tilde b) \\ 
\label{e:U_trev}
& {\mathscr T} U_{\pm}(a,b) {\mathscr T}^\dagger = U_{\mp}(-\tilde a,-\tilde b) \, ,
\end{align}
where the proofs for Eqs.~\eqref{e:U_herm},~\eqref{e:U_parity},~\eqref{e:U_trev} follow from the application of the transformation properties for the gluon field $A^\mu$ to the definition of the gauge link and its staple-like version (see Eq.~\eqref{e:par_transp_psi},~\eqref{e:staple_links}):
\begin{equation}
\label{e:A_transf}
{A^\mu(\xi)}^\dagger = A^\mu(\xi) \, , \ \ \ \ \
{\mathscr P} A^\mu(\xi) {\mathscr P}^\dagger = \tilde{A}^\mu(\tilde{\xi}) \, , \ \ \ \ \
{\mathscr T} A^\mu(\xi) {\mathscr T}^\dagger = \tilde{A}^\mu(-\tilde{\xi}) \, .
\end{equation}

The previous equations mean that hermiticity and the parity transformation do not mix with the structure of the gauge link, in particular with the associated path. On the contrary, time reversal does, as it changes a past-pointing Wilson line into a future-pointing one, and vice-versa. This has profound consequences at the level of hadron structure.

\section{Combining discrete and gauge symmetries}
\label{s:discrete_gauge}


We now revisit the definition for the quark-quark correlator for a spin 1/2 hadron given in Eq.~\eqref{e:qq_unint_corr}. Including the Wilson line to make the matrix element gauge invariant we have:
\begin{equation}
\Phi^{[U]}_{ij}(k,P,S) = \frac{1}{(2\pi)^4}\int d^4\xi\ e^{i\,k\cdot \xi}  
\langle P,S \vert \overline \psi_j(-\xi/2)\, U(-\xi/2,\xi/2)\, \psi_i(\xi/2) \vert P,S \rangle \, ,
\label{e:qq_unint_corr_link}
\end{equation}
where the superscript $[U]$ represents the dependence on the path of the Wilson line in spacetime.

Taking into account also the gauge links, the conditions imposed on the quark-quark correlator by applying hermiticity, parity, and time-reversal symmetry on the fields and the states are:
\begin{eqnarray}
\label{e:qq_d_herm}
\mbox{Hermiticity:} &\quad \Phi^{[\pm] \dagger}(k;P,S) = \gamma^0 \Phi^{[\pm]}(k;P,S) \gamma^0 \\
\label{e:qq_d_parity}
\mbox{Parity:} &\quad \Phi^{[\pm]}(k;P,S) = \gamma^0 \Phi^{[\pm]}(\tilde{k};\tilde{P},-\tilde{S}) \gamma^0 \\
\label{e:qq_d_trev}
\mbox{Time reversal:} &\quad \Phi^{[\pm] *}(k;P,S) = i\gamma^1\gamma^3 \Phi^{[\mp]}(\tilde{k};\tilde{P},\tilde{S}) i \gamma^1\gamma^3 \ ,
\label{e:qq_transf_link}
\end{eqnarray}
where the superscripts $[\pm]$ refer to the staple-like gauge links (see Eq.~\eqref{e:staple_links}) and the tilde refers to quantities transformed according to parity, see Eq.~\eqref{e:P_rules},~\eqref{e:T_rules}.

The interplay between the time reversal symmetry and the gauge link generates relations between the $\Phi^{[+]}$ and the $\Phi^{[-]}$ correlators, which allow one to define time reversal-odd (T-odd) and time reversal-even (T-even) combinations of $\Phi^{[\pm]}$. 
The cases of the $\gamma^+$ and $i\sigma^{k+}\gamma_5$ projections will be discussed in Sec.~\ref{ss:gammaplus} and Sec.~\ref{ss:todd_bm} respectively. 
Parity, instead, does not mix with the gauge link structure. For this reason, it always poses {\em constraints} on correlators, preventing the possibility of defining P-odd structures. 

The impact of the charge-conjugation transformation on quark and gluon distributions taking into account the gauge links is discussed in~\cite{Boer:1998im,Boer:2016xqr} respectively.


\subsection{The Sivers distribution}
\label{ss:gammaplus}

Considering $\Gamma = \gamma^+$, the relevant structures in the TMD correlator are:
\begin{equation}
\label{e:TMD_corr_gp}
\Phi^{[U]}(x,\vec{k}_T) = \frac{1}{2}\, f_1^{[U]}(x,k_T^2)\, \slash{n}_+ 
					+ \frac{1}{2M}\, f_{1T}^{[U] \perp}(x,k_T^2)\, \epsilon_T^{\alpha\beta} {S_T}_\alpha {k_T}_\beta\, \slash{n}_+ \, ,
\end{equation}
where $f_1^{[U]}$ and $f_{1T}^{\perp [U]}$ are, respectively, the path-dependent unpolarized and Sivers TMD PDFs, which are real functions. The tensor $\epsilon_{T}$ is defined in App.~\ref{a:lc_coordinates}.

By projecting along $\gamma^+$ the time-reversal relation in Eq.~\eqref{e:qq_d_trev}, we obtain:
\begin{equation}
\label{e:Trev_gp_proj1}
\Tr\bigg[ \Phi^{[+]*}(k,P,S)\, \gamma^+ \bigg] = 
\Tr\bigg[ \Phi^{[-]}(\tilde{k},\tilde{P},\tilde{S})\, (-i\gamma_5 C) \gamma^+ (-i\gamma_5 C) \bigg] \, ,
\end{equation}
which, by virtue of Eq.~\eqref{e:T_rules}, becomes:
\begin{equation}
\label{e:Trev_gp_proj2}
\Tr\bigg[ \Phi^{[+]*}(k,P,S)\, \gamma^+ \bigg] = 
\Tr\bigg[ \Phi^{[-]}(\tilde{k},\tilde{P},\tilde{S})\, {\gamma^+}^\prime \bigg] \, ,
\end{equation}
where the prime indicates the transformation under parity or time reversal. 
Keeping in mind that also the light-cone basis vectors change under parity and time reversal transformations (see Eq.~\eqref{e:P_rules},~\eqref{e:T_rules}) and considering the unpolarized part of the TMD correlator defined in Eq.~\eqref{e:TMD_corr_gp}, we get:
\begin{equation}
\label{e:unp_Trev_1}
\Tr\bigg[ f_1^{[+]}(x,k_T^2)\, \slash{n}_+\, \gamma^+ \bigg] = 
\Tr\bigg[ f_1^{[-]}(x,k_T^2)\, {\slash{n}_+}^\prime\, {\gamma^+}^\prime \bigg] \, ,
\end{equation}
which implies that $f_1^{[+]}(x,k_T^2) = f_1^{[-]}(x,k_T^2)$. \\

Considering the spin-dependent part of the TMD correlator, instead, we get:
\begin{equation}
\label{e:pol_Trev_1}
\Tr\bigg[ f_{1T}^{\perp [+]}(x,k_T^2)\, \epsilon^{\mu\nu\alpha\beta}\, {n_+}_\mu {n_-}_\nu {S_T}_\alpha {k_T}_\beta \,  
\slash{n}_+\, \gamma^+ \bigg] = 
(-)^4\, \Tr\bigg[ f_{1T}^{\perp [-]}(x,k_T^2)\, \epsilon^{\mu\nu\alpha\beta}\, {n_-}_\mu {n_+}_\nu {S_T}_\alpha {k_T}_\beta \, 
{\slash{n}_+}^\prime\, {\gamma^+}^\prime \bigg] \, ,
\end{equation}
which, thanks to the total antisymmetry of the $\epsilon$ tensor, implies:
\begin{equation}
\label{e:pol_Trev_2}
f_{1T}^{\perp [+]}(x,k_T^2)\, \epsilon^{\mu\nu\alpha\beta}\, {n_+}_\mu {n_-}_\nu {S_T}_\alpha {k_T}_\beta =
- f_{1T}^{\perp [-]}(x,k_T^2)\, \epsilon^{\nu\mu\alpha\beta}\, {n_+}_\nu {n_-}_\mu {S_T}_\alpha {k_T}_\beta \, ,
\end{equation}
namely 
\begin{equation}
\label{e:pol_Trev_3}
f_{1T}^{\perp [+]} = -f_{1T}^{\perp [-]} \, ,
\end{equation}
which is the well-known sign-change relation for the Sivers TMD PDF between the future and past-pointing gauge link configurations, namley between SIDIS-like and Drell-Yan-like processes. If the Sivers function is non-zero for a specific link configuration, than the symmetries of QCD predict that it should have the opposite value using the conjugated path in spacetime. 
It is important to notice that without the path-dependence of the distributions we would just obtain $f_{1T}^{\perp} = -f_{1T}^{\perp} = 0$. This argument applies to all the T-odd TMD PDFs, not only to the Sivers function. 

At the correlator level it is possible to define the T-even and the T-odd combinations of the two gauge link configurations:
\begin{align}
\label{e:T_corr}
& \Phi_{[\gamma^+]}^{\text{T-even}}(k,P,S) = \frac{1}{2} \bigg[ \Phi_{[\gamma^+]}^{[+]}(k,P,S) + \Phi_{[\gamma^+]}^{[-]}(k,P,S) \bigg] \, , \\ 
& \Phi_{[\gamma^+]}^{\text{T-odd}}(k,P,S) = \frac{1}{2} \bigg[ \Phi_{[\gamma^+]}^{[+]}(k,P,S) - \Phi_{[\gamma^+]}^{[-]}(k,P,S) \bigg] \, ,
\end{align}
which satisfy
\begin{align}
\label{e:T_corr_eigen}
& \Phi_{[\gamma^+]}^{\text{T-even}}(k,P,S)^* = \Phi_{[{\gamma^+}^\prime]}^{\text{T-even}}(\tilde{k},\tilde{P},\tilde{S}) \, , \\ 
& \Phi_{[\gamma^+]}^{\text{T-odd}}(k,P,S)^* = - \Phi_{[{\gamma^+}^\prime]}^{\text{T-odd}}(\tilde{k},\tilde{P},\tilde{S}) \, .
\end{align}
Moreover, we can also write:
\begin{equation}
\label{e:pm_even_odd}
\Phi_{[\gamma^+]}^{[\pm]}(k,P,S) = \Phi_{[\gamma^+]}^{\text{T-even}}(k,P,S) \pm \Phi_{[\gamma^+]}^{\text{T-odd}}(k,P,S) \, , 
\end{equation}
and thus one realizes that the unpolarized TMD PDF is the T-even part of $\Phi^{[\pm]}_{[\gamma^+]}$ whereas the Sivers TMD PDF is part of the T-odd part of $\Phi^{[\pm]}_{[\gamma^+]}$.  
At this point it is useful to introduce Tab.~\ref{t:TMD_PDFs_color} and~\ref{t:TMD_FFs_color}, which are an improved version of Tab.~\ref{t:TMD_PDFs} and~\ref{t:TMD_FFs} that classify the TMDs according to the polarization of the quark and of the parent hadron, and introduce a color code to mark the property of each function under the time reversal symmetry.

\subsection{The Boer-Mulders distribution}
\label{ss:todd_bm}

The case of the T-odd quark TMD PDF $h_1^\perp$ (Boer-Mulders distribution) (Tab.~\ref{t:tw2_TMD_PDFs_color}) is similar to the Sivers one, but related to a different Dirac projection of the correlator.
The projection of the correlator related to transversely polarized quarks is $\Phi^{[i\sigma^{k+}\gamma_5]}$.
Tracing the TMD quark distribution correlator with $i\sigma^{k+}\gamma_5$ and introducing time reversal transformations we get:
\begin{equation}
T \ \ : \ \ \Phi_{[i\sigma^{k+}\gamma_5]}^{[+]} (p,P,S;n_{\pm}) = - \Phi_{[\overline{i\sigma^{k+}\gamma_5}]}^{[-]} (\bar{p},\bar{P},\bar{S};\bar{n}_{\pm})^* \ . 
\label{e:T_isigmagamma5}
\end{equation}
As in Sec.~\ref{ss:gammaplus} the interchange of the staple links allows the decomposition of $\Phi_{[i\sigma^{k+}\gamma_5]}$ into T-even and T-odd pieces:
\begin{align}
& \Phi_{[i\sigma^{k+}\gamma_5]}^{\text{T-even}} (p,P,S) = \frac{1}{2} \bigg[ \Phi_{[i\sigma^{k+}\gamma_5]}^{[+]} (p,P,S) - \Phi_{[i\sigma^{k+}\gamma_5]}^{[-]} (p,P,S) \bigg] \\
& \Phi_{[i\sigma^{k+}\gamma_5]}^{\text{T-odd}} (p,P,S) = \frac{1}{2} \bigg[ \Phi_{[i\sigma^{k+}\gamma_5]}^{[+]} (p,P,S) + \Phi_{[i\sigma^{k+}\gamma_5]}^{[-]} (p,P,S) \bigg] \ ,
\label{e:teven_todd_forms_BM}
\end{align}
Note the different signs with respect to the Sivers case.
$h_1$, $h_{1L}^\perp$, $h_{1T}^\perp$ are the functions entering the T-even part of $\Phi_{[i\sigma^{k+}\gamma_5]}$ (Tab.~\ref{t:tw2_TMD_PDFs_color}). The Boer-Mulder function $h_1^\perp$ enters the T-odd part and, as the Sivers, is subject to a sign change between SIDIS and Drell-Yan:
\begin{equation}
h_1^{\perp [+]} (x,p_T^2) = - h_1^{\perp [-]} (x,p_T^2) \ .
\label{e:procdep_bm}
\end{equation} 

The sign-change relation can also be obtained from the interplay between the combined $PT$ symmetries and the gauge symmetry, as discussed for example in Ref.~\cite{Collins:2011zzd}. \\

For the discussion of time-reversal odd effects beyond the leading twist and in fragmentation functions vs PDFs, we refer the reader to Ref.~\cite{Boer:2003cm}. \\

Last but not least, we note that this argument is not affected by the specific gauge group used to define the gauge link. It can be the $SU(3)$ group for QCD (in the case of a colored quark extracted from a QCD bound state) or the $U(1)$ group of QED (in the case of an electrically charged particle -e.g. an electron- extracted from a QED bound state)~\cite{Manna:2021nep}.


\renewcommand{\arraystretch}{1.5}
\begin{table}
\centering

\begin{subtable}{0.48\textwidth}
\centering
\begin{tabular}{c|c|c|c|c|}
\multicolumn{2}{c}{}&\multicolumn{3}{c}{quark pol.}\\
\cline{2-5}
& & U & L & T \\ \cline{2-5}
\multirow{3}{*}{\rotatebox{90}{nucleon pol.}} &{U} & ${\color{black} f_1}$   & & { \color{red}$h_{1}^\perp$} \\
\cline{2-5}
& {L} & &${\color{black} g_{1L}}$ &  {\color{blue} $h_{1L}^\perp$} \\
\cline{2-5}
& {T} & ${\color{red}f_{1T}^\perp }$ &  {\color{blue} $g_{1T}$} &  {\color{black} $h_{1T}$}, {\color{blue}$h_{1T}^{\perp}$} \\
\cline{2-5} 
\end{tabular}
\caption{Twist-2 TMD PDFs}
\label{t:tw2_TMD_PDFs_color}
\end{subtable}
\hfill
\begin{subtable}{0.48\textwidth}
\centering
\begin{tabular}{c|c|c|c|c|}
\multicolumn{2}{c}{}&\multicolumn{3}{c}{quark pol.}\\
\cline{2-5}
& & U & L & T \\ \cline{2-5}
\multirow{3}{*}{\rotatebox{90}{nucleon pol.}} 
&{U} & ${\color{blue} f^{\perp}}$   &${\color{red} g^{\perp}}$ & { \color{red}$h$}, {\color{black}$e$} \\
\cline{2-5}
& {L} & ${\color{red}f_{L}^\perp }$ &${\color{blue} g_{L}^{\perp}}$ & {\color{black}$h_L$}, {\color{red}$e_L$}  \\
\cline{2-5}
& {T} & {\color{red} $f_T'$, $f_{T}^\perp$} &  {\color{black} $g_T'$}, {\color{blue} $g_{T}^{\perp}$} &  {\color{blue}$h_T$, $h_{T}^{\perp}$},  {\color{red} $e_T$,  $e_{T}^\perp $} \\
\cline{2-5}
\end{tabular}
\caption{Twist-3 TMD PDFs}
\label{t:tw3_TMD_PDFs_color}
\end{subtable}

\caption{TMD PDFs at twist-2 and twist-3. Functions in black survive
transverse momentum integration. Functions in red are T-odd.}
\label{t:TMD_PDFs_color}
\end{table}
\renewcommand{\arraystretch}{1.5}
\begin{table}
\centering

\begin{subtable}{0.48\textwidth}
\centering
\begin{tabular}{c|c|c|c|c|}
\multicolumn{2}{c}{}&\multicolumn{3}{c}{quark pol.}\\
\cline{2-5}
& & U & L & T \\ \cline{2-5}
\multirow{3}{*}{\rotatebox{90}{hadron pol.}} &{U} & ${\color{black} D_1}$   & & { \color{red}$H_{1}^\perp$} \\
\cline{2-5}
& {L} & &${\color{black} G_{1L}}$ &  {\color{blue} $H_{1L}^\perp$} \\
\cline{2-5}
& {T} & ${\color{red}D_{1T}^\perp }$ &  {\color{blue} $G_{1T}$} &  {\color{black} $H_{1}$}, {\color{blue}$H_{1T}^{\perp}$} \\
\cline{2-5} 
\end{tabular}
\caption{Twist-2 TMD FFs}
\label{t:tw2_TMD_FFs_color}
\end{subtable}
\hfill
\begin{subtable}{0.48\textwidth}
\centering
\begin{tabular}{c|c|c|c|c|}
\multicolumn{2}{c}{}&\multicolumn{3}{c}{quark pol.}\\
\cline{2-5}
& & U & L & T \\ \cline{2-5}
\multirow{3}{*}{\rotatebox{90}{hadron pol.}} 
&{U} & ${\color{blue} D^{\perp}}$   &${\color{red} G^{\perp}}$ & { \color{magenta}$H$}, {\color{black}$E$} \\
\cline{2-5}
& {L} & ${\color{red}D_{L}^\perp }$ &${\color{blue} G_{L}^{\perp}}$ & {\color{black}$H_L$}, {\color{magenta}$E_L$}  \\
\cline{2-5}
& {T} & {\color{magenta} $D_{T}$}, {\color{red} $D_{T}^\perp$} &  {\color{black} $G_{T}$}, {\color{blue} $G_{T}^{\perp}$} &  {\color{blue}$H_T$, $H_{T}^{\perp}$},  {\color{red} $E_T$,  $E_{T}^\perp $} \\
\cline{2-5}
\end{tabular}
\caption{Twist-3 TMD FFs}
\label{t:tw3_TMD_FFs_color}
\end{subtable}

\caption{TMD FFs at twist-2 and twist-3. 
Functions in black and magenta survive transverse momentum integration. Functions in red and magenta are T-odd.}
\label{t:TMD_FFs_color}

\end{table}

\section{Exercises}
\label{s:ex_symm}

\begin{itemize}

\item[1)] Solve the parallel transport equation~\eqref{e:par_transp} to get the expression of a generic Wilson line used in Eq.~\eqref{e:par_transp_psi}

\item[2)] Derive the expression of the gauge link in the case of Drell-Yan $U^{[-]}$, following the steps outlined in Sec.~\ref{ss:eikonal_q} for the case of $U^{[+]}$ (see Eqs.~\eqref{e:DIS_link} and~\eqref{e:DY_link})

\item[3)] Show that the staple gauge link $[\pm]$ transforms into the $[\mp]$ one under time reversal transformations, whereas the parity transformation sends $[\pm]$ in $[\pm]$ (see Sec.~\ref{ss:transf_link})

\item[4)] Prove Eqs.~\eqref{e:qq_d_herm},~\eqref{e:qq_d_parity},~\eqref{e:qq_d_trev} for the transformation of the gauge-invariant correlation function $\Phi^{[U]}(k,P,S)$, using the results discussed in Sec.~\ref{s:discrete} and Sec.~\ref{ss:transf_link}.

\end{itemize}

\clearpage
\part{Phenomenology}
\label{p:pheno}

The way 3D hadron structure in momentum space is modified changing the relevant renormalization scales in the associated operators is referred to as TMD evolution. The evolution equations which govern TMD evolution are a byproduct of the factorization theorems for the involved physical observables.  
For an overview of collinear factorization we refer to~\cite{Collins:1989gx,Brock:1993sz} and references therein. 
Factorization of observables in terms of TMD distributions is a lively research topic. As comprehensive references we indicate, for example, Refs.~\cite{Collins:1984kg,Collins:2011zzd,Scimemi:2019mlf} and references therein. 
Effective theories are widely employed to address factorization and evolution. Introductory material is available in Ref.~\cite{Becher:2014oda}, via the MIT online \href{https://www.edx.org/course/effective-field-theory-mitx-8-eftx}{lecture series} and \href{https://courses.edx.org/c4x/MITx/8.EFTx/asset/notes_EFT.pdf}{notes} and the \href{https://courses.edx.org/c4x/MITx/8.EFTx/asset/notes_scetnotes.pdf}{TASI lecture notes} for 2013 and 2014. 

The partonic transverse momentum is physically associated with the emissions of gluons from the considered parton. Accordingly, and as we will see in the following, TMD distributions receive both perturbative and non-perturbative contributions. It is of utmost importance to distinguish between these two contributions and to understand in which kinematic regions TMDs are dominated by non-perturbative (or, intrinsic) effects and in which other regions they are dominated by perturbatively calculable contributions. In those regions, the predictive power of the TMD formalism is maximal. 

In Sec.~\ref{s:TMD_evo} we review the evolution equations of TMDs and their solution. 
In Sec.~\ref{s:break_pt} we explicitly show how perturbation theory breaks down at large $b_T$. 
Finally, in Sec.~\ref{s:saddle} we explore quantitatively in which kinematic regions the TMD formalism is most predictive and in which regions the non-perturbative corrections dominate, relying on the saddle-point approximation.  
Most of the following is taken from Ref.~\cite{Grewal:2020hoc}.

\section{TMD evolution}
\label{s:TMD_evo}

Let's consider an unpolarized TMD PDF for a parton with flavor $a$, 
\begin{equation}
F_a(x, k_T^2; \mu, \zeta)\,, 
\end{equation}
which carries the collinear momentum fraction $x$ of the parent hadron collinear momentum and has a transverse momentum $k_T$ with respect to the hadron's momentum. The scales $\mu$ and $\zeta$ are the ultraviolet (UV) renormalization and rapidity regularization scales, respectively~\cite{Collins:2011zzd}. 
The $k_T$ dependence of the TMD PDF probed at the hard collision is a combination of the parton's intrinsic transverse momentum and the amount of $k_T$ generated by the parton shower. 
Rather than studying the $k_T$-dependence of the TMDs directly in momentum space, 
it can be advantageous to study the TMDs in their Fourier transformed form ($b_T$-space), defined as~\cite{Collins:1984kg}:
\begin{align}
F_a(x, b_T^2; \mu, \zeta) = 
\int d^2 {\vect k}_T\, e^{i\vect{k}_T \cdot \vect{b}_T}\, 
F_a(x, k_T^2; \mu, \zeta) \, . 
\end{align} 
When $b_T$ is small, much less than $1/\Lambda_{\rm QCD}$, the scale dependence of the TMDs' $b_T$-dependence is perturbatively calculable. Otherwise it is non-perturbative, as the $b_T$ dependence itself (see Sec.~\ref{s:break_pt}).  
Once we understand the TMD PDF in the $b_T$-space, we then Fourier transform it back into the momentum space:
\begin{align}
F_a(x, k_T^2; \mu, \zeta) 
& = \int \frac{d^2\vect{b}_T}{(2\pi)^2} \,e^{-i \vect{k}_T \cdot \vect{b}_T} F_a(x, b_T^2; \mu, \zeta) 
\nonumber \\
& = \frac{1}{2\pi} \int_0^{\infty} db_T \, b_T J_0(k_Tb_T) F_a(x, b_T^2; \mu, \zeta)\, .
\label{eq:F-b-space}
\end{align}
The zero-th order Bessel function $J_0$ emerges from the angular part of the integral and the absence of any dependence on the azimuthal angle of the transverse momentum ${\bm k}_T$ in the unpolarized case. Polarized TMDs feature a Bessel function of different order.  
The above Fourier transform would require the information of the 
$F_a(x, b_T^2; \mu, \zeta)$ for the entire $b_T\in [0,\infty)$ region. 
If the Fourier transform is dominated by the information of TMDs at small $b_T$, we will have a good predictive power for $F_a(x, k_T^2; \mu, \zeta)$ in all relevant $k_T$ regions, modulo the knowledge of the standard collinear PDFs. 
On the other hand, if the Fourier transform is sensitive to the large $b_T$ region, 
the distribution in momentum space will largely depend on the chosen non-perturbative models.

\subsection{Evolution equations}
\label{e:evo_equations}

The QCD evolution equations of the TMDs take the following form~\cite{Grewal:2020hoc}:
\begin{align}
\label{e:evoeq_1}
\frac{d\, \ln\, F_a(x, b_T^2; \mu, \zeta)}{d\, \ln\, \zeta} & = - D\left(b_T\mu, \alpha_s(\mu)\right)\,,
\\
\label{e:evoeq_2}
\frac{d\, \ln\, F_a(x, b_T^2; \mu, \zeta)}{d\ln \mu} & = \gamma_F\left(\alpha_s(\mu), \frac{\zeta}{\mu^2}\right)\,,
\\
\label{e:evoeq_3}
\frac{d\, D \left(b_T\mu, \alpha_s(\mu)\right)}{d\, \ln\, \mu^2} & = \frac{1}{2} \gamma_K\left(\alpha_s(\mu)\right)\,,
\\
\label{e:evoeq_4}
\frac{d\,\gamma_F \left(\alpha_s(\mu), \frac{\zeta}{\mu^2}\right)}{d\, \ln\, \zeta} & = - \gamma_K\left(\alpha_s(\mu)\right)\,.
\end{align}
$D\left(b_T\mu, \alpha_s(\mu)\right)$ is the Collins-Soper evolution kernel\footnote{The relation between the Collins--Soper kernel $\tilde{K}$ defined in Ref.~\cite{Collins:2011zzd} and the function $D$ introduced in Eq.~\eqref{e:evoeq_1} is $\tilde{K}=-2D$.}, and $\gamma_F\left(\alpha_s(\mu), \frac{\zeta}{\mu^2}\right)$ is the anomalous dimension of the operator defining the TMD PDF. The last equation is obtained from the fact that the differential order in $\zeta$ and in $\mu$ for $F_a(x, b_T^2; \mu, \zeta)$ is interchangeable, i.e.,
\begin{equation}
\frac{d}{d\, \ln\, \zeta}\frac{d}{d\, \ln\, \mu}\ln\, F_a(x, b_T^2; \mu, \zeta) = \frac{d}{d\, \ln\, \mu} \frac{d}{d\, \ln\, \zeta} \ln\, F_a(x, b_T^2; \mu, \zeta)\, ,
\end{equation}
so long as $F_a(x, b_T^2; \mu, \zeta)$ are differentiable in both $\mu$ and $\zeta$ in the kinematic regime that we are interested in.

In the perturbative region where $1/b_T\gg \Lambda_{\rm QCD}$ (small $b_T$), one can compute the evolution kernels in the above evolution equations in perturbation theory. 
For example, for a quark TMD PDF:
\begin{align}
\label{e:D_expansion}
D\left(b_T\mu, \alpha_s(\mu)\right) &= C_F \sum_{n=1}\left(\frac{\alpha_s}{4\pi}\right)^n \sum_{k=0}^n d^{(n,k)}\, \ln^k\left(\frac{\mu^2 b_T^2}{4e^{-2\gamma_E}}\right),
\\
\label{e:gammaK_gammacuso}
\gamma_K\left(\alpha_s(\mu)\right) &= \Gamma_{\rm cusp} \left(\alpha_s(\mu)\right),
\\
\label{e:gammaF_rel}
\gamma_F\left(\alpha_s(\mu), \frac{\zeta}{\mu^2}\right) &= \Gamma_{\rm cusp} \left(\alpha_s(\mu)\right) \ln\left(\frac{\mu^2}{\zeta}\right) + \gamma\left(\alpha_s(\mu)\right),
\end{align}
where $\gamma_E$ is the Euler constant. $\Gamma_{\rm cusp} \left(\alpha_s(\mu)\right)$ and $\gamma\left(\alpha_s(\mu)\right)$ are the cusp and non-cusp anomalous dimensions, respectively. 
They generally have the expansion 
\begin{equation}
\label{e:gammacusp_expansion}
\Gamma_{\rm cusp} \left(\alpha_s(\mu)\right) = 
\sum_{n=1}\Gamma_{n-1}\left(\frac{\alpha_s}{4\pi}\right)^n \, , 
\end{equation}
likewise for the non-cusp. For a quark TMD PDF, one has $\Gamma_0 = 4C_F,~\gamma_0 = 6C_F$, etc. At the same time, we have $d^{(1,1)} = 2\Gamma_0,~d^{(1,0)} = 0$, etc. 
At the first non-trivial order, we have
\begin{align}
\label{e:CS_first}
D\left(b_T\mu, \alpha_s(\mu)\right) &= \frac{\alpha_s}{2\pi} C_F \ln\left(\frac{\mu^2 b_T^2}{4e^{-2\gamma_E}}\right),
\\
\gamma_K\left(\alpha_s(\mu)\right) &= \frac{\alpha_s}{\pi} C_F,
\\
\gamma_F\left(\alpha_s(\mu), \frac{\zeta}{\mu^2}\right) &= \frac{\alpha_s}{\pi} C_F \left[ \ln\left(\frac{\mu^2}{\zeta}\right) + \frac{3}{2}\right]. 
\end{align}
The higher-order expressions, and the expressions for gluon TMD PDF can be found, e.g., in Ref.~\cite{Echevarria:2016scs}.

\subsection{Logarithmic accuracy}
\label{ss:log_expansion}

The $\gamma_F$ function contains products of powers of the coupling constant $\alpha_s$ and a logarithm $L \eqdef \ln\, \zeta/\mu^2$. 
In the perturbative regime, the coupling is small and the log can be significantly large. For this reason, it is possible to organize the contributions to Eq.~\eqref{e:gammaF_rel} assuming that $\alpha_s\, L \sim 1$:
\begin{align}
\label{e:andim_TMD_expLogs}
\gamma_F[\alpha_s(\mu),\zeta/\mu^2] \sim\ & \alpha_s L + \alpha_s^2 L + \alpha_s^3 L + \cdots \ \ \ \text{[cusp terms]} \\
\nonumber & + \alpha_s + \alpha_s^2 + \alpha_s^3 + \cdots \ \ \ \text{[non-cusp terms]}  \\
\nonumber \sim\ & \underbrace{\alpha_s L}_{\tiny{\text{LL}}} + \underbrace{(\alpha_s + \alpha_s^2 L)}_{\tiny{\text{NLL}}} + \underbrace{(\alpha_s^2 + \alpha_s^3 L)}_{\tiny{\text{NNLL}}} + \cdots \\
\nonumber \sim\ & 1 + \alpha_s + \alpha_s^2 + \cdots \ .
\end{align}
Thus, the $\gamma_F$ function can be organized according to logarithmic accuracy, which is equivalent to a power series in $\alpha_s$. 
The first terms in Eq.~\eqref{e:andim_TMD_expLogs} is the leading log (LL), the second is the next-to-leading log (NLL), the third the next-to-next-to-leading log (NNLL), etc.
Eq.~\eqref{e:andim_TMD_expLogs} further shows that $\Gamma_{\text{cusp}}$ is required at an order in $\alpha_s$ increased by one with respect to $\gamma$.


Increasing the logarithmic accuracy in the solution of the evolution equations increases both the precision and the accuracy of the calculation of TMD observables~\cite{Bacchetta:2022awv}.

\subsection{Solution}
\label{ss:solution_evo}

Solving the evolution equations, one can obtain the evolved TMD PDF as:
\begin{equation}
\label{e:pert_TMDPDF_bT}
F_a(x,b_T^2;\mu,\zeta) = F_a(x,b_T^2;\mu_0,\zeta_0)\, 
\exp \bigg[ \int_{\mu_0}^{\mu} \frac{d\mu'}{\mu'}  \gamma_F \left( \alpha_s(\mu'), \frac{\zeta}{\mu'^2} \right) \bigg]\ 
\bigg( \frac{\zeta}{\zeta_0} \bigg)^{-D\left(b_T\mu_0, \alpha_s(\mu_0)\right)} \ ,
\end{equation}
where $\mu_0$ and $\zeta_0$ are the initial values for the renormalization scales. 
Integrating Eq.~\eqref{e:evoeq_4} from $\mu^2$ to $\zeta$, one obtains:
\begin{equation}
\gamma_F \left( \alpha_s(\mu), \frac{\zeta}{\mu^2} \right) =  - \ln\left(\frac{\zeta}{\mu^2}\right) \gamma_K\left(\alpha_s(\mu)\right) + \gamma_F \left( \alpha_s(\mu), 1 \right),
\end{equation}
and thus we have
\begin{align}
F_a(x,b_T^2;\mu,\zeta) & = F_a(x,b_T^2;\mu_0,\zeta_0)
\nonumber \\
& \times \exp \Bigg\{ -\bigg[\int_{\mu_0}^{\mu} \frac{d\mu'}{\mu'}   \left( \ln\left(\frac{\zeta}{\mu'^2}\right) \gamma_K\left(\alpha_s(\mu')\right) -\gamma_F \left( \alpha_s(\mu'), 1 \right) \right) 
 + D\left(b_T\mu_0, \alpha_s(\mu_0)\right) \ln\left( \frac{\zeta}{\zeta_0}\right) \bigg]\Bigg\}\,.
\end{align}
Finally when both $\mu_0$ and $\zeta_0$ are in the perturbative region, the TMD PDF $F_a$ at the input scales $\mu_0$ and $\zeta_0$ can be re-factorized onto collinear PDFs $f_b(x, \mu_0)$ via an operator product expansion (OPE) at low $b_T$~\cite{Collins:2011zzd}:
\begin{align}
\label{e:input_TMDPDF}
F_a(x,b_T^2;\mu_0,\zeta_0) & = 
\sum_b C_{a/b}(x,b_T^2,\mu_0,\zeta_0) \otimes f_b(x,\mu_0) 
+ {\mathcal O}_a(b_T\, \Lambda_{\rm QCD})
\\ \nonumber 
& =  
\sum_b \int_x^1 \frac{d\hat x}{\hat x}\, 
C_{a/b}\bigg(\frac{x}{\hat x},b_T^2,\mu_0,\zeta_0\bigg) f_b\left(\hat x,\mu_0\right) 
+ {\mathcal O}_{a}(b_T\, \Lambda_{\rm QCD})
\, .
\end{align}

In practice, one typically chooses the input values as $\zeta_0 = \mu_0^2  = \mu_b^2$, where
\begin{equation}
\label{e:def_mub}
\mu_b = \frac{2e^{-\gamma_E}}{b_T} \, ,
\end{equation}
in order to eliminate the potentially large logarithms in the coefficient functions $C_{a/b}$ and improve the convergence of perturbation theory (see in particular Sec.~\ref{s:break_pt}). 
At the same time, one usually chooses $\mu$ and $\zeta$ to be associated with the hard scale $Q$, such as the invariant mass of the lepton pair in the Drell-Yan process, $pp\to[\gamma^*\to] \ell^+\ell^- + X$, or the one of the virtual photon in SIDIS:
\begin{equation}
\zeta = \mu^2 = Q^2 \, .
\end{equation}
Thus in phenomenological implementations of the TMDs, one typically writes the TMD PDF in Eq.~\eqref{e:pert_TMDPDF_bT} in the following form: 
\begin{align}
F_a(x,b_T^2; Q, Q^2) & = F_a(x,b_T^2;\mu_b,\mu_b^2)
\nonumber\\
& \times \exp \Bigg\{ - \bigg[ \int_{\mu_b}^{Q} \frac{d\mu'}{\mu'}   \left( \ln\left(\frac{Q^2}{\mu'^2}\right) \gamma_K\left(\alpha_s(\mu')\right) - \gamma_F \left( \alpha_s(\mu'), 1 \right) \right)  + D\left(c, \alpha_s(\mu_b)\right) \ln\left( \frac{Q^2}{\mu_b^2}\right) \bigg] \Bigg\}\, ,
\label{e:F_pert}
\end{align}
where $F_a(x,b_T^2;\mu_b,\mu_b^2)$ is obtained through Eq.~\eqref{e:input_TMDPDF}.

\section{Breakdown of perturbation theory}
\label{s:break_pt}

We now illustrate explicitly how perturbation theory breaks down in the calculation of the TMD distributions and of the Collins-Soper kernel, working at one loop accuracy.

\subsection{TMD distributions}
\label{ss:break_pt_TMDPDF}

The quark-to-quark coefficient in Eq.~\eqref{e:input_TMDPDF} at one loop reads~\cite{Collins:2011zzd,Echevarria:2016scs,Bacchetta:2013pqa}:
\begin{align}
\label{e:Cqq_1loop}
C_{q/q}(x,b_T;\mu,\zeta) & =
\delta(1-x)
+
\frac{\alpha_s(\mu)\, C_F}{2\pi}
\Bigg[
(1-x) + \delta(1-x) 
\left(
-\frac{1}{2} L_\mu^2 +
L_\mu\, \ln\, \frac{\zeta}{\mu^2} -\frac{\pi^2}{12}
\right)
- L_\mu\, P_{qq}(x)
\Bigg]
+ {\mathcal O}\big( \alpha_s(\mu)^2 \big) \, ,
\end{align}
where
\begin{equation}
L_\mu = \ln \left(\frac{\mu^2 b_T^2}{b_0^2}\right) \, ,
  \qquad 
b_0 = 2 e^{-\gamma_E},
\end{equation}
and
\begin{equation}
P_{qq}(x)
=
\left(\frac{1+x^2}{1-x}\right)_+
+\frac{3}{2}\delta(1-x) \, .
\end{equation}
The quark-from-gluon channel starts at one loop:
\begin{equation}
C_{q/g}(x,b_T;\mu,\zeta)
=
\frac{\alpha_s(\mu)\, T_R}{2\pi}\, 
\Big[
2x\, (1-x) - L_\mu\, P_{qg}(x)
\Big]
+ {\mathcal O}\big( \alpha_s(\mu)^2 \big) \, ,
\end{equation}
with
\begin{equation}
P_{qg}(x) = x^2 + (1-x)^2.
\end{equation}

Let's now consider the scenario in which we choose a generic value for $\mu$ and consider the large $b_T$ region. In this limit: 
\begin{equation}
L_\mu = \ln \left(\frac{\mu^2 b_T^2}{b_0^2}\right) \gg 1 \, .
\end{equation}
The perturbative expansion is effectively in powers of
$\alpha_s L_\mu^2$ and $\alpha_s L_\mu$ rather than $\alpha_s$ alone. In particular, the ratio between the one-loop and tree-level contributions in the TMD PDF at fixed $\mu$ and large $b_T$ behaves as:
\begin{equation}
\frac{F^{(1)}}{F^{(0)}} \sim \frac{\alpha_s(\mu)\, C_F}{2\pi} L_\mu^2 \, .
\end{equation}
When the logarithm grows such that $\alpha_s L_\mu^2 > 1$, the first order correction $F^{(1)}$ becomes larger that the leading term $F^{(0)}$ and fixed-order perturbation theory breaks down.\\

In order to circumvent the problem of large logarithms, one could fix the scale $\mu = \mu_b \equiv b_0/b_T$ and automatically obtain $L_{\mu_b} = 0$. 
The $qq$ coefficient then simplifies to:
\begin{align}
\label{e:Cqq_1loop_mub}
C_{q/q}(x,b_T;\mu_b) & =
\delta(1-x)
+
\frac{\alpha_s(\mu_b)\, C_F}{2\pi} \,  
\bigg[ (1-x) - \delta(1-x)\, \frac{\pi^2}{16} \bigg]  
+ {\mathcal O}\big( \alpha_s(\mu_b)^2 \big) \, ,
\end{align}
However, the coupling is now evaluated at the scale $\mu_b = b_0/b_T$. For large $b_T$, this scale becomes soft and eventually approaches $\Lambda_{\rm QCD}$, so that $\alpha_s(\mu_b)$ becomes large and perturbation theory fails. The same applies to the $qg$ coefficient. 

Beyond logarithmic effects, the OPE itself in Eq.~\eqref{e:input_TMDPDF} contains power corrections of the form
${\mathcal O}(b_T \Lambda_{\rm QCD})$
which are not suppressed when $b_T \Lambda_{\rm QCD} \sim 1$. 
Therefore, the perturbative matching onto collinear PDFs is valid only if $\alpha_s(1/b_T) \ll 1$ and $b_T \Lambda_{\rm QCD} \ll 1$, namely at small $b_T$.

\subsection{Collins--Soper kernel}
\label{ss:break_pt_CSk}

The same issue appears in the Collins--Soper evolution kernel $D$, which at one loop is given by (see Eq.~\eqref{e:CS_first}, Refs.~\cite{Collins:2011zzd,Echevarria:2011epo}):
\begin{equation}
D(b_T;\mu) =
\frac{\alpha_s(\mu)}{2\pi}\, C_F\, L_\mu 
+ {\mathcal O}\big( \alpha_s(\mu)^2 \big) \, .
\end{equation}
As for TMD distributions, at large $b_T$ either large logarithms appear at fixed $\mu$, or the coupling becomes nonperturbative because of the scale choice $\mu = \mu_b$. 
The result is the same: perturbation theory breaks and one needs non-perturbative corrections at large $b_T$. \\

Thus, the applicability region of perturbation theory at leading twist are defined by the following observations:
\begin{itemize}
\item at fixed $\mu$, large $b_T$ implies the presence of large logarithms $L_\mu$,
\item the choice $\mu = \mu_b \sim 1/b_T$ removes these logarithms but drives the coupling into the nonperturbative regime,
\item power corrections of the form ${\mathcal O}(b_T \Lambda_{\rm QCD})$ invalidate the leading-twist expansion of TMDs onto collinear distributions.
\end{itemize}

Summarizing: in the kinematic region where $Q$ is large ($Q\gg \Lambda_{\rm QCD}$) and in the small-$b_T$ region, one can rely on the perturbative result in Eq.~\eqref{e:F_pert} to obtain the information for the TMD PDF $F_a(x,b_T^2; Q, Q^2)$. 
However, in the large-$b_T$ region perturbative results are no longer reliable
and non-perturbative effects become relevant.
In particular, non-perturbative corrections are needed for the matching of the TMD distribution onto the collinear PDFs (see Sec.~\ref{ss:break_pt_TMDPDF}) and for the Collins-Soper kernel (see Sec.~\ref{ss:break_pt_CSk}). These corrections are usually referred to as the intrinsic part of the TMD distribution and the non-perturbative part of TMD evolution. 
Several proposals have been introduced to study the TMDs and their evolution at large $b_T$. It is not the purpose of these lecture notes to discuss them here, but the interested reader can look, for example, at Refs.~\cite{Bacchetta:2022awv,Scimemi:2019cmh}.

\section{Non-perturbative content} 
\label{s:saddle}

Let's consider again the unpolarized TMD PDF as a Fourier transform of the $b_T$-space distribution:
\begin{equation}
F_a(x, k_T^2; Q, Q^2) = \frac{1}{2\pi} \int_0^{\infty} db_T \, b_T J_0(k_Tb_T) F_a(x, b_T^2; Q, Q^2)\,,
\label{eq:F-b-space-Q}
\end{equation}
where we have set $\zeta=\mu^2=Q^2$ in Eq.~\eqref{eq:F-b-space}. As discussed, if $F_a(x, b_T^2; Q, Q^2)$ in the $b_T$-space is dominated by the small-$b_T$ behavior, the integration on the right-hand side, and thus $F_a(x, k_T^2; Q, Q^2)$ in $k_T$-space, will be mainly controlled by perturbative physics. On the contrary, if 
$F_a(x, b_T^2; Q, Q^2)$ is very sensitive to the large-$b_T$ behavior, non-perturbative physics will play an important role in the behavior of the TMD PDF $F_a(x, k_T^2; Q, Q^2)$ in momentum space. Understanding the TMD PDF $F_a(x, k_T^2; Q, Q^2)$ in momentum $k_T$-space, i.e., whether it is more dominated by perturbative (small-$b_T$) or non-perturbative (large-$b_T$) physics, is very important in order to investigate the predictive power of the TMD formalism.

\subsection{Saddle point approximation}
\label{ss:saddle_method}

Following Refs.~\cite{Collins:1984kg,Qiu:2000hf}, we use the saddle-point method to pinpoint if and how the integration on the right-hand side of Eq.~\eqref{eq:F-b-space-Q} is dominated by the small-$b_T$ region. The saddle-point approximation, or the method of steepest descent, is often used to approximate an integral when the integrand has the form of $e^{-c\, S(b_T)}$, where $c$ is a constant and $S$ a smooth function of the variable $b_T$. As the negative exponential function is rapidly decreasing, one only needs to look at the contribution from where the exponent is at its minimum. Since the TMD PDF in $b_T$-space follows such a form, see Eq.~\eqref{e:F_pert}, it is natural to apply the saddle-point approximation to analyze the TMD PDF~\cite{Grewal:2020hoc}. We mainly concentrate on the case where $k_T=0$. In such a case, $J_0(k_Tb_T)=1$ and no oscillations are present.  When $k_T>0$, the Bessel function $J_0(k_Tb_T)$ further suppresses the large-$b_T$ region of the $b_T$ integration, and the validity of the result presented in the following is even stronger. At $k_T=0$, one has:
\begin{equation}
F_a(x, k_T^2=0; Q, Q^2) = \frac{1}{2\pi} \int_0^{\infty} db_T \, b_T  F_a(x, b_T^2; Q, Q^2) 
=\frac{1}{4\pi} \int_{-\infty}^\infty d\left(\ln b_T^2\right)  \exp\left[ \ln \left(b_T^2\, F_a(x, b_T^2; Q, Q^2) \right)\right]\,,
\label{eq:integral}
\end{equation}
and thus the integral is dominated by a saddle point at $b_T^{sp}$, which is determined by~\cite{Collins:1984kg,Grewal:2020hoc}:
\begin{equation}
\frac{d}{db_T} \bigg\{ \ln \bigg[ b_T^2\, F_a(x,b_T^2; Q, Q^2) \bigg] \bigg\}_{\begin{subarray}{l} b_T = b_T^{sp} \\  \end{subarray}} = 0\,.
\label{eq:saddle-point}
\end{equation}
In the following, we study in details the kinematic dependence of the saddle point $b_T^{sp}$, in particular the most relevant $x$ and $Q$ dependencies: 
\begin{equation}
b_T^{sp} \equiv b_T^{sp}(x, Q) \, .
\end{equation}
The approximation relates the integral over $b_T$ in Eq.~\eqref{eq:integral} to the evaluation of the integrand at the saddle point $b_T^{sp}$. 
When the saddle point is small, $b_T^{sp} \ll  1/\Lambda_{\rm QCD}$, i.e., well in the perturbative region, then one would expect the TMD PDF $F_a(x, k_T^2; Q, Q^2)$ to be mainly controlled by the perturbative physics (always modulo the collinear PDFs). On the contrary, if $b_T^{sp}$ is large, i.e. $b_T^{sp} \gtrsim 1/\Lambda_{\rm QCD}$, the large-$b_T$ non-perturbative contributions are very important and one has to understand/constrain them well, in order to have a full understanding of the TMD PDF. In other words, we use the information on the saddle point $b_T^{sp}$ as an indication of the predictive power of the TMD formalism.

\subsection{Leading-log solution}
\label{ss:LL_saddle_point}

We use the perturbative contribution to $F_a(x, b_T^2; Q, Q^2)$ to compute the saddle-point $b_T^{sp}$. Plugging the perturbative expression in Eq.~\eqref{e:F_pert} into Eq.~\eqref{eq:saddle-point}, one obtains:
\begin{align}
\frac{d}{db_T} \Bigg\{ & \bigg[\int_{\mu_b}^{Q}\frac{d\mu'}{\mu'}   \left( \ln\left(\frac{Q^2}{\mu'^2}\right) \gamma_K\left(\alpha_s(\mu')\right) - \gamma_F \left( \alpha_s(\mu'), 1 \right) \right)  + D\left(c, \alpha_s(\mu_b)\right) \ln\left( \frac{Q^2}{\mu_b^2}\right) 
\nonumber\\
&- \ln \left(b_T^2 \right)
- \ln \bigg[  \sum_b C_{a/b}(x,b_T^2,\mu_b,\mu_b^2) \otimes f_b(x,\mu_b) \bigg] \Bigg\}\Bigg|_{b_T = b_T^{sp}} = 0\,.
\label{e:sp_equation}
\end{align}
In general, one can evaluate the saddle point $b_T^{sp}$ of the TMD PDF by solving numerically the above equation. 
However, at the leading logarithmic (LL) accuracy where one keeps the leading order (LO) result in $\Gamma_{\rm cusp}$ and in the coefficient functions $C_{a/b}$, one can solve analytically the above equation and obtain the following simple results:
\begin{equation}
\label{e:sp_equation_LL}
\frac{1}{2} \ln \left(\frac{Q^2}{\mu_b^{\star\,2}}\right)\ \Gamma_0 \ \frac{\alpha_s(\mu_b^\star)}{4\pi} = 1 - {\cal X}(x,\mu_b^\star) \, , \ \ \ \ \ \ \ \ \ {\cal X}(x,\mu) = \frac{d}{d \ln \mu^2} \ln f_a(x,\mu) \, ,
\end{equation}
where we have introduced $\mu_b^\star = c/b_T^{sp}$.  
The function ${\cal X}(x,\mu)$ quantifies the impact of the DGLAP evolution on the position of the saddle point. Its sign changes according to the value of the light-cone fraction $x$ and determines the $x$-dependence of the saddle point.

The saddle point for the resummed contribution to the Drell-Yan cross section differential with respect to the transverse momentum of the lepton pair has been discussed in Refs.~\cite{Collins:1984kg,Parisi:1979se}. In that treatment the effect of the $x$-dependence was neglected.  
In our treatment, neglecting the $x$-dependence corresponds to setting ${\cal X}=0$. Accordingly, the solution of Eq.~\eqref{e:sp_equation_LL} reads:
\begin{equation}
\label{e:bT_sp_appr_as_0}
b_T^{sp\, (0)} = \frac{c}{\Lambda_{\rm QCD}} \left(  \frac{Q}{\Lambda_{\rm QCD}}  \right)^{-  \Gamma_0 / \left(\Gamma_0 + 8\pi \beta_0 \right)  } \, , 
\ \ \ \ \ \ \ \ \beta_0 = \frac{11 C_A - 4 T_f\, n_f }{12 \pi} \, ,
\end{equation}
where $\beta_0$ is the one-loop coefficient of the QCD beta function~\cite{ParticleDataGroup:2016lqr}, and $n_f$ is the number of active flavors. The expression for $b_T^{sp\, (0)}$ is analogous to the one presented in Refs.~\cite{Collins:1984kg,Parisi:1979se}. It follows the usual wisdom that the larger the value of $Q$ is, the smaller $b_T^{sp\, (0)}$ is, and thus the perturbative contributions to the observable play a more important role. 
%
%
By including the contribution of ${\cal X}$, the solution to Eq.~\eqref{e:sp_equation_LL} acquires an $x$-dependence: 
\begin{equation}
\label{e:bT_sp_appr_as}
b_T^{sp} = \frac{c}{\Lambda_{\rm QCD}} \left(  \frac{Q}{\Lambda_{\rm QCD}}  \right)^{-  \Gamma_0 / \left[\Gamma_0 + 8\pi \beta_0 \left(1 - {\cal X}(x,\mu_b^\star) \right)  \right]  } \ . 
\end{equation}
Note that the right-hand side of Eq.~\eqref{e:bT_sp_appr_as} depends on $b_T^{sp}$ through $\mu_b^\star$, and thus Eq.~\eqref{e:sp_equation_LL} needs to be solved by iterations. A legitimate choice for the first iteration is to evaluate ${\cal X}$ at $b_T^{sp\, (0)}$. Comparing Eq.~\eqref{e:bT_sp_appr_as} with \eqref{e:bT_sp_appr_as_0}, one observes that if ${\cal X} > 0$ ($<0$), one would have $b_T^{sp} < b_T^{sp\,(0)}$ ($b_T^{sp} > b_T^{sp\,(0)}$). 
%
%
Apart from the gluon case at $\mu\lesssim 1$~GeV, the function ${\cal X}$ is positive for $x \lesssim 0.1$ and negative for $x \gtrsim 0.1$. 
Thus its effect is to reduce the value of the saddle point $b_T^{sp}$ with respect to the solution $b_T^{sp\, (0)}$ for $x \lesssim 0.1$ and to increase it for $x \gtrsim 0.1$. Because of this, for the same $Q$ value but at smaller $x$, the perturbative contribution (from small-$b_T$ region) plays a more important role for the TMD PDF. This means that in general, away from the limiting cases, the TMD PDF is more perturbatively dominated at large $Q$ and small $x$. On the other hand, the TMD PDF is more dominated by the non-perturbative contribution at small $Q$ and large $x$. This suggests that even for a moderately large $Q$, the TMD PDF at large $x$ could become quite sensitive to the non-perturbative contributions, due to the $x$-dependence of the ${\cal X}$ function. 
Following this argument, the sweet spot for investigating non-perturbative TMD effects would be a kinematics in which $x$ and $Q$ are sufficiently large. In this way, the $Q$-dependent power corrections to the factorization theorems for physical observables are kept under control, while the large $x$ region retains good sensitivity to the non-perturbative effects.

\section{Exercises}
\label{s:ex_evo}

\begin{itemize}

\item[1)] Derive the functional dependence of the saddle point for a TMD distribution as a function of $x$, $Q$.

\item[2)] Repeat the same calculation for a TMD fragmentation function.

\item[3)] Find the matching coefficients for the helicity and transversity TMD PDFs, $g_1(x,k_T^2)$ and $h_1(x,k_T^2)$, in Ref.~\cite{Bacchetta:2013pqa} and show that the perturbative calculations are not reliable at large $b_T$ (as for the unpolarized distribution).


\end{itemize}

\clearpage
\part{Summary and outlook}
\label{p:summary}

These notes have presented an introductory path to transverse-momentum imaging of hadrons, with the aim of providing graduate students and newcomers with a conceptual map of the field rather than a comprehensive review. 
Starting from inclusive deep-inelastic scattering, we introduced the description of hadron structure in terms of structure functions and collinear parton distribution functions, and then generalized this framework to include the transverse motion of partons through transverse-momentum-dependent (TMD) distributions. 
Semi-inclusive deep-inelastic scattering was presented as the natural process to access these quantities experimentally, while the role of discrete and gauge symmetries was discussed as part of the theoretical foundation of the rich spin-momentum correlations encoded in TMDs. 
Finally, we briefly introduced the concepts of TMD evolution and the interplay between perturbative and non-perturbative dynamics that underlies modern phenomenological analyses. \\

Many important topics have necessarily been omitted or only briefly mentioned. Among them are generalized parton distributions and the relation between momentum- and coordinate-space imaging~\cite{Diehl:2003ny,Diehl:2015uka}, generalized TMDs and Wigner distributions~\cite{Lorce:2025aqp}, small-$x$ dynamics, gluon tomography, lattice-QCD calculations, higher-order perturbative calculations, and the growing role of effective field theories in the derivation of factorization theorems~\cite{Boussarie:2023izj}. Likewise, the discussion of phenomenology has focused on the solution of the evolution equations and the associated non-perturbative effects rather than on the existing broad range of current global analyses and experimental measurements~\cite{Bacchetta:2022awv,Bacchetta:2024qre,Bacchetta:2025ara,Moos:2023yfa,Moos:2025sal}. \\

The coming years promise to be particularly exciting for hadron tomography. Precision measurements at the upgraded Jefferson Lab~\cite{Accardi:2023chb}, AMBER~\cite{Adams:2676885}, LHCSpin~\cite{LHCspin:2025lvj}, and especially at the future Electron-Ion Colliders~\cite{AbdulKhalek:2021gbh,Anderle:2021wcy} will considerably expand the available experimental information, probing both quark and gluon transverse structure over an unprecedented kinematic range. At the same time, continued developments in perturbative QCD, lattice calculations, global analyses, and effective field theory are steadily improving the theoretical description of TMD observables. \\

Ultimately, the study of transverse-momentum-dependent distributions is part of the broader effort to understand how the complex structure of hadrons emerges from Quantum Chromodynamics. Combining precise experimental measurements with increasingly sophisticated theoretical tools offers the prospect of constructing a quantitative three-dimensional picture of hadrons and hadronization in momentum space, providing new insights into the dynamics of confinement and the origin of the observable properties of strongly interacting matter.

\clearpage

\begin{appendices}


\addcontentsline{toc}{part}{Appendices}

\addtocontents{toc}{\string\tocdepth@munge}

\section{Light-cone coordinates}
\label{a:lc_coordinates}
An introduction to light-cone coordinates is given in Ref.~\cite{Collins:1997nm}.
Given a Cartesian basis $\{\hat{e}_0,\hat{e}_1,\hat{e}_2,\hat{e}_3\}$ we can introduce a light-cone basis as:
\begin{equation}
\label{e:lc_basis}
n_+ \equiv \frac{1}{\sqrt{2}} (\hat{e}_0 + \hat{e}_3),\ \ \ n_- \equiv \frac{1}{\sqrt{2}} (\hat{e}_0 - \hat{e}_3),\ \ \ \hat{e}_T \equiv (\hat{e}_1,\hat{e}_2) 
\end{equation}
such that
\begin{equation}
\label{e:basis_products}
n_+ \cdot n_- = 1, \ \ \ n_+^2=n_-^2=0 \ .
\end{equation}
Light-cone coordinates for a generic vector $V$ are:
\begin{align}
\label{e:lc_coord}
& V^+ = \frac{1}{\sqrt{2}} (V^0+V^3) \equiv V \cdot n_- \\
& V^- = \frac{1}{\sqrt{2}} (V^0-V^3) \equiv V \cdot n_+ \\
& V_T = (V^1,V^2) \ ,
\end{align}
where $\{ 0,1,2,3 \}$ are the standard Cartesian coordinates. 
In particular
\begin{equation}
\label{e:mink_vs_eucl_vec}
V_T^2 = - {\bf V}_T^2 \ .
\end{equation}
The symmetric projector onto the transverse plane is defined as:
\begin{equation}
\label{e:gT_npnm}
g_T^{\mu\nu} \equiv g^{\mu\nu} - n_+^{\{\mu} n_-^{\nu\}} \ .
\end{equation}
In components $g_T^{11}=g_T^{22}=-1$.
Accordingly, the scalar product between two 4-vectors reads:
\begin{equation}
a \cdot b
=
a^+ b^- + a^- b^+ - {\bf a}_T \cdot {\bf b}_T \, .
\end{equation}
The antisymmetric projector onto the transverse plane is:
\begin{equation}
\label{e:epsT}
\epsilon_T^{\mu\nu} \equiv \epsilon^{\rho\sigma\mu\nu} n_{+\rho} n_{-\sigma} = \epsilon^{-+\mu\nu} \equiv \epsilon^{n_+n_-\mu\nu}  \ ,
\end{equation}
We choose $\epsilon^{0123}=1$ so that $\epsilon_T^{12}=-\epsilon_T^{21}=1$.
We also use the following notation:
\begin{equation}
\label{e:vector_contr}
a^{b\nu} \equiv a^{\mu\nu}b_\mu \ .
\end{equation}
An example of convention for light-cone coordinates in SCET is given in Ref.~\cite{Becher:2014oda}.\\

\section{Spin}
\label{a:spin}

The description of the spin is inherited from the formalism of quantum mechanics. A particle of spin $J$ is described as a combination of tensors up to rank $2J$. 
For example, the spin state of a spin 1/2 particle is described by a rank 1 tensor (a spin vector $S^\mu$) with the formalism of the spin density matrix:
\begin{equation}
\label{e:rho_spin12}
\rho = \frac{1}{2} \left( {\mathbbm 1} + \vect{S} \cdot {\bf \sigma} \right) \, .
\end{equation}
The covariant spin 4-vector is defined as:
\begin{equation}
\label{e:Smu}
S^\mu = \big( 0, \vect{S}^i \big) = (0, \vect{S}_T^x, \vect{S}_T^y, S_L) \, ,
\end{equation}
where $\vect{S}_T$ is the 2-dimensional transverse spin vector, and $S_L$ is the longitudinal spin component (we implicitly choose $\hat{z}$ as the longitudinal direction).
Accordingly, one has
\begin{equation}
S^2 = -1,
\qquad
P \cdot S = 0 \, ,
\end{equation}
and that the covariant spin $S^\mu$ can be written in terms of the hadronic momentum $P^\mu$~\cite{Mulders:2016pln,Mulders:2015ggi}, which identifies the leading light cone direction (see App.~\ref{a:lc_coordinates}):
\begin{equation}
\label{e:Smu_lc_dec}
S^\mu \approx S_L \frac{P^\mu}{M} + S_T^\mu \, .
\end{equation}

For a spin-1 target the density matrix involves a spin vector and a spin tensor:
\begin{equation}
\label{e:rho_spin12}
\rho = \frac{1}{3} \left( {\mathbbm 1} + 
						  \frac{3}{2}\, \vect{S}^i\, \vect{\Sigma}^i +
						  3\, \vect{T}^{ij}\, \vect{\Sigma}^{ij}
				   \right) \, ,
\end{equation}
where $\vect{\Sigma}^i$ and $\vect{\Sigma}^{ij}$ are the 3D Pauli matrices and their rank-2 generalization,
$S^\mu$ is the spin vector and 
$T^{\mu\nu}$ the spin tensor. For further details related to the physics of the spin 1 targets and the description of its components, we refer to Refs.~\cite{Bacchetta:2000jk,Bacchetta:2002xd,Boer:2016xqr,Cosyn:2024drt,Kumano:2024fpr,Zhao:2025vol}.
A spin-1 target (e.g. the deuteron) has more degrees of freedom compared to a spin-1/2 target (e.g. nucleon). There exists unique spin-momentum correlations~\cite{Bacchetta:2000jk,Ninomiya:2017ggn,Kumano:2024fpr} in spin-1 states that do not appear in the nucleon and can be experimentally studied in (Semi-Inclusive) DIS in fixed-target configuration, for example at Jefferson Lab.

\section{Dirac matrices}
\label{aa:dirac_m}

Dirac matrices~\cite{Peskin:1995ev} satisfy the anticommutation relation:
\begin{equation}
\label{e:dirac_anticomm}
\{\gamma^\mu,\gamma^\nu\} = 2g^{\mu\nu} \mathbb{I} \ ,
\end{equation}
where $\mathbb{I}$ is the identity in Dirac space. A fifth matrix is defined as:
\begin{equation}
\label{e:gamma5}
\gamma_5 \equiv i \gamma^0 \gamma^1 \gamma^2 \gamma^3 \ .
\end{equation}
Furthermore, an antisymmetric structure can be defined in the following way:
\begin{equation}
\label{e:sigma_dirac}
\sigma^{\mu\nu} \equiv \frac{i}{2}\ [\gamma^\mu, \gamma^\nu] \ . 
\end{equation}

\section{Anticommutation relations}
\label{ss:anticomm}

The anticommutation relation for the fermion fields expressed in light-cone coordinates reads~\cite{Mannheim:2020rod}:
\begin{equation}
\label{e:anticomm_psi_psibar}
\big\{ \psi_\alpha(x), \overline{\psi}_\beta(y) \big\} = \big[i \slash{\partial} + m \big]_{\alpha\beta}\,\, i\, \Delta(x-y) \, 
\end{equation}
where $i\Delta(x-y)$ is the unequal-light-front time commutator of two scalar fields which satisfy the Klein-Gordon equation. 

For space-like separations $(x-y)^2<0$, causality requires that $i\Delta(x-y) = 0$. Thus the fermion fields in Eq.~\eqref{e:anticomm_psi_psibar} anticommute. 
For timelike distances $(x-y)^2=0$, instead, the scalar commutator takes the (singular) form~\cite{Mannheim:2020rod}:
\begin{equation}
\label{e:Delta_light_like}
i\Delta(x-y)_{\vert_{(x-y)^2=0}} = -\frac{i}{2\pi}\, \epsilon(x^+ - y^+)\, \delta\big( (x-y)^2 \big)\, ,  
\end{equation}
where $\epsilon(z)=\theta(z) - \theta(-z)$ and $\theta$ is the step function with $\theta(0^+)=\theta(0^-)=1/2$~\cite{Mannheim:2019lss}. 

For a collinear correlator, such as $\Phi(x)$, the momentum integrations force $\xi^+=\vect{\xi}_T=0$. Thus the non-locality is light-like, $\xi^2=0$. 
In this case, Eq.~\eqref{e:Delta_light_like} means that strictly speaking the Dirac fields do not anticommute. However, the anticommutator produces an operator that gives rise to a disconnected diagram, which can be neglected in the calculation of PDFs (see Ref.~\cite{Collins:2011zzd}, Sec. 6.9.3) and effectively one can anticommute the two fermion fields in the definition of $\Phi(x)$.
For a TMD correlator, instead, one has only $\xi^+ = 0$, which implies that $\xi^2 = -|\vect{\xi}_T|^2$. If $\vect{\xi}_T \neq 0$ causality implies that the Dirac fields in the definition of the correlator anticommute. If $\vect{\xi}_T=0$, instead, one recovers the case of the collinear correlator.

\addtocontents{toc}{\string\tocdepth@restore}

\end{appendices}

\clearpage
\begin{acknowledgments}
Several discussions with the schools' participants are gratefully acknowledged. If you find typos or mistakes please let me know.  
I want to thank in particular ICTS and Nanjing University for their warm hospitality.  
These notes are the results of efforts spread across several years, for which support has been received by multiple institutions and funding agencies. Thus, I acknowledge the support received from 
the U.S. Department of Energy, Office of Science, Office of Nuclear Physics, contract no. 
DE-AC05-06OR23177 and 
contract no. DE-AC02-06CH11357, 
the European Commission through the Marie Sklodowska-Curie Action SQuHadron (grant agreement ID: 795475), and 
the European Union ``Next Generation EU'' program through the Italian PRIN 2022 grant n. 20225ZHA7W.
This research was also supported in part by the International Centre for Theoretical Sciences (ICTS) for participating in the program - International School and Workshop on Probing Hadron Structure at the Electron-Ion Collider (code:ICTS/QEICIII2024/01). 
Last but not least, I want to thank my family and my colleagues for their patience during the (lengthy) preparation of these notes. 
\end{acknowledgments}

\bibliographystyle{JHEP}

\bibliography{lecture_notes}

\providecommand{\href}[2]{#2}\begingroup\raggedright\begin{thebibliography}{100}

\bibitem{Diehl:2015uka}
M.~Diehl, \emph{{Introduction to GPDs and TMDs}},
  \href{http://dx.doi.org/10.1140/epja/i2016-16149-3}{\emph{Eur. Phys. J. A}
  {\bf 52} (2016) 149}, [\href{http://arxiv.org/abs/1512.01328}{{\tt
  1512.01328}}].

\bibitem{Diehl:2003ny}
M.~Diehl, \emph{{Generalized parton distributions}},
  \href{http://dx.doi.org/10.1016/j.physrep.2003.08.002}{\emph{Phys. Rept.}
  {\bf 388} (2003) 41--277}, [\href{http://arxiv.org/abs/hep-ph/0307382}{{\tt
  hep-ph/0307382}}].

\bibitem{particles9030069}
C.~Mezrag, \emph{A step-by-step introduction to generalised parton
  distributions},
  \href{http://dx.doi.org/10.3390/particles9030069}{\emph{Particles} {\bf 9}
  (2026) }.

\bibitem{Lorce:2025aqp}
C.~Lorc{\'e}, A.~Metz, B.~Pasquini and P.~Schweitzer, \emph{{Parton
  Distribution Functions and their Generalizations}},  7, 2025.
\newblock \href{http://arxiv.org/abs/2507.12664}{{\tt 2507.12664}}.

\bibitem{Collins:2011zzd}
J.~Collins, \emph{{Foundations of perturbative QCD}}.
\newblock Cambridge University Press, 2013.

\bibitem{Boussarie:2023izj}
R.~Boussarie et~al., \emph{{TMD Handbook}},
  \href{http://arxiv.org/abs/2304.03302}{{\tt 2304.03302}}.

\bibitem{Jaffe:1996zw}
R.~L. Jaffe, \emph{{Spin, twist and hadron structure in deep inelastic
  processes}},  in \emph{{The spin structure of the nucleon. Proceedings,
  International School of Nucleon Structure, 1st Course, Erice, Italy, August
  3-10, 1995}}, pp.~42--129, 1996.
\newblock \href{http://arxiv.org/abs/hep-ph/9602236}{{\tt hep-ph/9602236}}.

\bibitem{Mulders:2015ggi}
P.~J. Mulders, \emph{{Transverse-momentum distributions and beyond: setting up
  the nucleon tomography}}.
\newblock Unpublished lecture notes -
  \href{https://www.nat.vu.nl/~mulders/tmdreview-vs3.pdf}{hyperlink to file},
  2015.

\bibitem{Roberts:2015lja}
C.~D. Roberts, \emph{{Three Lectures on Hadron Physics}},
  \href{http://dx.doi.org/10.1088/1742-6596/706/2/022003}{\emph{J. Phys. Conf.
  Ser.} {\bf 706} (2016) 022003}, [\href{http://arxiv.org/abs/1509.02925}{{\tt
  1509.02925}}].

\bibitem{Bacchetta:2012aln}
A.~Bacchetta, \emph{{Transverse momentum distributions}}.
\newblock Unpublished lecture notes -
  \href{https://drive.google.com/file/d/1wU7zH9FXWaYL0OA4XQbiVKvwLOy5LEjk/view?usp=sharing}{hyperlink
  to file}, 2012.

\bibitem{Muta:2010xua}
T.~Muta, \emph{{Foundations of Quantum Chromodynamics: An Introduction to
  Perturbative Methods in Gauge Theories, (3rd ed.)}}, vol.~78 of \emph{World
  scientific Lecture Notes in Physics}.
\newblock World Scientific, Hackensack, N.J., 2010.

\bibitem{Peskin:1995ev}
M.~E. Peskin and D.~V. Schroeder, \emph{{An Introduction to quantum field
  theory}}.
\newblock 1995.

\bibitem{Collins:1981uw}
J.~C. Collins and D.~E. Soper, \emph{{Parton Distribution and Decay
  Functions}},
  \href{http://dx.doi.org/10.1016/0550-3213(82)90021-9}{\emph{Nucl. Phys.} {\bf
  B194} (1982) 445--492}.

\bibitem{Brock:1993sz}
{\scshape CTEQ} collaboration, R.~Brock et~al., \emph{{Handbook of perturbative
  QCD: Version 1.0}},
  \href{http://dx.doi.org/10.1103/RevModPhys.67.157}{\emph{Rev. Mod. Phys.}
  {\bf 67} (1995) 157--248}.

\bibitem{Bacchetta:2004jz}
A.~Bacchetta, U.~D'Alesio, M.~Diehl and C.~A. Miller, \emph{{Single-spin
  asymmetries: The Trento conventions}},
  \href{http://dx.doi.org/10.1103/PhysRevD.70.117504}{\emph{Phys. Rev. D} {\bf
  70} (2004) 117504}, [\href{http://arxiv.org/abs/hep-ph/0410050}{{\tt
  hep-ph/0410050}}].

\bibitem{Collins:1997nm}
J.~C. Collins, \emph{{Light cone variables, rapidity and all that}},
  \href{http://arxiv.org/abs/hep-ph/9705393}{{\tt hep-ph/9705393}}.

\bibitem{Metz:2016swz}
A.~Metz and A.~Vossen, \emph{{Parton Fragmentation Functions}},
  \href{http://dx.doi.org/10.1016/j.ppnp.2016.08.003}{\emph{Prog. Part. Nucl.
  Phys.} {\bf 91} (2016) 136--202},
  [\href{http://arxiv.org/abs/1607.02521}{{\tt 1607.02521}}].

\bibitem{Scimemi:2019mlf}
I.~Scimemi, \emph{{A short review on recent developments in TMD factorization
  and implementation}},
  \href{http://dx.doi.org/10.1155/2019/3142510}{\emph{Adv. High Energy Phys.}
  {\bf 2019} (2019) 3142510}, [\href{http://arxiv.org/abs/1901.08398}{{\tt
  1901.08398}}].

\bibitem{Roberts:2012sv}
C.~D. Roberts, \emph{{Strong QCD and Dyson-Schwinger Equations}}, {\emph{IRMA
  Lect. Math. Theor. Phys.} {\bf 21} (2015) 355--458},
  [\href{http://arxiv.org/abs/1203.5341}{{\tt 1203.5341}}].

\bibitem{Horn:2016rip}
T.~Horn and C.~D. Roberts, \emph{{The pion: an enigma within the Standard
  Model}}, \href{http://dx.doi.org/10.1088/0954-3899/43/7/073001}{\emph{J.
  Phys. G} {\bf 43} (2016) 073001},
  [\href{http://arxiv.org/abs/1602.04016}{{\tt 1602.04016}}].

\bibitem{Ji:2020ena}
X.~Ji, F.~Yuan and Y.~Zhao, \emph{{What we know and what we
  don{\textquoteright}t know about the proton spin after 30 years}},
  \href{http://dx.doi.org/10.1038/s42254-020-00248-4}{\emph{Nature Rev. Phys.}
  {\bf 3} (2021) 27--38}, [\href{http://arxiv.org/abs/2009.01291}{{\tt
  2009.01291}}].

\bibitem{Dudek:2012vr}
J.~Dudek et~al., \emph{{Physics Opportunities with the 12 GeV Upgrade at
  Jefferson Lab}},
  \href{http://dx.doi.org/10.1140/epja/i2012-12187-1}{\emph{Eur. Phys. J.} {\bf
  A48} (2012) 187}, [\href{http://arxiv.org/abs/1208.1244}{{\tt 1208.1244}}].

\bibitem{Accardi:2012qut}
A.~Accardi et~al., \emph{{Electron Ion Collider: The Next QCD Frontier -
  Understanding the glue that binds us all}},
  \href{http://arxiv.org/abs/1212.1701}{{\tt 1212.1701}}.

\bibitem{AbdulKhalek:2021gbh}
R.~Abdul~Khalek et~al., \emph{{Science Requirements and Detector Concepts for
  the Electron-Ion Collider}: {EIC Yellow Report}},
  \href{http://dx.doi.org/10.1016/j.nuclphysa.2022.122447}{\emph{Nucl. Phys. A}
  {\bf 1026} (2022) 122447}, [\href{http://arxiv.org/abs/2103.05419}{{\tt
  2103.05419}}].

\bibitem{Anderle:2021wcy}
D.~P. Anderle et~al., \emph{{Electron-ion collider in China}},
  \href{http://dx.doi.org/10.1007/s11467-021-1062-0}{\emph{Front. Phys.
  (Beijing)} {\bf 16} (2021) 64701},
  [\href{http://arxiv.org/abs/2102.09222}{{\tt 2102.09222}}].

\bibitem{LHCspin:2025lvj}
{\scshape LHCspin} collaboration, A.~Accardi et~al., \emph{{LHCspin: a
  Polarized Gas Target for LHC}},  \href{http://arxiv.org/abs/2504.16034}{{\tt
  2504.16034}}.

\bibitem{Callan:1969uq}
C.~G. Callan, Jr. and D.~J. Gross, \emph{{High-energy electroproduction and the
  constitution of the electric current}},
  \href{http://dx.doi.org/10.1103/PhysRevLett.22.156}{\emph{Phys. Rev. Lett.}
  {\bf 22} (1969) 156--159}.

\bibitem{Mulders:2016pln}
P.~J. Mulders, \emph{{Transverse momentum dependence in high-energy scattering
  processes}}.
\newblock Unpublished lecture notes -
  \href{http://www.nat.vu.nl/~mulders/correlations-0new.pdf}{hyperlink to
  file}, 2016.

\bibitem{Bacchetta:2002xd}
A.~Bacchetta, \emph{{Probing the Transverse Spin of Quarks in Deep Inelastic
  Scattering}}.
\newblock PhD thesis, Vrije U., Amsterdam, 2002.
\newblock \href{http://arxiv.org/abs/hep-ph/0212025}{{\tt hep-ph/0212025}}.

\bibitem{Cutkosky:1960sp}
R.~E. Cutkosky, \emph{{Singularities and discontinuities of Feynman
  amplitudes}}, \href{http://dx.doi.org/10.1063/1.1703676}{\emph{J. Math.
  Phys.} {\bf 1} (1960) 429--433}.

\bibitem{Zwicky:2016lka}
R.~Zwicky, \emph{{A brief Introduction to Dispersion Relations and
  Analyticity}},  in \emph{{Quantum Field Theory at the Limits}: {from Strong
  Fields to Heavy Quarks}}, pp.~93--120, 2017.
\newblock \href{http://arxiv.org/abs/1610.06090}{{\tt 1610.06090}}.
\newblock \href{http://dx.doi.org/10.3204/DESY-PROC-2016-04/Zwicky}{DOI}.

\bibitem{Bacchetta:2000jk}
A.~Bacchetta and P.~J. Mulders, \emph{{Deep inelastic leptoproduction of
  spin-one hadrons}},
  \href{http://dx.doi.org/10.1103/PhysRevD.62.114004}{\emph{Phys. Rev. D} {\bf
  62} (2000) 114004}, [\href{http://arxiv.org/abs/hep-ph/0007120}{{\tt
  hep-ph/0007120}}].

\bibitem{Boer:2016xqr}
D.~Boer, S.~Cotogno, T.~van Daal, P.~J. Mulders, A.~Signori and Y.-J. Zhou,
  \emph{{Gluon and Wilson loop TMDs for hadrons of spin $\leq$ 1}},
  \href{http://arxiv.org/abs/1607.01654}{{\tt 1607.01654}}.

\bibitem{Kumano:2024fpr}
S.~Kumano, \emph{{Parton distribution functions and fragmentation functions of
  spin-1 hadrons}},
  \href{http://dx.doi.org/10.1140/epja/s10050-024-01411-6}{\emph{Eur. Phys. J.
  A} {\bf 60} (2024) 205}, [\href{http://arxiv.org/abs/2406.01180}{{\tt
  2406.01180}}].

\bibitem{Zhao:2025vol}
J.~Zhao, A.~Bacchetta, S.~Kumano, T.~Liu and Y.-j. Zhou, \emph{{Semi-inclusive
  deep inelastic scattering off a tensor-polarized spin-1 target}},
  \href{http://dx.doi.org/10.1007/JHEP12(2025)067}{\emph{JHEP} {\bf 12} (2025)
  067}, [\href{http://arxiv.org/abs/2508.06134}{{\tt 2508.06134}}].

\bibitem{Bacchetta:2006tn}
A.~Bacchetta, M.~Diehl, K.~Goeke, A.~Metz, P.~J. Mulders and M.~Schlegel,
  \emph{{Semi-inclusive deep inelastic scattering at small transverse
  momentum}},
  \href{http://dx.doi.org/10.1088/1126-6708/2007/02/093}{\emph{JHEP} {\bf 02}
  (2007) 093}, [\href{http://arxiv.org/abs/hep-ph/0611265}{{\tt
  hep-ph/0611265}}].

\bibitem{Mulders:1995dh}
P.~J. Mulders and R.~D. Tangerman, \emph{{The Complete tree level result up to
  order 1/Q for polarized deep inelastic leptoproduction}},
  \href{http://dx.doi.org/10.1016/0550-3213(95)00632-X}{\emph{Nucl. Phys. B}
  {\bf 461} (1996) 197--237}, [\href{http://arxiv.org/abs/hep-ph/9510301}{{\tt
  hep-ph/9510301}}].

\bibitem{Diehl:2005pc}
M.~Diehl and S.~Sapeta, \emph{{On the analysis of lepton scattering on
  longitudinally or transversely polarized protons}},
  \href{http://dx.doi.org/10.1140/epjc/s2005-02242-9}{\emph{Eur. Phys. J. C}
  {\bf 41} (2005) 515--533}, [\href{http://arxiv.org/abs/hep-ph/0503023}{{\tt
  hep-ph/0503023}}].

\bibitem{Accardi:2017pmi}
A.~Accardi and A.~Bacchetta, \emph{{Accessing the nucleon transverse structure
  in inclusive deep inelastic scattering}},
  \href{http://dx.doi.org/10.1016/j.physletb.2017.08.074}{\emph{Phys. Lett. B}
  {\bf 773} (2017) 632--638}, [\href{http://arxiv.org/abs/1706.02000}{{\tt
  1706.02000}}].

\bibitem{Hoodbhoy:1988am}
P.~Hoodbhoy, R.~L. Jaffe and A.~Manohar, \emph{{Novel Effects in Deep Inelastic
  Scattering from Spin 1 Hadrons}},
  \href{http://dx.doi.org/10.1016/0550-3213(89)90572-5}{\emph{Nucl. Phys. B}
  {\bf 312} (1989) 571--588}.

\bibitem{Cosyn:2024drt}
W.~Cosyn, B.~R. Tomei, A.~Sosa and A.~Zec, \emph{{Polarization options in
  inclusive DIS off tensor polarized deuteron}},
  \href{http://dx.doi.org/10.1140/epja/s10050-025-01511-x}{\emph{Eur. Phys. J.
  A} {\bf 61} (2025) 83}, [\href{http://arxiv.org/abs/2410.12764}{{\tt
  2410.12764}}].

\bibitem{Wilson:1969zs}
K.~G. Wilson, \emph{{Nonlagrangian models of current algebra}},
  \href{http://dx.doi.org/10.1103/PhysRev.179.1499}{\emph{Phys. Rev.} {\bf 179}
  (1969) 1499--1512}.

\bibitem{Wilson:1972ee}
K.~G. Wilson and W.~Zimmermann, \emph{{Operator product expansions and
  composite field operators in the general framework of quantum field theory}},
  \href{http://dx.doi.org/10.1007/BF01878448}{\emph{Commun. Math. Phys.} {\bf
  24} (1972) 87--106}.

\bibitem{Weinberg:1996kr}
S.~Weinberg, \emph{{The quantum theory of fields. Vol. 2: Modern
  applications}}.
\newblock Cambridge University Press, 8, 2013,
  \href{http://dx.doi.org/10.1017/CBO9781139644174}{10.1017/CBO9781139644174}.

\bibitem{Buffing:2015tja}
M.~G.~A. Buffing, \emph{{Color and TMD Universality in Hadronic Interactions}}.
\newblock PhD thesis, NIKHEF, Amsterdam, 2015-09-02.

\bibitem{Zimmermann:1972tv}
W.~Zimmermann, \emph{{Normal products and the short distance expansion in the
  perturbation theory of renormalizable interactions}},
  \href{http://dx.doi.org/10.1016/0003-4916(73)90430-2}{\emph{Annals Phys.}
  {\bf 77} (1973) 570--601}.

\bibitem{Diehl:1998sm}
M.~Diehl and T.~Gousset, \emph{{Time ordering in off diagonal parton
  distributions}},
  \href{http://dx.doi.org/10.1016/S0370-2693(98)00439-0}{\emph{Phys. Lett. B}
  {\bf 428} (1998) 359--370}, [\href{http://arxiv.org/abs/hep-ph/9801233}{{\tt
  hep-ph/9801233}}].

\bibitem{Gross:1971wn}
D.~J. Gross and S.~B. Treiman, \emph{{Light cone structure of current
  commutators in the gluon quark model}},
  \href{http://dx.doi.org/10.1103/PhysRevD.4.1059}{\emph{Phys. Rev. D} {\bf 4}
  (1971) 1059--1072}.

\bibitem{Accardi:2019luo}
A.~Accardi and A.~Signori, \emph{{Quark fragmentation as a probe of dynamical
  mass generation}},
  \href{http://dx.doi.org/10.1016/j.physletb.2019.134993}{\emph{Phys. Lett. B}
  {\bf 798} (2019) 134993}, [\href{http://arxiv.org/abs/1903.04458}{{\tt
  1903.04458}}].

\bibitem{Accardi:2020iqn}
A.~Accardi and A.~Signori, \emph{{On the connection between quark propagation
  and hadronization}},
  \href{http://dx.doi.org/10.1140/epjc/s10052-020-8380-1}{\emph{Eur. Phys. J.
  C} {\bf 80} (2020) 825}, [\href{http://arxiv.org/abs/2005.11310}{{\tt
  2005.11310}}].

\bibitem{Accardi:2009au}
A.~Accardi, A.~Bacchetta, W.~Melnitchouk and M.~Schlegel, \emph{{What can break
  the Wandzura-Wilczek relation?}},
  \href{http://dx.doi.org/10.1088/1126-6708/2009/11/093}{\emph{JHEP} {\bf 11}
  (2009) 093}, [\href{http://arxiv.org/abs/0907.2942}{{\tt 0907.2942}}].

\bibitem{Echevarria:2016scs}
M.~G. Echevarria, I.~Scimemi and A.~Vladimirov, \emph{{Unpolarized Transverse
  Momentum Dependent Parton Distribution and Fragmentation Functions at
  next-to-next-to-leading order}},
  \href{http://dx.doi.org/10.1007/JHEP09(2016)004}{\emph{JHEP} {\bf 09} (2016)
  004}, [\href{http://arxiv.org/abs/1604.07869}{{\tt 1604.07869}}].

\bibitem{Jaffe:1983hp}
R.~L. Jaffe, \emph{{Parton Distribution Functions for Twist Four}},
  \href{http://dx.doi.org/10.1016/0550-3213(83)90361-9}{\emph{Nucl. Phys. B}
  {\bf 229} (1983) 205--230}.

\bibitem{Boer:1998im}
D.~Boer, \emph{{Azimuthal asymmetries in hard scattering processes}}.
\newblock PhD thesis, Vrije U., Amsterdam, 1998.

\bibitem{Gamberg:2010uw}
L.~P. Gamberg, A.~Mukherjee and P.~J. Mulders, \emph{{A model independent
  analysis of gluonic pole matrix elements and universality of TMD
  fragmentation functions}},
  \href{http://dx.doi.org/10.1103/PhysRevD.83.071503}{\emph{Phys. Rev. D} {\bf
  83} (2011) 071503}, [\href{http://arxiv.org/abs/1010.4556}{{\tt 1010.4556}}].

\bibitem{Goeke:2005hb}
K.~Goeke, A.~Metz and M.~Schlegel, \emph{{Parameterization of the quark-quark
  correlator of a spin-1/2 hadron}},
  \href{http://dx.doi.org/10.1016/j.physletb.2005.05.037}{\emph{Phys. Lett. B}
  {\bf 618} (2005) 90--96}, [\href{http://arxiv.org/abs/hep-ph/0504130}{{\tt
  hep-ph/0504130}}].

\bibitem{Boer:1997nt}
D.~Boer and P.~J. Mulders, \emph{{Time reversal odd distribution functions in
  leptoproduction}},
  \href{http://dx.doi.org/10.1103/PhysRevD.57.5780}{\emph{Phys. Rev. D} {\bf
  57} (1998) 5780--5786}, [\href{http://arxiv.org/abs/hep-ph/9711485}{{\tt
  hep-ph/9711485}}].

\bibitem{Ralston:1979ys}
J.~P. Ralston and D.~E. Soper, \emph{{Production of Dimuons from High-Energy
  Polarized Proton Proton Collisions}},
  \href{http://dx.doi.org/10.1016/0550-3213(79)90082-8}{\emph{Nucl. Phys. B}
  {\bf 152} (1979) 109}.

\bibitem{Barone:2001sp}
V.~Barone, A.~Drago and P.~G. Ratcliffe, \emph{{Transverse polarisation of
  quarks in hadrons}},
  \href{http://dx.doi.org/10.1016/S0370-1573(01)00051-5}{\emph{Phys. Rept.}
  {\bf 359} (2002) 1--168}, [\href{http://arxiv.org/abs/hep-ph/0104283}{{\tt
  hep-ph/0104283}}].

\bibitem{Idilbi:2004vb}
A.~Idilbi, X.-d. Ji, J.-P. Ma and F.~Yuan, \emph{{Collins-Soper equation for
  the energy evolution of transverse-momentum and spin dependent parton
  distributions}},
  \href{http://dx.doi.org/10.1103/PhysRevD.70.074021}{\emph{Phys. Rev. D} {\bf
  70} (2004) 074021}, [\href{http://arxiv.org/abs/hep-ph/0406302}{{\tt
  hep-ph/0406302}}].

\bibitem{Ebert:2021jhy}
M.~A. Ebert, A.~Gao and I.~W. Stewart, \emph{{Factorization for azimuthal
  asymmetries in SIDIS at next-to-leading power}},
  \href{http://dx.doi.org/10.1007/JHEP06(2022)007}{\emph{JHEP} {\bf 06} (2022)
  007}, [\href{http://arxiv.org/abs/2112.07680}{{\tt 2112.07680}}].

\bibitem{Rodini:2023plb}
S.~Rodini and A.~Vladimirov, \emph{{Transverse momentum dependent factorization
  for SIDIS at next-to-leading power}},
  \href{http://dx.doi.org/10.1103/PhysRevD.110.034009}{\emph{Phys. Rev. D} {\bf
  110} (2024) 034009}, [\href{http://arxiv.org/abs/2306.09495}{{\tt
  2306.09495}}].

\bibitem{Anselmino:1995vq}
M.~Anselmino, M.~Boglione, J.~Hansson and F.~Murgia, \emph{{Polarized inclusive
  leptoproduction, lepton N --{\ensuremath{>}} h X, and the hadron helicity
  density matrix, rho(h): Possible measurements and predictions}},
  \href{http://dx.doi.org/10.1103/PhysRevD.54.828}{\emph{Phys. Rev. D} {\bf 54}
  (1996) 828--837}, [\href{http://arxiv.org/abs/hep-ph/9512379}{{\tt
  hep-ph/9512379}}].

\bibitem{Anselmino:1996qx}
M.~Anselmino, E.~Leader and F.~Murgia, \emph{{Single spin asymmetries in DIS}},
  \href{http://dx.doi.org/10.1103/PhysRevD.56.6021}{\emph{Phys. Rev. D} {\bf
  56} (1997) 6021--6024}, [\href{http://arxiv.org/abs/hep-ph/9610407}{{\tt
  hep-ph/9610407}}].

\bibitem{Anselmino:2005sh}
M.~Anselmino, M.~Boglione, U.~D'Alesio, E.~Leader, S.~Melis and F.~Murgia,
  \emph{{The general partonic structure for hadronic spin asymmetries}},
  \href{http://dx.doi.org/10.1103/PhysRevD.73.014020}{\emph{Phys. Rev. D} {\bf
  73} (2006) 014020}, [\href{http://arxiv.org/abs/hep-ph/0509035}{{\tt
  hep-ph/0509035}}].

\bibitem{vanDaal:2016glj}
T.~van Daal, \emph{{Quark and gluon TMD correlators in momentum and coordinate
  space}},  in \emph{{22nd International Symposium on Spin Physics}}, 12, 2016.
\newblock \href{http://arxiv.org/abs/1612.06585}{{\tt 1612.06585}}.

\bibitem{Boer:2008fr}
D.~Boer, \emph{{Angular dependences in inclusive two-hadron production at
  BELLE}}, \href{http://dx.doi.org/10.1016/j.nuclphysb.2008.06.011}{\emph{Nucl.
  Phys.} {\bf B806} (2009) 23--67}, [\href{http://arxiv.org/abs/0804.2408}{{\tt
  0804.2408}}].

\bibitem{Goeke:2003az}
K.~Goeke, A.~Metz, P.~V. Pobylitsa and M.~V. Polyakov, \emph{{Lorentz
  invariance relations among parton distributions revisited}},
  \href{http://dx.doi.org/10.1016/S0370-2693(03)00870-0}{\emph{Phys. Lett. B}
  {\bf 567} (2003) 27--30}, [\href{http://arxiv.org/abs/hep-ph/0302028}{{\tt
  hep-ph/0302028}}].

\bibitem{Meissner:2007rx}
S.~Meissner, A.~Metz and K.~Goeke, \emph{{Relations between generalized and
  transverse momentum dependent parton distributions}},
  \href{http://dx.doi.org/10.1103/PhysRevD.76.034002}{\emph{Phys. Rev. D} {\bf
  76} (2007) 034002}, [\href{http://arxiv.org/abs/hep-ph/0703176}{{\tt
  hep-ph/0703176}}].

\bibitem{Levelt:1993ac}
J.~Levelt and P.~J. Mulders, \emph{{Quark correlation functions in deep
  inelastic semiinclusive processes}},
  \href{http://dx.doi.org/10.1103/PhysRevD.49.96}{\emph{Phys. Rev. D} {\bf 49}
  (1994) 96--113}, [\href{http://arxiv.org/abs/hep-ph/9304232}{{\tt
  hep-ph/9304232}}].

\bibitem{Boer:2003cm}
D.~Boer, P.~J. Mulders and F.~Pijlman, \emph{{Universality of T odd effects in
  single spin and azimuthal asymmetries}},
  \href{http://dx.doi.org/10.1016/S0550-3213(03)00527-3}{\emph{Nucl. Phys. B}
  {\bf 667} (2003) 201--241}, [\href{http://arxiv.org/abs/hep-ph/0303034}{{\tt
  hep-ph/0303034}}].

\bibitem{Tangerman:1994bb}
R.~D. Tangerman and P.~J. Mulders, \emph{{Polarized twist - three distributions
  g(T) and h(L) and the role of intrinsic transverse momentum}},
  \href{http://arxiv.org/abs/hep-ph/9408305}{{\tt hep-ph/9408305}}.

\bibitem{Drell:1969jm}
S.~D. Drell, D.~J. Levy and T.-M. Yan, \emph{{A Theory of Deep Inelastic
  Lepton-Nucleon Scattering and Lepton Pair Annihilation Processes. 1.}},
  \href{http://dx.doi.org/10.1103/PhysRev.187.2159}{\emph{Phys. Rev.} {\bf 187}
  (1969) 2159--2171}.

\bibitem{Drell:1969wd}
S.~D. Drell, D.~J. Levy and T.-M. Yan, \emph{{A Theory of Deep Inelastic Lepton
  Nucleon Scattering and Lepton Pair Annihilation Processes. 2. Deep Inelastic
  electron Scattering}},
  \href{http://dx.doi.org/10.1103/PhysRevD.1.1035}{\emph{Phys. Rev. D} {\bf 1}
  (1970) 1035--1068}.

\bibitem{Drell:1969wb}
S.~D. Drell, D.~J. Levy and T.-M. Yan, \emph{{A Theory of Deep Inelastic
  Lepton-Nucleon Scattering and Lepton Pair Annihilation Processes. 3. Deep
  Inelastic electron-Positron Annihilation}},
  \href{http://dx.doi.org/10.1103/PhysRevD.1.1617}{\emph{Phys. Rev. D} {\bf 1}
  (1970) 1617--1639}.

\bibitem{Ito:2009zc}
T.~Ito, W.~Bentz, I.~C. Cloet, A.~W. Thomas and K.~Yazaki, \emph{{The NJL-jet
  model for quark fragmentation functions}},
  \href{http://dx.doi.org/10.1103/PhysRevD.80.074008}{\emph{Phys. Rev. D} {\bf
  80} (2009) 074008}, [\href{http://arxiv.org/abs/0906.5362}{{\tt 0906.5362}}].

\bibitem{Jakob:1997wg}
R.~Jakob, P.~J. Mulders and J.~Rodrigues, \emph{{Modeling quark distribution
  and fragmentation functions}},
  \href{http://dx.doi.org/10.1016/S0375-9474(97)00588-5}{\emph{Nucl. Phys. A}
  {\bf 626} (1997) 937--965}, [\href{http://arxiv.org/abs/hep-ph/9704335}{{\tt
  hep-ph/9704335}}].

\bibitem{Belitsky:1997ay}
A.~V. Belitsky, \emph{{Leading order analysis of twist-3 space- and time-like
  cut vertices in QCD}},
  \href{http://dx.doi.org/10.1142/S0217751X17300186}{\emph{Int. J. Mod. Phys.
  A} {\bf 32} (2017) 1730018}, [\href{http://arxiv.org/abs/hep-ph/9703432}{{\tt
  hep-ph/9703432}}].

\bibitem{Henneman:2001ev}
A.~A. Henneman, D.~Boer and P.~J. Mulders, \emph{{Evolution of transverse
  momentum dependent distribution and fragmentation functions}},
  \href{http://dx.doi.org/10.1016/S0550-3213(01)00557-0}{\emph{Nucl. Phys. B}
  {\bf 620} (2002) 331--350}, [\href{http://arxiv.org/abs/hep-ph/0104271}{{\tt
  hep-ph/0104271}}].

\bibitem{Kiselev:2025lpb}
V.~V. Kiselev, \emph{{Reciprocity in Heavy Quark Fragmentation Function}},
  \href{http://arxiv.org/abs/2511.17690}{{\tt 2511.17690}}.

\bibitem{Kartvelishvili:1977pi}
V.~G. Kartvelishvili, A.~K. Likhoded and V.~A. Petrov, \emph{{On the
  Fragmentation Functions of Heavy Quarks Into Hadrons}},
  \href{http://dx.doi.org/10.1016/0370-2693(78)90653-6}{\emph{Phys. Lett. B}
  {\bf 78} (1978) 615--617}.

\bibitem{Bentz:2016rav}
W.~Bentz, A.~Kotzinian, H.~H. Matevosyan, Y.~Ninomiya, A.~W. Thomas and
  K.~Yazaki, \emph{{Quark-Jet model for transverse momentum dependent
  fragmentation functions}},
  \href{http://dx.doi.org/10.1103/PhysRevD.94.034004}{\emph{Phys. Rev. D} {\bf
  94} (2016) 034004}, [\href{http://arxiv.org/abs/1603.08333}{{\tt
  1603.08333}}].

\bibitem{Politzer:1980me}
H.~D. Politzer, \emph{{Power Corrections at Short Distances}},
  \href{http://dx.doi.org/10.1016/0550-3213(80)90172-8}{\emph{Nucl. Phys. B}
  {\bf 172} (1980) 349--382}.

\bibitem{Pijlman:2006vm}
F.~Pijlman, \emph{{Single spin asymmetries and gauge invariance in hard
  scattering processes}},  other thesis, 1, 2006.

\bibitem{Bacchetta:2019qkv}
A.~Bacchetta, G.~Bozzi, M.~G. Echevarria, C.~Pisano, A.~Prokudin and M.~Radici,
  \emph{{Azimuthal asymmetries in unpolarized SIDIS and Drell-Yan processes: a
  case study towards TMD factorization at subleading twist}},
  \href{http://dx.doi.org/10.1016/j.physletb.2019.134850}{\emph{Phys. Lett. B}
  {\bf 797} (2019) 134850}, [\href{http://arxiv.org/abs/1906.07037}{{\tt
  1906.07037}}].

\bibitem{Vladimirov:2021hdn}
A.~Vladimirov, V.~Moos and I.~Scimemi, \emph{{Transverse momentum dependent
  operator expansion at next-to-leading power}},
  \href{http://dx.doi.org/10.1007/JHEP01(2022)110}{\emph{JHEP} {\bf 01} (2022)
  110}, [\href{http://arxiv.org/abs/2109.09771}{{\tt 2109.09771}}].

\bibitem{Bacchetta:2022awv}
{\scshape MAP (Multi-dimensional Analyses of Partonic distributions)}
  collaboration, A.~Bacchetta, V.~Bertone, C.~Bissolotti, G.~Bozzi, M.~Cerutti,
  F.~Piacenza et~al., \emph{{Unpolarized transverse momentum distributions from
  a global fit of Drell-Yan and semi-inclusive deep-inelastic scattering
  data}}, \href{http://dx.doi.org/10.1007/JHEP10(2022)127}{\emph{JHEP} {\bf 10}
  (2022) 127}, [\href{http://arxiv.org/abs/2206.07598}{{\tt 2206.07598}}].

\bibitem{Cahn:1978se}
R.~N. Cahn, \emph{{Azimuthal Dependence in Leptoproduction: A Simple Parton
  Model Calculation}},
  \href{http://dx.doi.org/10.1016/0370-2693(78)90020-5}{\emph{Phys. Lett. B}
  {\bf 78} (1978) 269--273}.

\bibitem{CLAS:2025asy}
{\scshape CLAS} collaboration, A.~Kripko et~al., \emph{{Multidimensional
  measurements of beam single-spin asymmetries in semi-inclusive deep-inelastic
  charged-kaon electroproduction off protons in the valence region}},
  \href{http://dx.doi.org/10.1103/2m4z-htrp}{\emph{Phys. Rev. C} {\bf 112}
  (2025) 055202}, [\href{http://arxiv.org/abs/2504.08580}{{\tt 2504.08580}}].

\bibitem{Tangerman:1994eh}
R.~D. Tangerman and P.~J. Mulders, \emph{{Intrinsic transverse momentum and the
  polarized Drell-Yan process}},
  \href{http://dx.doi.org/10.1103/PhysRevD.51.3357}{\emph{Phys. Rev. D} {\bf
  51} (1995) 3357--3372}, [\href{http://arxiv.org/abs/hep-ph/9403227}{{\tt
  hep-ph/9403227}}].

\bibitem{Mannheim:2020rod}
P.~D. Mannheim, P.~Lowdon and S.~J. Brodsky, \emph{{Comparing light-front
  quantization with instant-time quantization}},
  \href{http://dx.doi.org/10.1016/j.physrep.2020.09.001}{\emph{Phys. Rept.}
  {\bf 891} (2021) 1--65}, [\href{http://arxiv.org/abs/2005.00109}{{\tt
  2005.00109}}].

\bibitem{Signori:2016lvd}
A.~Signori, \emph{{Flavor and Evolution Effects in TMD Phenomenology}}.
\newblock PhD thesis, Vrije U., Amsterdam, 2016.

\bibitem{Buffing:2011mj}
M.~G.~A. Buffing and P.~J. Mulders, \emph{{Gauge links for transverse momentum
  dependent correlators at tree-level}},
  \href{http://dx.doi.org/10.1007/JHEP07(2011)065}{\emph{JHEP} {\bf 07} (2011)
  065}, [\href{http://arxiv.org/abs/1105.4804}{{\tt 1105.4804}}].

\bibitem{Manna:2021nep}
D.~Manna, A.~Signori and C.~Aidala, \emph{{Time reversal-odd effects in QCD and
  beyond}},
  \href{http://dx.doi.org/10.21468/SciPostPhysProc.8.169}{\emph{SciPost Phys.
  Proc.} {\bf 8} (2022) 169}, [\href{http://arxiv.org/abs/2107.13965}{{\tt
  2107.13965}}].

\bibitem{Collins:1989gx}
J.~C. Collins, D.~E. Soper and G.~F. Sterman, \emph{{Factorization of Hard
  Processes in QCD}},
  \href{http://dx.doi.org/10.1142/9789814503266_0001}{\emph{Adv. Ser. Direct.
  High Energy Phys.} {\bf 5} (1989) 1--91},
  [\href{http://arxiv.org/abs/hep-ph/0409313}{{\tt hep-ph/0409313}}].

\bibitem{Collins:1984kg}
J.~C. Collins, D.~E. Soper and G.~F. Sterman, \emph{{Transverse Momentum
  Distribution in Drell-Yan Pair and W and Z Boson Production}},
  \href{http://dx.doi.org/10.1016/0550-3213(85)90479-1}{\emph{Nucl. Phys. B}
  {\bf 250} (1985) 199--224}.

\bibitem{Becher:2014oda}
T.~Becher, A.~Broggio and A.~Ferroglia, \emph{{Introduction to Soft-Collinear
  Effective Theory}},
  \href{http://dx.doi.org/10.1007/978-3-319-14848-9}{\emph{Lect. Notes Phys.}
  {\bf 896} (2015) pp.1--206}, [\href{http://arxiv.org/abs/1410.1892}{{\tt
  1410.1892}}].

\bibitem{Grewal:2020hoc}
M.~Grewal, Z.-B. Kang, J.-W. Qiu and A.~Signori, \emph{{Predictive power of
  transverse-momentum-dependent distributions}},
  \href{http://dx.doi.org/10.1103/PhysRevD.101.114023}{\emph{Phys. Rev. D} {\bf
  101} (2020) 114023}, [\href{http://arxiv.org/abs/2003.07453}{{\tt
  2003.07453}}].

\bibitem{Bacchetta:2013pqa}
A.~Bacchetta and A.~Prokudin, \emph{{Evolution of the helicity and transversity
  Transverse-Momentum-Dependent parton distributions}},
  \href{http://dx.doi.org/10.1016/j.nuclphysb.2013.07.013}{\emph{Nucl. Phys. B}
  {\bf 875} (2013) 536--551}, [\href{http://arxiv.org/abs/1303.2129}{{\tt
  1303.2129}}].

\bibitem{Echevarria:2011epo}
M.~G. Echevarria, A.~Idilbi and I.~Scimemi, \emph{{Factorization Theorem For
  Drell-Yan At Low $q_T$ And Transverse Momentum Distributions
  On-The-Light-Cone}},
  \href{http://dx.doi.org/10.1007/JHEP07(2012)002}{\emph{JHEP} {\bf 07} (2012)
  002}, [\href{http://arxiv.org/abs/1111.4996}{{\tt 1111.4996}}].

\bibitem{Scimemi:2019cmh}
I.~Scimemi and A.~Vladimirov, \emph{{Non-perturbative structure of
  semi-inclusive deep-inelastic and Drell-Yan scattering at small transverse
  momentum}}, \href{http://dx.doi.org/10.1007/JHEP06(2020)137}{\emph{JHEP} {\bf
  06} (2020) 137}, [\href{http://arxiv.org/abs/1912.06532}{{\tt 1912.06532}}].

\bibitem{Qiu:2000hf}
J.-w. Qiu and X.-f. Zhang, \emph{{Role of the nonperturbative input in QCD
  resummed Drell-Yan $Q_{T}$ distributions}},
  \href{http://dx.doi.org/10.1103/PhysRevD.63.114011}{\emph{Phys. Rev. D} {\bf
  63} (2001) 114011}, [\href{http://arxiv.org/abs/hep-ph/0012348}{{\tt
  hep-ph/0012348}}].

\bibitem{Parisi:1979se}
G.~Parisi and R.~Petronzio, \emph{{Small Transverse Momentum Distributions in
  Hard Processes}},
  \href{http://dx.doi.org/10.1016/0550-3213(79)90040-3}{\emph{Nucl. Phys. B}
  {\bf 154} (1979) 427--440}.

\bibitem{ParticleDataGroup:2016lqr}
{\scshape Particle Data Group} collaboration, C.~Patrignani et~al.,
  \emph{{Review of Particle Physics}},
  \href{http://dx.doi.org/10.1088/1674-1137/40/10/100001}{\emph{Chin. Phys. C}
  {\bf 40} (2016) 100001}.

\bibitem{Bacchetta:2024qre}
{\scshape MAP (Multi-dimensional Analyses of Partonic distributions)}
  collaboration, A.~Bacchetta, V.~Bertone, C.~Bissolotti, G.~Bozzi, M.~Cerutti,
  F.~Delcarro et~al., \emph{{Flavor dependence of unpolarized quark transverse
  momentum distributions from a global fit}},
  \href{http://dx.doi.org/10.1007/JHEP08(2024)232}{\emph{JHEP} {\bf 08} (2024)
  232}, [\href{http://arxiv.org/abs/2405.13833}{{\tt 2405.13833}}].

\bibitem{Bacchetta:2025ara}
{\scshape MAP (Multi-dimensional Analyses of Partonic distributions)}
  collaboration, A.~Bacchetta, V.~Bertone, C.~Bissolotti, M.~Cerutti,
  M.~Radici, S.~Rodini et~al., \emph{{Neural-Network Extraction of Unpolarized
  Transverse-Momentum-Dependent Distributions}},
  \href{http://dx.doi.org/10.1103/csc2-bj91}{\emph{Phys. Rev. Lett.} {\bf 135}
  (2025) 021904}, [\href{http://arxiv.org/abs/2502.04166}{{\tt 2502.04166}}].

\bibitem{Moos:2023yfa}
V.~Moos, I.~Scimemi, A.~Vladimirov and P.~Zurita, \emph{{Extraction of
  unpolarized transverse momentum distributions from the fit of Drell-Yan data
  at N$^{4}$LL}}, \href{http://dx.doi.org/10.1007/JHEP05(2024)036}{\emph{JHEP}
  {\bf 05} (2024) 036}, [\href{http://arxiv.org/abs/2305.07473}{{\tt
  2305.07473}}].

\bibitem{Moos:2025sal}
V.~Moos, I.~Scimemi, A.~Vladimirov and P.~Zurita, \emph{{Determination of
  unpolarized TMD distributions from the fit of Drell-Yan and SIDIS data at
  N$^{4}$LL}}, \href{http://dx.doi.org/10.1007/JHEP11(2025)134}{\emph{JHEP}
  {\bf 11} (2025) 134}, [\href{http://arxiv.org/abs/2503.11201}{{\tt
  2503.11201}}].

\bibitem{Accardi:2023chb}
A.~Accardi et~al., \emph{{Strong interaction physics at the luminosity frontier
  with 22 GeV electrons at Jefferson Lab}},
  \href{http://dx.doi.org/10.1140/epja/s10050-024-01282-x}{\emph{Eur. Phys. J.
  A} {\bf 60} (2024) 173}, [\href{http://arxiv.org/abs/2306.09360}{{\tt
  2306.09360}}].

\bibitem{Adams:2676885}
B.~Adams et~al., \emph{{COMPASS++/AMBER: Proposal for Measurements at the M2
  beam line of the CERN SPS Phase-1: 2022-2024}}.
\newblock Proposal for measurements -
  \href{https://cds.cern.ch/record/2676885}{hyperlink to file}, 2019.

\bibitem{Ninomiya:2017ggn}
Y.~Ninomiya, W.~Bentz and I.~C. Clo{\"e}t, \emph{{Transverse-momentum-dependent
  quark distribution functions of spin-one targets: Formalism and covariant
  calculations}},
  \href{http://dx.doi.org/10.1103/PhysRevC.96.045206}{\emph{Phys. Rev. C} {\bf
  96} (2017) 045206}, [\href{http://arxiv.org/abs/1707.03787}{{\tt
  1707.03787}}].

\bibitem{Mannheim:2019lss}
P.~D. Mannheim, P.~Lowdon and S.~J. Brodsky, \emph{{Structure of light front
  vacuum sector diagrams}},
  \href{http://dx.doi.org/10.1016/j.physletb.2019.134916}{\emph{Phys. Lett. B}
  {\bf 797} (2019) 134916}, [\href{http://arxiv.org/abs/1904.05253}{{\tt
  1904.05253}}].

\end{thebibliography}\endgroup

\end{document}